\documentclass[11pt,a4paper]{article}
\usepackage{jheppub} 

\usepackage{tikz}
\usepackage[compat=1.1.0]{tikz-feynman}

\usepackage{mathtools}
\usepackage{bbm}
\usepackage{bm}
\usepackage{dsfont}
\usepackage{physics}
\usepackage{comment}
\usepackage{enumitem}
\setitemize[1]{leftmargin=*}
\setenumerate[1]{leftmargin=*}
\setdescription[1]{leftmargin=*}
\usepackage{subcaption}
\usepackage{booktabs}
\usepackage{acronym}
\usepackage{todonotes}

\newcommand{\hk}[1]{\textcolor{red}{[{\bf HK}: #1]}} % Helena

\newcommand{\dk}[1]{\textcolor{red}{[{\bf DK}: #1]}} % Daniil

\newcommand{\MLO}[1]{M^2_{\text{LO}#1}}
\newcommand{\MNLO}[1]{M^2_{\text{NLO}#1}}
\newcommand{\FLO}[1]{F_{\text{LO}#1}}
\newcommand{\FNLO}[1]{F_{\text{NLO}#1}}

\renewcommand{\sec}{section~}

\newcommand{\refe}{ref.~}
\newcommand{\refes}{refs.~}
\newcommand{\Refe}{Ref.~}

\newcommand{\eq}{eq.~}
\newcommand{\eqs}{eqs.~}
\newcommand{\Eq}{Eq.~}

\newcommand{\fig}{figure~}
\newcommand{\figs}{figures~}
\newcommand{\Fig}{Figure~}
\newcommand{\app}{appendix~}

\newcommand{\tableTag}{table~}

\newcommand{\spth}{$SU(4)/Sp(4)$ }
\newcommand{\soth}{$SU(4)/SO(4)$ }

\DeclareMathAlphabet\mathbfcal{OMS}{cmsy}{b}{n}

\title{NLO observables for QCD-like theories and application to pion dark matter}
\author[a]{Helena Kole\v{s}ov\'{a},}
\author[a]{Daniil Krichevskiy}
\author[b]{and Suchita Kulkarni}
\affiliation[a]{Department of Mathematics and Physics, 
University of Stavanger\\
Kristine Bonnevies vei 22, 4021 Stavanger, Norway
}
\affiliation[b]{Institute of Physics, NAWI Graz, University of Graz,\\ Universitätsplatz 5, 8010 Graz, Austria}
\emailAdd{helena.kolesova@uis.no}
\emailAdd{daniil.krichevskiy@uis.no}

\emailAdd{suchita.kulkarni@uni-graz.at}
\abstract{

QCD-like theories are of interest in various areas of beyond-Standard-Model phenomenology, including composite Higgs models or pionic dark matter. The effective field theories provide a framework for describing the dynamics of such strongly coupled gauge theories at low energies. In this work, we present next-to-leading order (NLO) expressions for masses, condensates, decay constants, and scattering amplitudes in the chiral expansion of QCD-like theories with $N_F=2$ fermions with non-degenerate masses in both real and pseudoreal representations of the gauge group. %These results offer a systematic approach for analyzing the impact of NLO corrections in such theories. 
We further apply the NLO formulas to fit existing lattice spectroscopic and scattering data for $Sp(N_c=4)$ gauge theory with $N_F=2$ fermions in fundamental representation, extracting the NLO low-energy constants of the theory. Using these fits, we refine the NLO formulas describing the $2\to 2$ pion self-interactions and confirm that the NLO contributions play a crucial role in determining the viable parameter space of pion dark matter scenarios like the strongly interacting massive particles (SIMP).
}
\keywords{Spontaneous Symmetry Breaking, Chiral Lagrangian, Models for Dark Matter}
\begin{document}
\maketitle
%%%%%

% https://jhep.sissa.it/jhep/help/JHEP/TeXclass/DOCS/JHEP-author-manual.pdf}{JHEP requirements 
%\begin{document}
%%%%%%%%%%%%%%%%%%%%%%%%%%%%%%%%%%%%%%%%%%

\section{Introduction}\label{sec:intro}

Although the Standard Model (SM) of particle physics is an extremely successful framework, it leaves several fundamental questions unanswered. Among these are the nature of dark matter in our Universe~\cite{Bertone:2004pz} or the so-called hierarchy problem~\cite{Gildener:1976ai} that raises the question of why the Higgs boson mass is much smaller than the Planck scale, despite quantum corrections that tend to drive it higher. One compelling approach to address the latter issue is to postulate that the Higgs is not an elementary particle, but rather a composite state arising from a new strongly interacting sector (see~\cite{Weinberg:1975gm,Susskind:1978ms,Kaplan:1983fs,Kaplan:1983sm} for early works or~\cite{Cacciapaglia:2020kgq} for a recent review). Furthermore, different beyond-Standard-Model (BSM) strongly coupled frameworks may give rise to dark matter candidates\footnote{Let us add as a curiosity that a strongly coupled dark matter candidate, the sexaquark, was considered even within SM~\cite{Farrar:2017eqq}, however, its viability region seems to be strongly narrowed down~\cite{Moore:2024mot}.} (see~\cite{
%review:
Kribs:2016cew,Cline:2021itd} for reviews) or might be even related to cosmological inflation (see, e.g., examples in~\cite{Aoki:2021skm,Cacciapaglia:2023kat}).
%in principle, a review on lattice results for BSM strongly coupled theories is also in Lee:2024xhb
%other pion/baryon DM:
%Hur:2007uz,Hur:2011sv,Frandsen:2011kt,Buckley:2012ky,Bai:2013xga,Cline:2013zca,Antipin:2015xia,GarciaGarcia:2015fol,Appelquist:2015yfa,Dienes:2016vei,Lonsdale:2017mzg,Davoudiasl:2017zws,Beauchesne:2019ato,Mies:2020mzw,Cheng:2021kjg,Butterworth:2021jto,Carmona:2024tkg,
%SIMP: 
%Hochberg:2014kqa,Hansen:2015yaa,Lee:2015gsa,Hochberg:2015vrg,Kamada:2017tsq,Hochberg:2018rjs,Hochberg:2018vdo,Berlin:2018tvf,Choi:2018iit,Katz:2020ywn,Kulkarni:2022bvh,Zierler:2022uez,Bernreuther:2023kcg,Braat:2023fhn,Pomper:2024otb,
%glueballs
%Boddy:2014yra,Forestell:2016qhc,Soni:2016gzf,Gross:2020zam,Carenza:2022pjd,Carenza:2023eua,Biondini:2024cpf,McKeen:2024trt} or might be even related to cosmological inflation~\cite{Aoki:2021skm,Cacciapaglia:2023kat}. %\hk{The last reference is rather a curiosity, I don't insist on including this, perhaps other more relevant references should be added here.}

In general, the theoretical description of strongly coupled theories is challenging, often the only possible first-principle treatment is enabled by numerical lattice calculations. An important exception to this rule is the case of low-energy interactions of the \mbox{(pseudo-)Goldstone} bosons appearing in the theory spectrum due to spontaneous breaking of a global symmetry. In SM Quantum Chromodynamics (QCD), such particles correspond to pions that arise from the breaking of the approximate $SU(2)_L \times SU(2)_R$ flavor symmetry transforming independently the two flavors of the left-handed and right-handed light quarks to its vector subgroup $SU(2)_V$ by the QCD vacuum.  The pattern of the spontaneous symmetry breaking then uniquely determines the shape of the low-energy effective field theory (EFT) describing the pion interactions~\cite{Callan:1969sn,Coleman:1969sm,Weinberg:1978kz,Gasser:1983yg,Gasser:1984gg,Pich:1995bw,Scherer:2012xha,Kogut:2000ek} which enables analytic calculations. Also in the context of BSM theories, low-energy interactions of dark pions are often relevant for the phenomenology studies. For example, when treating the pion dark matter freeze out, EFT description is typically sufficient~\cite{Kribs:2016cew,Cline:2021itd}.
%,Hansen:2015yaa,Lee:2015gsa,Hochberg:2015vrg,Kamada:2017tsq,Hochberg:2018rjs,Hochberg:2018vdo,Berlin:2018tvf,Choi:2018iit,Katz:2020ywn,Kulkarni:2022bvh,Zierler:2022uez,Bernreuther:2023kcg,Braat:2023fhn,Pomper:2024otb}. %Motivated by this fact, we calculate different observables related to dark pions within the EFT approach.

While SM QCD is based on the $SU(N_c=3)$ gauge group (with $N_c$ denoting the number of colors) and quarks in the fundamental representation, BSM strongly coupled theories might be based on other gauge groups or fermion representations. In particular, if dark quarks are in real or pseudoreal representations of the gauge group, the flavor symmetry among $N_F$ dark Dirac fermions is enlarged from $SU(N_F)_L \times SU(N_F)_R$ to $SU(2N_F)$. At low energies, this symmetry is spontaneously broken by the emergence of the quark condensate to $SO(2N_F)$ or $Sp(2N_F)$ in the real or pseudoreal case, respectively. Consequently, the low-energy EFT has to be modified compared to the QCD case where quarks are in a complex representation~\cite{Kogut:2000ek}. Dark quarks in (pseudo)real representations are relevant, e.g., for the minimal setup of the pion-dark-matter candidate usually dubbed ``strongly interacting massive particle'' (SIMP)~\cite{Hochberg:2014kqa} or for many composite Higgs models~\cite{Ferretti:2013kya, Witzel:2019jbe,Bellazzini:2014yua,Cacciapaglia:2020kgq}. Finally, let us remark that the original motivation to build an EFT for QCD-like theories~\cite{Kogut:2000ek} was related to the fact that for theories with quarks in (pseudo)real representations, the so-called ``sign problem''~\cite{Nagata:2021ugx} is absent. Consequently, lattice simulations can be performed also at finite chemical potential (see, e.g.,~\cite{Cotter:2012mb} for one of many examples), which is problematic for the case of real-world QCD.
%\dk{In real world QCD there also exist some more complicated methods which try to overcome this. Now it sounds like it is completely impossible... }

%Let us also remark that the QCD-like theories play an important role for the studies of the QCD phase diagram since if the quarks are in (pseudo)real representations, the so-called ``sign problem''~\cite{Nagata:2021ugx} is absent and lattice simulations can be performed also at finite chemical potential, unlike for the case of real-world QCD [references].  

In this work, we focus on the low-energy EFT for the case of (pseudo)real quarks mentioned above and we give the expressions for the quark condensates and pion masses, decay constants, and scattering amplitudes including the next-to-leading order (NLO) corrections. Since the EFT expansion parameter is proportional to dark-pion mass, higher-order corrections are important for the BSM scenarios with relatively large pion masses. Indeed, NLO corrections were shown to be crucial, e.g., for the phenomenology of SIMP dark matter~\cite{Hansen:2015yaa}. Moreover, higher-order corrections are necessary for interpolation between the limit of massless quarks that is most relevant for the composite Higgs models and the finite-mass regime where lattice simulations can be performed. Existing literature \cite{Bijnens:2009qm,Bijnens:2011fm} covers the EFT for (pseudo)real quarks at NLO and even next-to-next-to-leading order (NNLO); we extend the NLO results by assuming non-degenerate quark masses. This is motivated, e.g., for the SIMP-dark-matter scenarios where the viable parameter space might be substantially enlarged if different pion species have different masses~\cite{Hochberg:2014kqa}. Let us note that NLO corrections play an important role for the description of the pion mass splitting, as in some cases leading-order (LO) EFT predicts the pion masses to be degenerate even for split quark masses and the pion mass splitting is captured only by NLO corrections.

%We restrict ourselves to the most relevant case of two Dirac flavors, $N_F=2$. %\hk{Do we have some better further argument here? Larger $N_F$ would bring long formulas, but nothing genuinely new? Too many pions for larger $N_F$? Perhaps move the restriction $N_F=2$ after we speak about lattice?} Let us note that NLO corrections play also an important role for the pion mass splitting in the pseudoreal case, since even for distinct dark quark masses, the dark pion masses are predicted to be equal by the EFT at leading order (LO).

While we argue that the NLO corrections might be important for a reliable description of the BSM theories, the disadvantage is that new low-energy constants (LECs) need to be introduced at this order of EFT expansion. These constants differ for different gauge groups and fermion representations and their ignorance introduces additional uncertainty in the phenomenology results (see, e.g., the example of the NLO SIMP study~\cite{Hansen:2015yaa}). 
In the QCD case, the LEC can be determined from experiments~\cite{Colangelo:2001df}, however, for the BSM theories the determination of these constants relies only on lattice calculations. Since these are computationally expensive, the results for BSM theories started appearing only recently, e.g., part of the LECs for the case of $SU(N_c=2)\cong Sp(N_c=2)$ gauge theory with $N_F=2$ of quarks in fundamental representation were determined in Ref.~\cite{Arthur:2016dir}. In our work, we make use of the new lattice data presented in Refs.~\cite{Kulkarni:2022bvh,Dengler:2024maq} and we perform a fit of LEC for another example of a theory with pseudoreal quarks, namely, the $Sp(N_c=4)$ gauge theory with $N_F=2$ quarks in fundamental representation.\footnote{For our NLO results, we restrict ourselves to the minimal case of $N_F=2$ Dirac flavors that is most relevant, e.g., for dark matter phenomenology~\cite{Hochberg:2014kqa} and where the lattice data for the mass-split case are available~\cite{Kulkarni:2022bvh}. Lattice results for $SU(N_c=2)$ gauge theory with larger number of flavors were presented in~\cite{Amato:2015dqp}, but these consider the mass-degenerate case only.} This allows us to fully predict the non-relativistic pion scattering cross sections at NLO within the EFT framework that is an important input for dark-matter phenomenology studies: these dark-matter self-interactions may, on one hand, potentially explain the small-scale structure puzzles~\cite{Tulin:2017ara,Zhang:2025bju}, on the other hand, they are constrained by astrophysical observations~\cite{Randall:2008ppe,Robertson:2016xjh,Wittman:2017gxn}. Let us note that the LECs obtained in our analysis are applicable both to mass-degenerate and non-degenerate case.
%\dk{I suggest adding here the recent reference which supports the self interacting DM: \cite{Zhang:2025bju}}.

This text is structured as follows. In section~\ref{sec:EFT} we review the EFT construction while the NLO results for the case of pseudoreal and real representations are presented in sections~\ref{sec:Sp} and~\ref{sec:SO}, respectively. Subsequently, the values of the NLO LEC are obtained based on fits of lattice data in \sec\ref{sec:fit} and these results are applied in the context of pion dark matter in \sec\ref{sec:Sp4.Applications}. Finally, we present our conclusions and outlook for future work in \sec\ref{sec:conclusions} and we leave several technical details to appendices.

\section{Effective field theory}\label{sec:EFT}
In this section, we briefly summarize the construction of the EFT describing the low-energy interactions of dark pions mainly following \cite{Bijnens:2009qm}. For a thorough discussion of the Chiral EFT construction see e.g. \cite{Scherer:2012xha} or \cite{Kogut:2000ek} and for the discussion of Callan-Coleman-Wess-Zumino (CCWZ) scheme in different contexts, e.g.   \cite{Brauner:2024juy} and \cite{Arbuzov:2019rcl}.

\subsection{UV theory}
As a starting point for our discussion, let us consider a UV gauge theory with $N_F$ flavors of left-handed fermions and $N_F$ flavors of right-handed fermions (in total $2N_F$ Weyl degrees of freedom) transforming in a real or pseudoreal representation of the gauge group. Examples of such theories are the cases where the fermions transform in the fundamental representations of the $SO(N_c)$ or $Sp(N_c)$ gauge groups with $N_c$ being the number of colors.
The Lagrangian, enhanced by an external source, namely a (pseudo)scalar field $\mathcal{M}$ reads
\begin{align}\label{eq.EFT1}
\begin{split}
\mathcal{L} =& \overline{q}_{Li} i \gamma^\mu D_\mu q_{Li} + \overline{q}_{Ri} i \gamma^\mu D_\mu q_{Ri}
- \overline{q}_{Ri} \mathcal{M}_{ij} q_{Lj} - \overline{q}_{Li} \mathcal{M}_{ij}^\dagger q_{Rj},
\end{split}
\end{align}
where we assume summation over the flavor indices $i,j=1\ldots N_F$. In the case of massive fermions, $\mathcal{M}$ can play a role of a mass-matrix. The covariant derivative reads $D_{\mu }q=\partial_{\mu } q-iG^{a}_{\mu }t^{a}$, where $G_\mu^a$ are the gauge fields and $t^{a}$ are the Hermitian generators of the gauge group taken in the representation under which the fermions transform. The fermion fields $q_R$ and $q_L$ are vectors  in flavor space. The color and spinor indices are suppressed.

For (pseudo)real fermion representation there exists an (anti-)symmetric unitary matrix $\epsilon$ such that
\begin{align}\label{eq.EFT2}
\begin{split}
-t^{a\ast }=\epsilon^{-1}t^{a}\epsilon, \quad \epsilon^T=\eta\epsilon,
\end{split}
\end{align}
where $\eta$ is $(-)1$ for (pseudo)real representation.
As in~\cite{Kogut:2000ek,Bijnens:2009qm}, we define a new object that transforms as a right-handed spinor under Lorenz transformations and as the same representation of the gauge group as $q_L$:
\begin{align}\label{eq.EFT3}
\begin{split}
\tilde{q}_{R} =\epsilon C\bar{q}^{T}_{L},
\end{split}
\end{align}
where $C=i\gamma^{2} \gamma^{0}$ is a matrix acting in the Dirac space.  

Using the property~\eqref{eq.EFT2}, 
performing integration by parts, and taking into account that the fermionic fields are Grassmann-valued, the Lagrangian \eqref{eq.EFT1} can be rewritten as\footnote{The transpose operation in the first instance of $\hat{q}$ in the second %third - TODO: Daniil, please check!
term is an amendment to~\cite{Bijnens:2009qm}, where this transpose sign is absent.}
\begin{align}\label{eq.EFT4}
\begin{split}
\mathcal{L} =\bar{\hat{q} } i\gamma^{\mu } D_{\mu }\hat{q} -\frac{\eta }{2} \hat{q}^{T} \hat{\mathcal{M} }^\dagger \epsilon^{\ast } C\hat{q} -\frac{1}{2} \bar{\hat{q} } C\epsilon \hat{\mathcal{M} } \bar{\hat{q} }^{\hspace{0.1em}T}, 
\end{split}
\end{align}
where we used a vector $\hat{q}$ with $2N_F$ flavor components and a $2N_F\times2N_F$ matrix in the flavor space $\hat{\mathcal{M}}$ is defined as   
\begin{align}\label{eq.EFT5}
\begin{split}
\hat{q} =\begin{pmatrix}q_{R}\\ \tilde{q}_{R} \end{pmatrix} ,\quad  \hat{\mathcal{M} } =\left( \begin{matrix}0&\eta\mathcal{M} \\ \mathcal{M}^{T} &0\end{matrix} \right).  
\end{split}
\end{align}

In the absence of $\hat{\mathcal{M}}$, the classical Lagrangian is symmetric with respect to global $U(2N_F)$ transformations $\hat{q}\to g\hat{q}$.\footnote{The symmetry is enlarged in comparison to the standard QCD case with fermions in complex representation, where the flavor symmetry on the classical level is is $U(N_F)_L\times U(N_F)_R$.} In quantum theory the axial anomaly reduces the global symmetry group down to $SU(2N_F)$. To generate correlation functions with current insertions, $j^{a}_{\mu }=\bar{\hat{q} } \gamma_{\mu } T^{a}\hat{q} $, where $T^a$ are the Hermitian generators of $SU(2N_F)$, we can enhance the Lagrangian \eqref{eq.EFT4} with external vector fields which couple to the currents:
\begin{align}\label{eq.EFT4.1}
\begin{split}
\mathcal{L} =\bar{\hat{q} } i\gamma^{\mu } D_{\mu }\hat{q} -\frac{\eta }{2} \hat{q}^{T} \hat{\mathcal{M} }^{\dagger } \epsilon^{\ast } C\hat{q} -\frac{1}{2} \bar{\hat{q} } C\epsilon \hat{\mathcal{M} } \bar{\hat{q} }^{\hspace{0.1em}T} +\bar{\hat{q} } \gamma^{\mu } V_{\mu }\hat{q} ,
\end{split}
\end{align}
where $V_{\mu }=V^{a}_{\mu }T^{a}$. In order to make this Lagrangian invariant under the $SU(2N_F)$ symmetry, also the external fields $V_\mu$ and $\hat{\mathcal{M} }$ need to transform under this symmetry. This symmetry can even be made local if we impose:
\begin{align}\label{eq.EFT6}
\begin{split}
V_{\mu }\rightarrow gV_{\mu }g^{\dagger }+ig\partial_{\mu } g^{\dagger },\quad \hat{\mathcal{M} } \rightarrow g\hat{\mathcal{M} } g^{T}, \quad g\in SU(2N_F). 
\end{split}
\end{align}

\subsection{IR description}

\subsubsection{Symmetry breaking}
We assume that at low energies, the vacuum is characterized by non-zero quark condensate which is an order parameter for the spontaneous symmetry breaking of chiral symmetry: 
\begin{align}\label{eq.EFT7}
\begin{split}
\left< \bar{q} q\right>  \equiv \sum^{N_{F}}_{i=1} \left< \bar{q}_{Li} q_{Ri}+\bar{q}_{Ri} q_{Li}\right>  = \left< \frac{1}{2} \hat{q}^{T} J\epsilon^{\ast } C\hat{q} +\frac{1}{2} \bar{\hat{q} } C\epsilon J\bar{\hat{q}}^{\hspace{0.1em}T} \right>  
\neq0,
\end{split}
\end{align}
with
\begin{align}\label{eq.EFT8}
\begin{split}
J=\begin{pmatrix}0&\eta \mathbf{1}_{N_{F}} \\ \mathbf{1}_{N_{F}} &0\end{pmatrix}, 
\end{split}
\end{align}
where $\mathbf{1}_{N_{F}}$ is an $N_F\times N_F$ unit matrix. Using the notation of \eq\eqref{eq.EFT5}, the quark condensate for individual flavor combinations can be written as:
\begin{align}\label{eq.EFT9}
\begin{split}
\langle \hat{q}^{T}_{i} \epsilon^{\ast } C\hat{q}_{j} \rangle =\frac{\left< \bar{q}_{L} q_{R}\right>  }{N_{F}} J_{ij}.
\end{split}
\end{align}

The condensate is only invariant under a subgroup of $SU(2N_F)$ such that
\begin{align}\label{eq.EFT10}
\begin{split}
J=h^{T}Jh,\quad h\in H\subset G=SU\left( 2N_{F}\right)  ,\quad \begin{cases}H=SO\left( 2N_{F}\right)  ,&\eta =1\\ H=Sp\left( 2N_{F}\right)  ,&\eta =-1\end{cases}. 
\end{split}
\end{align}
For real representations, the Goldstone manifold is given by $SU(2N_F)/SO(2N_F)$ with $N_F(2N_F+1)-1$ Goldstone modes, while for pseudoreal representations, the coset space is $SU(2N_F)/Sp(2N_F)$ with $N_F(2N_F-1)-1$ independent modes. In the $N_F=2$ case relevant for this paper, this corresponds to $N_\pi=9$ pions in the real case and $N_\pi=5$ in the pseudoreal one.
Using \eqref{eq.EFT9} we can separate the generators of the group $G$ into broken part $X^a$ and the unbroken part $Q^a$ (see \app\ref{appendixSU4gen} for the explicit form of the generators in the $N_F=2$ case):
\begin{align}\label{eq.EFT10b}
\begin{split}
JQ^{a}+Q^{aT}J=0,\quad JX^{a}-X^{aT}J=0.
\end{split}
\end{align}
%%%%

Note that the symmetry is explicitly broken in case of massive fermions. If we consider a local infinitesimal transformation with $g\approx 1+i\theta^{a} \left( x\right)  T^{a}$, then by virtue of the Gell-Mann-Levi method, we can find the expression for the divergence of the currents $j^{a}_{\mu }$:
\begin{align}\label{eq.EFT6.1}
\begin{split}
\partial^{\mu } j^{a}_{\mu }  =\frac{i}{2} \left( \hat{q}^{T} \left( T^{aT}\hat{\mathcal{M} } +\hat{\mathcal{M} } T^{a}\right)  C\epsilon^{\ast } \hat{q} -\  \bar{\hat{q} } \left( T^{a}\hat{\mathcal{M} } +\hat{\mathcal{M} } T^{aT}\right)  C\epsilon \bar{\hat{q} }^{T} \right). 
\end{split}
\end{align}
From here it is evident that for a mass-degenerate fermions with mass $m$, i.e., $\mathcal{M}=m\mathbf{1}_{N_{F}}$ and $\hat{\mathcal{M}}=mJ$, the generators are again split in the same subsets of broken and unbroken ones.

The EFT for the Goldstone modes can be written in terms of a rotated vacuum $U$:
\begin{align}\label{eq.EFT11}
\begin{split}
U=uJu^T=u^2J\to gUg^T,
\end{split}
\end{align}
where $u$ is the parametrization of the coset space 
\begin{align}\label{eq.EFT12}
\begin{split}
u=e^{\frac{i}{\sqrt{2} F} \pi^{a} X^{a}} \in G/H.
\end{split}
\end{align}
Here pion decay constant $F$ with mass dimension $1$ is introduced in order to make the pseudoscalar Goldstone fields $\pi^a$ dimensionful. The fields $\pi^a$ will interchangeably be referred to as dark pions or dark mesons in this text, and the adjective 
``dark'' will often be omitted for brevity. 
The generators are normalized as $\left< X^{a}X^{b}\right>  =\delta^{ab} $, where $\left< \ldots \right> $ denotes a trace in the flavor space.\footnote{In our conventions the decay constant for QCD pions would have a value $F_\pi\approx93 \ \text{MeV}$. }

In CCWZ construction \cite{Coleman:1969sm, Callan:1969sn} matrix $u$ is directly used to build the Lagrangian. Under the broken group $G$ it is transformed non-linearly as $u\rightarrow guh$, where $h \in H$ is, in general, a nonlinear function of $u$ and $g$. However, given the transformation rule \eqref{eq.EFT6}, one can construct term which is, in fact, an element of the Lie algebra of the group $G$ and thus can be expanded in broken and unbroken parts:
\begin{align}\label{eq.EFT13}
\begin{split}
u^\dagger \left( \partial_\mu - i V_\mu \right) u \equiv \Gamma_\mu - \frac{i}{2} u_\mu, 
\quad 
\Gamma_\mu = \Gamma_\mu^a Q^a, 
\quad 
u_\mu = u_\mu^a X^a.
\end{split}
\end{align}
In our work, the unbroken part $u_\mu$ is used to construct the effective Lagrangian for pions as in \cite{Bijnens:2009qm}. It can be read off from \eqref{eq.EFT13} using \eqref{eq.EFT10}:
\begin{align}\label{eq.EFT14}
\begin{split}
 u_{\mu }&=i\left( u^{\dagger }\left( \partial_{\mu } -iV_{\mu }\right)  u-u\left( \partial_{\mu } +iJV^{T}_{\mu }J^{T}\right)  u^{\dagger }\right) .
\end{split}
\end{align}
Another building block for the ChPT Lagrangian are the spurion fields $\chi_\pm$:
\begin{align}
\chi_{\pm } &=u^{\dagger }\chi J^{T}u^{\dagger }\pm uJ\chi^{\dagger } u\,, \label{eqEFT16}
\end{align}
where $\chi=2B_0\hat{\mathcal{M}}$. $B_0$ is a low-energy constant with dimension of mass whose physical interpretation — relating it to the quark condensate — is given in \sec\ref{secCondensatesSP(4)}. Objects \eqref{eq.EFT14}-\eqref{eqEFT16} transform homogeneously under $G$, i.e., $u_{\mu }\rightarrow hu_{\mu }h^{\dagger }$ and $\chi_{\pm }\rightarrow h\chi_{\pm }h^{\dagger }$. Note that $u_\mu$ and $\chi_{-}$ are constructed to have negative parity, while $\chi_{+}$ has positive parity.

\subsubsection{Power counting}
Our goal is to describe the interactions of dark pions at low energies, hence, we assume that the pion momenta $p$ (and also masses $M_\pi$, see below) are much lower than the cutoff scale $4\pi F_\pi$ determined by the pion decay constant. This allows us to introduce a consistent expansion scheme. Since the powers of momenta, e.g., in the scattering amplitude are related to the number of derivatives in the corresponding interaction Lagrangian, the effective Lagrangian for the Goldstone bosons is organized by the number of derivatives which in 4-dimensional space-time can only be even:
\begin{align}\label{eq.EFT17}
\begin{split}
\mathcal{L}_\text{ChPT} = \sum^{\infty}_{k=1} \mathcal{L}_{2k} \left[ \pi \right]\,.
\end{split}
\end{align}
When calculating different observables at given order in chiral expansion, the contributions of different Feynman diagrams can then be classified in the following way (see, e.g.,~\cite{Brauner:2024juy} for details).

In the chiral limit, a diagram $\Gamma$ evaluates to a homogeneous function of the external momentum $\sim\mathcal{O}(p^{\text{deg}\,\Gamma})$, where $\text{deg}\,\Gamma$ is the degree of the function (or the so-called chiral dimension). Its value is calculated as
\begin{align}\label{eq.EFT18}
\begin{split}
 \text{deg}\,\Gamma=2+\left( n-2\right)  N_{L}+\sum^{\infty }_{k=1} 2\left( k-1\right)  N_{2k},
\end{split}
\end{align}
where $n$ is the number of space-time dimensions, $N_L$ is the number of independent loops, $N_{2k}$ is the number of vertices originating from $\mathcal{L}_{2k}$. Higher degree implies stronger suppression at low energies. At LO~$\equiv \mathcal{O}(p^2)$, one calculates the diagrams with $\text{deg}\,\Gamma = 2$ while at NLO~$\equiv \mathcal{O}(p^4)$, also the diagrams with $\text{deg}\,\Gamma = 4$ contribute.

When explicit symmetry breaking by fermion masses is introduced, the pions are no longer massless and their masses appear in the propagators. Consequently, the diagrams are no longer homogeneous functions of the external momenta. However, small fermion masses $m_q$  can be assigned chiral dimension 2, which leads to $\text{deg}\,M_\pi=1$ as $M_\pi\propto\sqrt{m_q}$. This allows for inclusion of the terms with mass insertions to $\mathcal{L}_\text{ChPT}$ in a consistent way.

As the external vector fields appear in the covariant derivative (see \eqref{eq.EFT14}), chiral dimension $\text{deg}\,V_\mu=1$ is assigned to it. In summary, we arrive at $\text{deg}\,\chi_\pm = 2$, and $\text{deg} \, u_\mu=1$ that allows us to build the LO and NLO Lagrangians in the next section. 

\subsubsection{LO and NLO Lagrangians}
At each order, all terms consistent with chiral symmetry, Lorentz invariance, and other relevant symmetries are included in the Lagrangian. Based on the power counting mentioned in the previous section, the LO $\mathcal{O}(p^2)$ Lagrangian reads
\begin{align}\label{eq.EFT15}
 \mathcal{L}_{2} \equiv \mathcal{L}_\text{LO} =\frac{F^2}{4} \left< u_{\mu }u^{\mu }+\chi_{{}+} \right>\,.
\end{align}
Further, the part of the NLO $\mathcal{O}(p^4)$ Lagrangian relevant for our calculations\footnote{If one has to consider dynamical vector fields, then it is necessary to incorporate three additional terms which are built from the field tensor $V_{\mu \nu }=\partial_{\mu } V_{\nu }-\partial_{\nu } V_{\mu }-i\left[ V_{\mu },V_{\nu }\right]$, see \refe\cite{Bijnens:2009qm}. In total, $\mathcal{L}_\text{NLO}$ then contains 13 low energy constants.}reads~\cite{Bijnens:2009qm}
\begin{equation}\label{eq.EFT16}
    \begin{split}
     \mathcal{L}_{4} \equiv \mathcal{L}_\text{NLO} &=L_{0}\left< u^{\mu }u^{\nu }u_{\mu }u_{\nu }\right>  +L_{1}\left< u^{\mu }u_{\mu }\right>  \left< u^{\nu }u_{\nu }\right>  +L_{2}\left< u^{\mu }u^{\nu }\right>  \left< u_{\mu }u_{\nu }\right>  +L_{3}\left< u^{\mu }u_{\mu }u^{\nu }u_{\nu }\right>  \\ &+L_{4}\left< u^{\mu }u_{\mu }\right>  \left< \chi_{+} \right>  +L_{5}\left< u^{\mu }u_{\mu }\chi_{+} \right>  +L_{6}\left< \chi_{+} \right>^{2}  +L_{7}\left< \chi_{-} \right>^{2}  +\frac{1}{2} L_{8}\left< \chi^{2}_{+} +\chi^{2}_{-} \right> \\ &+H_{2}\left< \chi \chi^{\dagger } \right>,
\end{split}
\end{equation}
where $L_0\dots L_8, H_2$ are a priori unknown low energy constants (LECs) which encode short-distance dynamics. The NLO Lagrangian in Eq.~\eqref{eq.EFT16} is valid for both real and pseudoreal theories, as well as for a theory with complex fermions — provided that, in the latter case, a correct embedding of the $N_F \times N_F$ matrices into $2N_F \times 2N_F$ matrices is performed.
It is worth noting that the higher-order Lagrangians at $\mathcal{O}(p^6)$ and $\mathcal{O}(p^8)$ have been discussed in detail for the complex case in Refs.~\cite{Bijnens:1999sh} and \cite{Bijnens:2018lez}, which contain 115 and 1862 terms, respectively. The $\mathcal{O}(p^6)$ Lagrangian for the (pseudo)real case has the same overall structure, although it may include a certain number of redundant terms~\cite{Bijnens:2011xt}. To the best of our knowledge, the $\mathcal{O}(p^8)$ Lagrangian has not yet been discussed in the literature in the context of (pseudo)real theories.
 
When calculating observables at NLO$\equiv \mathcal{O}(p^4)$ precision, 1-loop diagrams built from the $\mathcal{L}_2$ interactions will contribute according to formula~\eqref{eq.EFT18}. Such diagrams lead to divergencies that are absorbed via renormalization of the LECs entering the $\mathcal{L}_4$ Lagrangian. %makes it possible to absorb all the divergencies which appear from the loop diagrams originating from the LO Lagrangian. 
We define the renormalized LECs in the $\overline{\text{MS} }$ scheme as
\begin{equation}\label{eq.EFT17a}
    \begin{split}
L_{i}=L^{r}_{i}(\mu)+\frac{\Gamma_{i} }{32\pi^{2} } R,
\end{split}
\end{equation}
with 
\begin{equation}\label{eq.EFT18a}
\begin{split}
R=\frac{2}{n-4} -\log \left( 4\pi \right)  +\gamma_{E} -1,
\end{split}
\end{equation}
where $\gamma_{E} =-\Gamma^{\prime } (1)$ is the Euler-Mascheroni constant.\footnote{Note that this renormalization scheme is standard in the ChPT context (see \cite{Scherer:2012xha}); however, in other applications the ``-1'' term is absent.} The values of $\Gamma_i$ for complex, real and pseudoreal fermions are given in \cite{Bijnens:2009qm}  and cited here in table \ref{tab:LECs_table}.

We use Lagrangians~\eqref{eq.EFT15} and~\eqref{eq.EFT16} to derive NLO formulas for several observables presented in sections~\ref{sec:Sp} and~\ref{sec:SO}. Let us introduce the following chiral expansion for physical observables:
%In this section we are presenting the results for the observables at NLO. We denote the theory with  $2N_F = 4$  pseudoreal fermions as  $SU(4)/Sp(4)$, and the corresponding real-fermion case as  $SU(4)/SO(4)$.
%The results are presented for non-degenerate quark masses, but the theory is still expected to approximately respect the flavor  $Sp(4)$  or  $SO(4)$ symmetry.
%The chiral expansion of some physical observable typically reads
\begin{equation}\label{eqChExp}
\begin{split}
  \mathcal{O}_\text{phys}=\mathcal{O}_\text{LO}+\mathcal{O}_\text{NLO}+\mathcal{O}_\text{NNLO}+\dots,
\end{split}
\end{equation}
where the expansion parameter is $p^2/(4\pi F_\pi)^2\sim M_\pi^2/(4\pi F_\pi)^2\lesssim 1$. 
%\dk{which sign is better here?} 
%In this paper we denote the observable calculated at the NLO ($\mathcal{O}(p^4)$) level as $\mathcal{O}_4$.
The NLO corrections contain logarithmic contributions from the loops as well as terms with LECs. Let us note that $M_\pi \equiv M_\text{phys}$ and $F_\pi \equiv F_\text{phys}$ denote the physical values of the pion mass and decay constant in this text.

%The NLO LECs are dimensionless parameters with typical values of the order $10^{-4}-10^{-3}$. The fits of the QCD NLO LECs can be found in e.g. \cite{Amoros:2001cp}.
%The results for the $SU(4)/Sp(4)$ and $SU(4)/SO(4)$ theories are presented in sections~\ref{sec:Sp} and~\ref{sec:SO}, respectively.

\renewcommand{\arraystretch}{1.5}
\begin{table}[h]
\centering
\begin{tabular}{c c c c c c c c c c c}
\hline
Theory & $L_0$ & $L_1$ & $L_2$ & $L_3$ & $L_4$ & $L_5$ & $L_6$ & $L_7$ & $L_8$ & $H_2$ \\
\hline
$SU(4)/Sp(4)$ & $-\frac{1}{24}$ & $\frac{1}{32}$ & $\frac{1}{16}$ & $\frac{1}{6}$ & $\frac{1}{16}$ & $\frac{1}{4}$ & $\frac{5}{128}$ & $0$ & $0$ & $0$ \\
$SU(4)/SO(4)$ & $\frac{1}{8}$ & $\frac{1}{32}$ & $\frac{1}{16}$ & $0$ & $\frac{1}{16}$ & $\frac{1}{4}$ & $\frac{5}{128}$ & $0$ & $\frac{1}{8}$ & $\frac{1}{4}$ \\
\hline
\end{tabular}
\caption{Coefficients $\Gamma_i$ for the NLO LECs of $SU(4)/Sp(4)$ and $SU(4)/SO(4)$ theories reproduced from \cite{Bijnens:2009qm}.}
\label{tab:LECs_table}
\end{table}
\renewcommand{\arraystretch}{1.0}

\subsection{Mass splitting}

When building the chiral Lagrangian, we treated $\hat{\mathcal{M}}$ as a spurion field transforming under chiral transformations to ensure chiral symmetry. Now we replace $\hat{\mathcal{M}}$ by its expectation value which is nothing but the fermion mass matrix. We are interested in the mass non-generate case with $N_F=2$, i.e., the mass matrix reads 
\begin{align}\label{eq.EFT17b}
\begin{split}
\mathcal{M} =\begin{pmatrix}m_{u}&0\\ 0&m_{d}\end{pmatrix}, 
\end{split}
\end{align}
where we employed the notation for the quark masses inspired by QCD. Non-zero quark masses break the chiral symmetry explicitly. 
As mentioned above, when the two quark masses are equal, the remaining symmetry is $SO(4)$ and $Sp(4)$ in the real and pseudoreal cases, respectively. On the other hand, for non-degenerate masses, further symmetry breaking is introduced. The breaking patterns for non-degenerate quark masses for the pseudoreal and real cases considered here are well established~\cite{Kulkarni:2022bvh,Pomper:2024otb} and will be confirmed by our calculation of the pion masses in sections~\ref{sec:SpMass} and~\ref{sec:SOMass}. For pseudoreal representations with 2 non-degenerate flavours, the pion five-plet under the $Sp(4)$ group breaks into a four-plet plus a singlet under the $SU(2)\times SU(2)$ left-over symmetry~\cite{Kulkarni:2022bvh}. Similarly for a real representation, the nine-plet under $SO(4)$ group breaks into a four-plet, two doublets and a singlet under $SO(2)\times SO(2)$ left-over symmetry~\cite{Pomper:2024otb}. Let us note that for the mass non-degenerate case, also~\eq\eqref{eq.EFT9} holds only approximately. 

\section{NLO results for $SU(4)/Sp(4)$ theories}\label{sec:Sp}

In this section, we apply the EFT tools introduced above to the calculation of different observables in theories with quarks in pseudoreal representations of the gauge group at the NLO~$\equiv \mathcal{O}(p^4)$ precision (see formula~\eqref{eqChExp}). We introduce general strategies for such calculations and, when appropriate, mention the differences with the case of real representations, so that details can be later omitted in \sec\ref{sec:SO}.

\subsection{Masses}\label{sec:SpMass}
Upon expanding  $\mathcal{L}_{2}$, the contribution at  $O(\pi^2)$  takes the form:
\begin{equation}\label{eq.Sp4.31}
\begin{split}
    \mathcal{L}^{2\pi }_{2} =\frac{1}{2} \sum^{N_{\pi }}_{i=1} \left( \partial_{\mu } \pi^{i} \right)^{2}  -\sum^{N_{\pi }}_{i=1} \frac{\MLO{,i}}{2} \left( \pi^{i} \right)^{2}.
\end{split}
\end{equation}
%Consequently, 
where
\begin{equation}\label{eq:MLOSp}
M^2\equiv \MLO{,k} =B_0(m_u+m_d),\quad \forall k\,,
\end{equation}
i.e., in the \spth theory, the LO (tree-level) masses are equal for all the pion species. 
%while in the $SO(4)$ the splitting appears already at the LO. 
The $k$-th physical mass is given by a pole of a full propagator of $k$-th field \cite{Scherer:2012xha}:
\begin{equation}\label{eq.Sp4.32}
\begin{split}
     i\bigtriangleup_{k} =\frac{i}{p^{2}-
     \MLO{,k}-\Sigma_{k} \left( p^{2}\right)  +i\epsilon },
\end{split}
\end{equation}
where the self-energy operator $\Sigma_{k}$ can be found as a sum of all one-particle irreducible diagrams at the given order of perturbation theory. The  %physical 
pion masses at the corresponding order are then defined from the equation
\begin{equation}\label{eq.Sp4.33}
\begin{split}
    M^{2}_{\pi ,k}-\MLO{,k}-\Sigma_{k} \left(p^2 = M^{2}_{\pi ,k}\right)  =0\,.
\end{split}
\end{equation}
At $O(p^4)$, the self-energy has the following structure:
\begin{equation}\label{eq.Sp4.34}
\begin{split}
    \Sigma_{k}\left( p^{2}\right)  =A_{k}+B_{k}p^{2},
\end{split}
\end{equation}
and then, using \eqref{eq.Sp4.33}, we can deduce:
\begin{equation}\label{eq.Sp4.35}
\begin{split}
    \MNLO{,k}=\MLO{,k}B_{k} +A_{k}.
\end{split}
\end{equation}

\begin{figure}[h]
    \centering
    \includegraphics[width=0.5\textwidth]{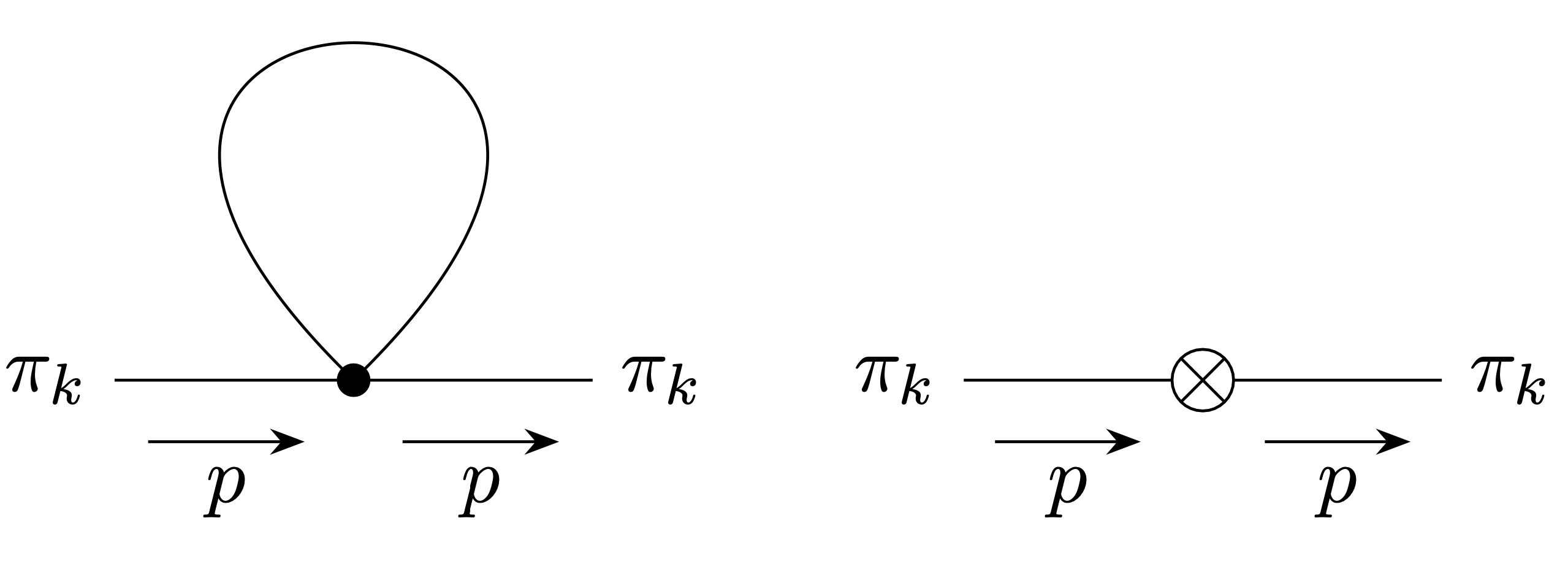}
    \caption{Diagrams which contribute to $-i\Sigma_k(p^2)$ at NLO. In this and following figures, we adhere to the \cite{Bijnens:2009qm} notation: dot vertex comes from $\mathcal{L}_{\text{LO}}$, while crossed circle vertex comes from the $\mathcal{L}_{\text{NLO}}$ Lagrangian.}
    \label{fig:SelfEnergyNLO}
\end{figure}

At order $O(p^4)$, there exist only two diagrams with chiral dimension $\text{deg}\,\Gamma=4$ (see figure \ref{fig:SelfEnergyNLO}).
We can deduce $A_k$ and $B_k$ contributions and derive physical masses at NLO using~\eqref{eq.Sp4.35}.
Note that, inside the loops, in principle, all the pion species propagate so the sum over the pion species should be performed.

For $i=1,2,4,5$ the NLO correction to the mass reads
\begin{align}\label{eq.Sp4.42}
\MNLO{,i}= \frac{M^{4}}{F^{2}} \left(-32L^{r}_{4}-8L^{r}_{5}+64L^{r}_{6}+16L^{r}_{8}+\frac{3}{64\pi^{2} } \log \left( \frac{M^{2}}{\mu^{2} } \right)  \right) 
\end{align}
where the numbering of the pion species follows our choice of generators, see \app\ref{appendixSU4gen}.
The NLO mass of the third pion is different:
\begin{equation}\label{eq.Sp4.42.1}
\begin{split}
     \MNLO{,3}&= \frac{M^{4}}{F^{2}} \Biggl( -32L^{r}_{4}-8L^{r}_{5}\\ &+64L^{r}_{6}+64L^{r}_{7}\left( \frac{1-r}{1+r} \right)^{2}   +32L^{r}_{8}\frac{1+r^{2}}{\left( 1+r\right)^{2}  }  +\frac{3}{64\pi^{2} } \log \left( \frac{M^{2}}{\mu^{2} } \right)  \Biggr) ,
\end{split}
\end{equation}
where $r\equiv m_d/m_u$.

The multiplet structure here corresponds to the results presented in \cite{Kulkarni:2022bvh}.
We observe that the  mass splitting appears at NLO and, as expected, vanishes in the limit $m_u\to m_d$ where the result matches the mass-degenerate expression from \refe\cite{Bijnens:2009qm}.
The mass hierarchy is defined by the relative hierarchies of the LECs $L_7^r$ and $L_8^r$: if $L_8^r<-4L_7^r$ then $\MNLO{,3}<\MNLO{,i}$, $i\neq 3$.

\begin{comment}
    \begin{figure}
    \centering
\feynmandiagram [horizontal=a to b, layered layout] {
  a[particle=\(\pi_k\)] -- [ momentum'={\(p\)}] b [ dot] -- [out=135, in=45, loop, min distance=3cm] b--[ momentum'={\(p\)}] c[particle=\(\pi_k\)],
};
\feynmandiagram [horizontal=a to b, layered layout] {
  a[particle=\(\pi_k\)]--[ momentum'={\(p\)}] b [crossed dot]  -- [ momentum'={\(p\)}]c[particle=\(\pi_k\)],
};
    \caption{Diagrams which contribute to $-i\Sigma_k(p^2)$ at NLO. In this and following figures, we adhere to the \cite{Bijnens:2009qm} notation: dot vertex comes from $\mathcal{L}_{\text{LO}}$, while crossed circle vertex comes from the $\mathcal{L}_{\text{NLO}}$ Lagrangian.}
    \label{fig:SelfEnergyNLO}
\end{figure}
\end{comment}

\subsection{Condensates}\label{secCondensatesSP(4)}
\begin{figure}[h]
    \centering
    \includegraphics[width=0.35\textwidth]{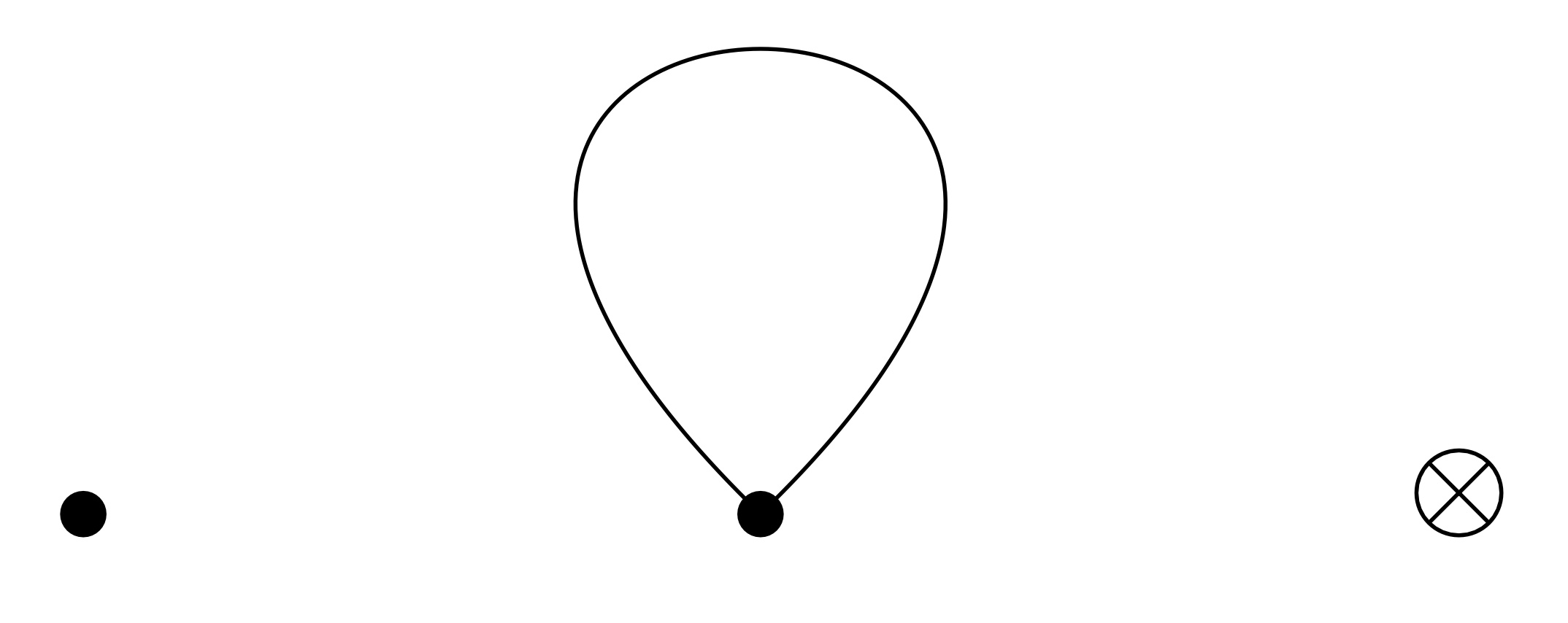}
   \caption{Feynman diagrams which contribute to the quark condensates at LO and NLO. %Dot denotes the insertion which comes from $\mathcal{O}(p^2)$ Lagrangian, while crossed circle denotes the insertion which comes from  $\mathcal{O}(p^4)$ Lagrangian.
    }
    \label{fig:CondNLO}
\end{figure}

To calculate the values of the quark condensates one has to calculate the effective potential, $V_{\text{eff} }$, and compute the derivative of its minimum value with respect to the corresponding source, namely
\begin{equation}\label{condSp-1.1}
\left< \bar{u} u\right>=\frac{\partial V_{\text{eff} }  }{\partial m_{u}}\left( \pi =0\right),\quad \left< \bar{d} d\right>  =\frac{\partial V_{\text{eff} }  }{\partial m_{d}}\left( \pi =0\right).  
\end{equation}
According to \eqref{eq.EFT7}, the total condensate reads 
\begin{equation}\label{condSp-2}
\left< \bar{q} q\right>  =\left< \bar{u} u\right>  +\left< \bar{d} d\right>.  
\end{equation}
As we work up to the order $O(p^4)$ we include the diagrams shown in figure \ref{fig:CondNLO} in the calculation of the VEV. These diagrams come from the $O(\pi^0)$ and $O(\pi^2)$ parts of $\mathcal{L}_{\text{LO}}$  and $O(\pi^0)$ part of $\mathcal{L}_{\text{NLO}}$. The loops generate a non-analytic correction to the quark condensate.

\begin{comment}
\begin{figure}
    \centering
\feynmandiagram  {b [ dot] };
\feynmandiagram  {
   b [ dot] -- [out=135, in=45, loop, min distance=3cm] b
};
\feynmandiagram  { b [crossed dot] };
    \caption{Feynman diagrams which contribute to the quark condensates at LO and NLO. %Dot denotes the insertion which comes from $\mathcal{O}(p^2)$ Lagrangian, while crossed circle denotes the insertion which comes from  $\mathcal{O}(p^4)$ Lagrangian.
    }
    \label{fig:CondNLO}
\end{figure}
\end{comment}

At leading order, the condensates are the same:
\begin{align}
\left< \bar{u} u\right>_\text{LO}  &=\left< \bar{d} d\right>_\text{LO} =-2B_{0}F^{2},\label{condSp1}
 \end{align}
but they receive a splitting at the NLO:\footnote{Let us note that in~\cite{Bijnens:2009qm}, the LO contribution is given by
%\begin{equation}\label{condSp4}
    $\left< \bar{q} q\right>_\text{LO}  =\left< \bar{u} u\right>_\text{LO}  +\left< \bar{d} d\right>_\text{LO} =-N_{F}B_{0}F^{2}=-2B_{0}F^{2}$
and also their NLO contribution differs by a factor of $2$ from our result in the mass-degenerate limit, $m_u=m_d$, %the results \eqref{condSp1}-\eqref{condSp3} 
possibly due to a mismatch in the definition of the quark condensate. 
%\end{equation}
The same applies to the results for the \soth theory. However, our results are consistent with the values of condensates presented in \cite{Kogut:2000ek}, \cite{Kulkarni:2022bvh} and \cite{Pomper:2024otb}. 
%We checked for consistency the condensates in the $SU(N_F)\times SU(N_F)/SU(N_F)$ theory and got the same result as presented in~\cite{Bijnens:2009qm}.
In view of the discrepancies with~\cite{Bijnens:2009qm}, we present the explicit derivation of the LO quark condensate in appendix \ref{AppCondensate}}.
\begin{align}
\left< \bar{u} u\right>_\text{NLO}  &=\left< \bar{u} u\right>_\text{LO} \left[  \frac{8 B_0 m_u}{F^2} \left( H_2^r + 2 L_8^r \right) + \frac{M^2}{F^2} \left( 64 L_6^r - \frac{5}{64\pi^2} \log \left( \frac{M^2}{\mu^2}\right) \right) \right] ,\label{condSp2} \\   \langle \overline{d}d \rangle_\text{NLO}  &= \left< \bar{d} d\right>_\text{LO} \left[  \frac{8 B_0 m_d}{F^2} \left( H_2^r + 2 L_8^r \right) + \frac{M^2}{F^2} \left( 64 L_6^r - \frac{5}{64\pi^2} \log \left( \frac{M^2}{\mu^2}\right) \right) \right]. \label{condSp3}
\end{align}

\subsection{Decay constants}
\begin{figure}[h]
    \centering
    \includegraphics[width=0.8\textwidth]{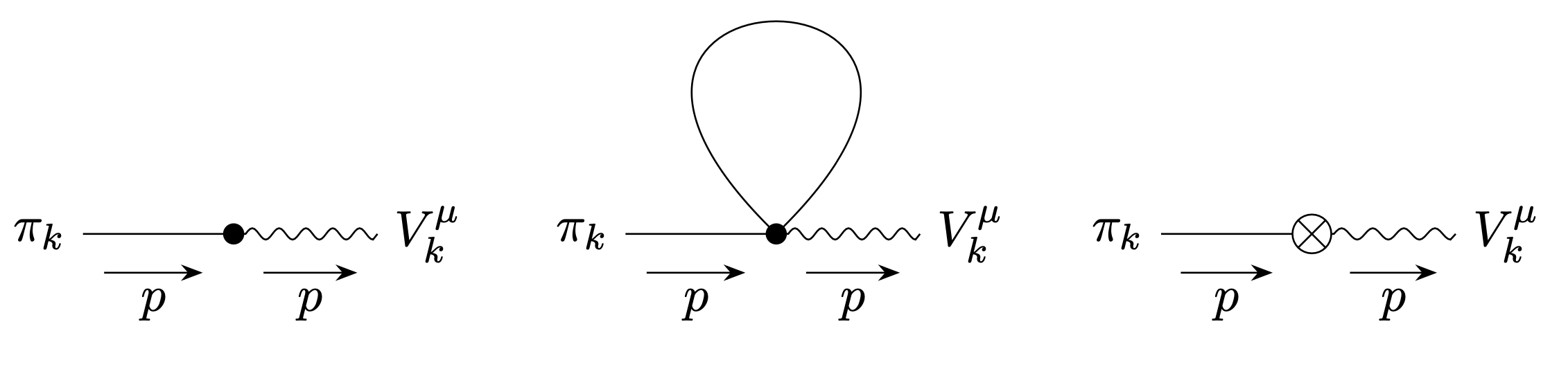}
   \caption{Diagrams which contribute to $\mathcal{M}_{\pi_kV_k}$ determining the pion decay constant at LO and NLO. %Dot vertex comes from $O(p^2)$ Lagrangian, while crossed circle vertex comes from  $O(p^4)$ Lagrangian.
}
    \label{fig:DacayConstantNLO}
\end{figure}

The physical pseudoscalar decay constants $F_{\pi,k}$ are defined from the matrix elements of currents which correspond to broken generators
\cite{Amoros:1999dp}:
\begin{equation}\label{eq.Sp4.43}
\begin{split}
\langle 0|A^{k}\left( 0\right)_{\mu }  |\pi^{k} \left( p\right)  \rangle =i  \ \sqrt[]{2} p_{\mu }F_{\pi,k},\  \  \  k=1...N_{\pi }.
\end{split}
\end{equation}
Accordingly, the pion decay constant can be calculated as
\begin{comment}
   \begin{equation}\label{eq.Sp4.44}
\begin{split}
F_{\pi,a}=\sqrt[]{Z_{a}} F^{\text{1PI} }_{a}=F^{1\text{PI} }_{\text{LO} ,a}+\left( \frac{1}{2} Z_{\text{NLO} ,a}F^{1\text{PI} }_{\text{LO} ,a}+F^{1\text{PI} }_{N\text{LO} ,a}\right)  +\mathcal{O} \left( p^{6}\right),  
\end{split}
\end{equation} 
\end{comment}
\begin{equation}\label{eq.Sp4.44}
\begin{split}
F_{\pi ,k}=\sqrt[]{Z_{k}} \mathcal{M}_{\pi_kV_k},  
\end{split}
\end{equation} 
where $\mathcal{M}_{\pi_kV_k}$ is the  2-point vertex function with an external pion $\pi_k$ and an external vector $V_\mu^k$ which sources the corresponding broken current $A_\mu^k$ (see figure \ref{fig:DacayConstantNLO}). Further, $Z_{k}$ is the wave function renormalization of the pion calculated as a residue of the full pion propagator.  It is related to the self-energy correction via
\begin{equation}\label{eq.Sp4.46}
\begin{split}
Z_k =\left( 1-\frac{d\Sigma_k }{dp^{2}} \left( p^{2}=M^{2}_{\pi ,k}\right)  \right)^{-1}=1+Z_{\text{NLO},k}+\mathcal{O}(p^6).
\end{split}
\end{equation}
In the \spth theory the NLO wave function renormalization is the same for all pions, while in the \soth case there is a splitting.  

The LO decay constant is simply 
$\FLO{,k}=F$ for any $k$ for both real and pseudoreal quarks. At NLO the result for \spth theory reads:
\begin{equation}\label{eq.Sp4.48}
\begin{split}
\FNLO{}= \frac{M^{2}}{F}\left( 4\left( 4L^{r}_{4}+L^{r}_{5}\right)  -\frac{1}{16\pi^{2} } \log \left( \frac{M^{2}}{\mu^{2} } \right)  \right). 
\end{split}
\end{equation} 
There is no splitting at NLO and this expression matches the corresponding
result from \refe\cite{Bijnens:2009qm}. For the \soth theory, there is splitting already at NLO as will be shown in \sec\ref{sec:SO}.   

\begin{comment}
\begin{figure}
    \centering
\feynmandiagram [horizontal=a to b, layered layout] {
  a[particle=\(\pi_k\)]--[ momentum'={\(p\)}] b [ dot]  -- [ photon,momentum'={\(p\)}]c[particle=\(  V^{\mu}_k \)],
};
\feynmandiagram [horizontal=a to b, layered layout] {
  a[particle=\(\pi_k\)] -- [ momentum'={\(p\)}] b [ dot] -- [out=135, in=45, loop, min distance=3cm] b--[ photon,momentum'={\(p\)}] c[particle=\(V^{\mu}_k\)],
};
\feynmandiagram [horizontal=a to b, layered layout] {
  a[particle=\(\pi_k\)]--[ momentum'={\(p\)}] b [crossed dot]  -- [ photon,momentum'={\(p\)}]c[particle=\(V^{\mu}_k\)],
};
\caption{Diagrams which contribute to $\mathcal{M}_{\pi_kV_k}$ determining the pion decay constant at LO and NLO. %Dot vertex comes from $O(p^2)$ Lagrangian, while crossed circle vertex comes from  $O(p^4)$ Lagrangian.
}
    \label{fig:DacayConstantNLO}
\end{figure}
\end{comment}
  
\subsection{Meson scattering}\label{secScat}

\begin{figure}[h]
    \centering
    \includegraphics[width=0.9\textwidth]{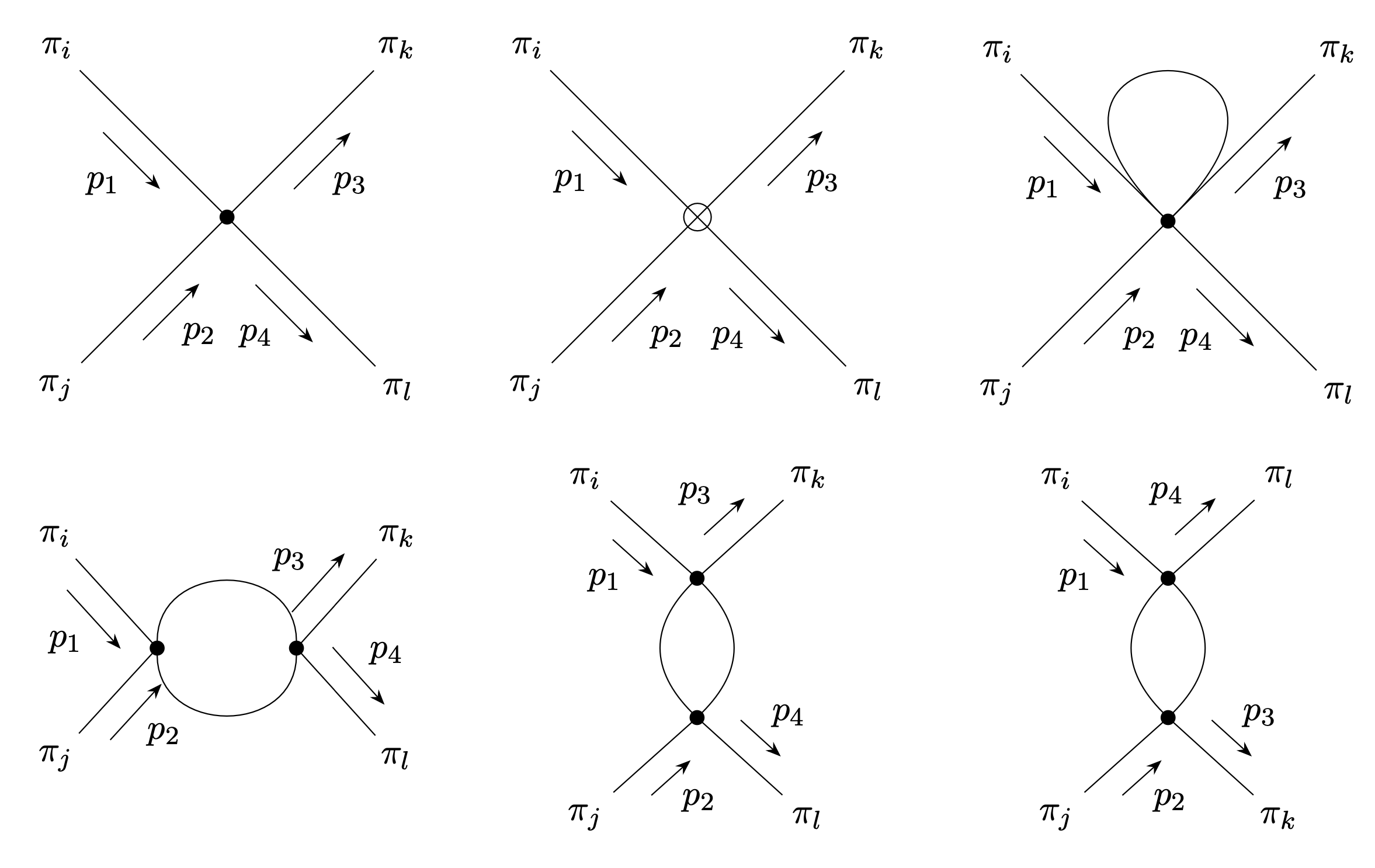}
    \caption{Diagrams which contribute to the $2\to2$ scattering at LO and NLO. %Dot vertex comes from $O(p^2)$ Lagrangian, while crossed circle vertex comes from  $O(p^4)$ Lagrangian.
    }
    \label{fig:2to2_scattering}
\end{figure}

The $2\to2$ scattering amplitude for a process $\pi_{i} (p_{1})+\pi_{j} (p_{2})\rightarrow \pi_{k} (p_{3})+\pi_{l} (p_{4})$ is calculated as

\begin{equation}\label{eq.Sp4.Scat.1}
\begin{split}
\mathcal{M}^{ij\rightarrow kl} &=\langle \pi_k(p_3)\pi_l(p_4)|\pi_i(p_1)\pi_j(p_2) \rangle 
\\&=\sqrt[]{Z_{i}}\ \sqrt[]{Z_{j}} \ \sqrt[]{Z_{k}} \ \sqrt[]{Z_{l}} \;  [
\text{sum of amputated 4-point diagrams}
]^{ij\to kl}\, .
\end{split}
\end{equation} 
When working at NLO precision, the terms up to the order $\mathcal{O}\left(M^4/F^4\right)$ will be included on the right-hand side. The relevant diagrams are shown in figure \ref{fig:2to2_scattering}. The amplitude is a function of Mandelstam variables
\begin{equation}\label{eq.Sp4.Scat.11}
s=\left( p_{1}+p_{2}\right)^{2}  ,\quad t=\left( p_{1}-p_{3}\right)^{2}  ,\quad u=\left( p_{1}-p_{4}\right)^{2}, 
\end{equation} 
that satisfy
\begin{equation}\label{eq.Sp4.Scat.11.1}
\  s+t+u=M^{2}_{\pi ,i}+M^{2}_{\pi,j}+M^{2}_{\pi ,k}+M^{2}_{\pi ,l}.
\end{equation} 
The expansion parameter $M/F$ can be traded for the ratio of the physical pion mass and decay constant that are more relevant for the phenomenology studies. This is done by inverting the formulas for masses and decay constants as explained in \cite{Bijnens:2009qm,Bijnens:2011fm}.
For example, in the \spth theory, parameters $M$ and $F$ can be expressed  in terms of $M_{\pi,1}$ and $F_\pi$ if we invert \eqref{eq.Sp4.42} and \eqref{eq.Sp4.48}:
\begin{align}
 F_{}&=F_\pi\Bigg( 1 +\frac{M^2_{\pi,1}}{F^2_\pi} \Bigg[ -4\left( 4L^{r}_{4}+L^{r}_{5}\right)  +\frac{1  }{16\pi^{2} }\log \left( \frac{M^2_{\pi,1}}{\mu^{2} } \right) \Bigg] + \mathcal{O}\left(\frac{M^4_{\pi,1}}{F^4_\pi}\right) \Bigg),   \label{eq.Sp4.Scat.3}\\ 
 M^{2}_{{}  }&=M^2_{\pi,1}\Bigg( 1\nonumber\\&+ \frac{M^2_{\pi,1}}{F^2_\pi}\Bigg[ 32L^{r}_{4}+8L^{r}_{5}-64L^{r}_{6}-16L^{r}_{8}-\frac{3}{64\pi^{2} } \log \left( \frac{M^2_{\pi,1}}{\mu^{2} } \right)  \Bigg] + \mathcal{O}\left(\frac{M^4_{\pi,1}}{F^4_\pi} \right)\Bigg) . \label{eq.Sp4.Scat.4}
\end{align}
 We can also express $M^2_{\pi,3}$ in terms of $M^2_{\pi,1}$ as follows:
\begin{equation}\label{eq.Sp4.Scat.5}
\begin{split}
M^2_{\pi,3}=M^2_{\pi,1}\left( 1+\frac{M^2_{\pi,1}}{F^2_\pi} \left[ 16\left( 4L^{r}_{7}+L^{r}_{8}\right)  \left( \frac{1-r}{1+r} \right)^{2}    \right] + \mathcal{O}\left(\frac{M^4_{\pi,1}}{F^4_\pi}\right) \right).  
\end{split}
\end{equation} 
In these expressions which are understood to be valid up to NLO we can replace the NLO mass and decay constant by the physical ones as the difference would only appear at NNLO. When imposing the on-shell conditions on the external momenta in the diagrams, we also replace $p^2$ by the square of the appropriate physical mass.%When the external momenta in the diagrams are put on-shell we also use the square of physical pion masses: $p^2 = M_\pi^2$. 

We observe that in \spth theory, the 4-point amplitude is non-zero only if the four external pions are either all of the same type or include two pairs of identical species. Let us note that this is not true for the \soth theory. 

If  $m_u = m_d$ and consequently $M^{2}_{\pi,3}=M^{2}_{\pi,1}$,  the scattering amplitude for degenerate masses from \cite{Bijnens:2011fm} is reproduced.\footnote{We confirm the amendment noted in \cite{Hansen:2015yaa}: specifically, the sign of the $\pi_{16}$ term in the expression for $\alpha_2$ in the “Two-color” case should be negative rather than positive (see table 5 in \cite{Bijnens:2011fm}). No discrepancies have been found by us for the real or “Adjoint” case.}
Even if $m_u \neq m_d$, the NLO amplitude is the same as presented in \cite{Bijnens:2011fm} in case $i,j,k,l\neq 3$ (with their $M^2_\text{phys}$ corresponding to our $M^2_{\pi,1}$). However, when two or four pions are of the third flavor, the non-degenerate case amplitude gets a correction. %\footnote{In $SU(4)/Sp(4)$ theory the 4-point amplitude is non-zero only if the four external pions are either all the same type or form two identical pairs of different types. This is not true for $SU(4)/SO(4)$-theory.} 
We find that the explicit quark mass dependence of the NLO amplitude arises solely through the combination $(m_d-m_u)^2/(m_d+m_u)^2$, which can be expressed in terms of $M^{2}_{{\pi,3}}$ and $M^{2}_{\pi,1}$ using \eqref{eq.Sp4.Scat.5}:
\begin{equation}\label{eq.Sp4.Scat.5.1}
  \left( \frac{1-r}{1+r} \right)^{2}  =\frac{1}{16\left( 4L^{r}_{7}+L^{r}_{8}\right)  } \left( \frac{M^{2}_{\pi ,3}}{M^{2}_{\pi ,1}} -1\right)  \left( \frac{M^{2}_{\pi ,1}}{F^{2}_{\pi }} \right)^{-1}  +\mathcal{O} \left( \frac{M^{2}_{\pi ,1}}{F^{2}_{\pi }} \right).  
\end{equation}
\begin{comment}
   We see that due to the factor $\left(M^{2}_{\pi,1}/F^{2}_{\pi} \right)^{-1}$ in \eqref{eq.Sp4.Scat.5.1} the NLO term which contains $\left( (1-r)/(1+r) \right)^{2} $ modifies the LO (but not the NLO) part of the amplitude.
\end{comment}

The LO parts of the modified amplitudes read
\begin{align}
 \mathcal{M}^{ii\rightarrow 33}_{\text{LO} } &=\frac{M^{2}_{\pi ,1}+2s-t-u}{6F^{2}_{\pi }} ,\quad i\neq3, \label{eq.LOii33}\\
\mathcal{M}^{33\rightarrow 33}_{\text{LO} } &=\frac{M^{2}_{\pi ,1}}{2F^{2}_{\pi }} ,\label{eq.LO3333}
\end{align}
and the NLO parts are
\begin{align}\label{eq.NLOii33}
F^{4}_{\pi }\mathcal{M}^{ii\rightarrow 33}_{\text{NLO} } &=\frac{M^{2}_{\pi ,3}-M^{2}_{\pi ,1}}{3F^{2}_{\pi }} \nonumber \\
& -\frac{1}{1152\pi^{2} } \bigg( M^{4}_{\pi,1 }+18\, s^{2}-10\, s\, t+3\, t^{2}-10\, s\, u-4\, t\, u+3\, u^{2} \nonumber \\&-M^{2}_{\pi,1 }(4\, M^{2}_{\pi ,3}-11\, s+t+u)+M^{2}_{\pi ,3}(s+t+u)\bigg)  \nonumber \\
&+2 \left( 2  M^{4}_{\pi,1 }  +2  M^{4}_{\pi,3 }  -s^{2}+t^{2}+2  M^{2}_{\pi,1 }(s-t-u)+2  M^{2}_{\pi ,3}(s-t-u)+u^{2}\right)L^{r}_{0}\nonumber \\
& -2  \left( 2  M^{2}_{\pi,1 }-s\right) \left( -2  M^{2}_{\pi ,3}+s\right)(4L^{r}_{1}+L_3^r)  \nonumber \\
& +4  \left( 2  M^{4}_{\pi,1 }  +4  M^{2}_{\pi,1 }  M^{2}_{\pi ,3}+2  M^{4}_{\pi ,3}  +t^{2}+u^{2}-2  M^{2}_{\pi,1 }(t+u)-2  M^{2}_{\pi ,3}(t+u)\right)L^{r}_{2} \nonumber \\
& -16M^{2}_{\pi,1 }\left( M^{2}_{\pi,1 }+M^{2}_{\pi ,3}-s\right)  (4L^{r}_{4}+L^{r}_{5}) +8M^{4}_{\pi,1 }\left( 4L^{r}_{6}+L^{r}_{8}\right)  \nonumber \\
&+\frac{1}{24 } \left( -5M^{4}_{\pi,1 }  +3s(2s-t-u)+2M^{2}_{\pi,1 }(s+t+u)\right)  \bar{J} (M^{2}_{\pi,1 },M^{2}_{\pi,1 },s)\nonumber \\   
&+\frac{1}{144t} \bigg( -12M^{6}_{\pi ,3}  +6M^{4}_{\pi,1 }  (4M^{2}_{\pi ,3}+3t)\nonumber \\
&-M^{2}_{\pi,1 }(12M^{4}_{\pi ,3}  +8M^{2}_{\pi ,3}t+t(7s+19t-17u))\nonumber\\
&+t(2M^{4}_{\pi ,3}  +6t(t-u)+M^{2}_{\pi ,3}(s+t+u))\bigg)  \bar{J} (M^{2}_{\pi,1 },M^{2}_{\pi,1 },t) \nonumber\\   
&+\frac{1}{144u} \bigg( -12M^{6}_{\pi,1}  +6M^{4}_{\pi,1 }  (4M^{2}_{\pi ,3}+3u) \nonumber\\
&-M^{2}_{\pi,1}(12M^{4}_{\pi ,3}  +8M^{2}_{\pi ,3}u+u(7s-17t+19u))\nonumber\\ &+u(2M^{4}_{\pi ,3} +6u(-t+u)+M^{2}_{\pi ,3}(s+t+u))\bigg)  \bar{J} (M^{2}_{\pi,1 },M^{2}_{\pi,1 },u)\nonumber \\ 
&+\frac{1}{5760\pi^{2} } \bigg( -111M^{4}_{\pi,1 }+3M^{2}_{\pi,1 }\left( 24M^{2}_{\pi ,3}-13s+7(t+u)\right)   \nonumber\\
&-5( 2M^{4}_{\pi ,3}  +M^{2}_{\pi ,3}(s+t+u)\nonumber\\
&+3\left( 6s^{2}+(t-u)^{2}-3s(t+u)\right)  )  \bigg)  \log \left( \frac{M^{2}_{\pi,1 }}{\mu^{2} } \right), \quad i\neq 3  
\end{align}
and
\begin{align}\label{eq.NLO3333}
&F^{4}_{\pi }\mathcal{M}^{33\rightarrow 33}_{\text{NLO} } = 2\frac{M^{2}_{\pi ,3}-M^{2}_{\pi ,1}}{F^{2}_{\pi }} + \frac{1}{384\pi^{2} } \Bigg( -13M^{4}_{\pi,1 }+4M^{2}_{\pi,1 }(s+t+u)\nonumber \\ 
&+8\left( 2M^{4}_{\pi ,3}-s^{2}-t^{2}+tu-u^{2}+s(t+u)-M^{2}_{\pi ,3}(s+t+u)\right)  \Bigg)  \nonumber \\ 
&+2\left( 12M^{4}_{\pi ,3}+s^{2}+t^{2}+u^{2}-4M^{2}_{\pi ,3}(s+t+u)\right)  \left( L^{r}_{0}+4L^{r}_{1}+4L^{r}_{2}+L^{r}_{3}\right)  \nonumber \\ 
&+4M^{2}_{\pi,1 }(-6M^{2}_{\pi ,3}+s+t+u)(4L^{r}_{4}+L^{r}_{5})+24(4L^{r}_{6}+L^{r}_{8})M^{4}_{\pi,1 }\nonumber \\
&+\frac{1}{72} \left( 13M^{4}_{\pi ,1}-4(2M^{2}_{\pi ,3}-3s)(2M^{2}_{\pi ,3}+2s-t-u)+4M^{2}_{\pi,1 }(-5s+t+u)\right)  \bar{J} (M^2_{\pi,1}, M^2_{\pi,1}, s)\nonumber \\
&+\frac{1}{72} \left( 13M^{4}_{\pi,1 }-4(2M^{2}_{\pi ,3}-3t)(2M^{2}_{\pi ,3}+2t-s-u)+4M^{2}_{\pi,1 }(-5t+s+u)\right)  \bar{J} (M^2_{\pi,1},M^2_{\pi,1}, t) \nonumber\\
&+\frac{1}{72} \left( 13M^{4}_{\pi,1 }-4(2M^{2}_{\pi ,3}-3u)(2M^{2}_{\pi ,3}+2u-s-t)+4M^{2}_{\pi,1 }(-5u+s+t)\right)  \bar{J} (M^2_{\pi,1},M^2_{\pi,1}, u)\nonumber \\
&+\frac{1}{1920\pi^{2} } \bigg( -193M^{4}_{\pi,1 }+4M^{2}_{\pi,1 }\left( 16M^{2}_{\pi ,3}+9(s+t+u)\right) \nonumber \\
& +40( 2M^{4}_{\pi ,3}-s^{2}-t^{2}+tu-u^{2}+s(t+u)\nonumber\\
&-M^{2}_{\pi ,3}(s+t+u))  \bigg)  \log \left( \frac{M^{2}_{\pi,1 }}{\mu^{2} } \right), 
\end{align}
where the loop-function $\bar{J} \left( m^{2}_{1},m^{2}_{2},p^{2}\right)  $ is given by \cite{Bijnens:2011fm}:
\begin{equation}\label{eq:IntegralJbar1}
    (32\pi^2) \bar{J}(m_1^2, m_2^2, p^2) = 2 + \left( -\frac{\Delta}{p^2} + \frac{\Sigma}{\Delta} \right) \ln \frac{m_1^2}{m_2^2} - \frac{\nu}{p^2} \ln \frac{(p^2 + \nu)^2 - \Delta^2}{(p^2 - \nu)^2 - \Delta^2},
\end{equation}
with
\begin{align}\label{eq:IntegralJbar2}
    \Delta &= m_1^2 - m_2^2, \\
    \Sigma &= m_1^2 + m_2^2, \\
    \nu^2 &= p^4 + m_1^4 + m_2^4 - 2p^2m_1^2 - 2p^2m_2^2 - 2m_1^2m_2^2.
\end{align}
The function $\bar{J}(m_1^2, m_2^2, p^2)$ develops an imaginary part for $p^2 \geq (m_1 + m_2)^2$. In these NLO formulas, we kept the masses $M^{2}_{\pi ,1}$ and $M^{2}_{\pi ,3}$ coming from the on-shell momenta, but, in principle, they can be set equal at this order. If this is done, then the only differences from the corresponding amplitudes in \cite{Bijnens:2011fm} are the first terms in \eqref{eq.NLOii33} and \eqref{eq.NLO3333}.
\begin{comment}
Finally, the modification is due to the factor \eqref{eq.Sp4.Scat.5.1} and on-shell relation $s+t+u=\sum_{i=1}^4M^2_{\pi,i}$ where the difference between the pion masses matters if appears in the LO part. The total change can be easily captured by modifying the function 
\begin{equation}
  B_{\text{LO} }(s,t,u)=\frac{M^{2}_{\pi }}{F^{2}_{\pi }} \left( -\frac{t}{2M^{2}_{\pi }} +1\right)  
\end{equation}
which appears in \cite{Bijnens:2011fm} (see Appendix  \ref{AppendixSIMP}):
\begin{align}
    & i,j,k,l\neq 3: \ B_\text{LO}(s,t,u)=\frac{M^{2}_{\pi}}{F^{2}_{\pi}} \left( 1-\frac{t}{2M^{2}_{\pi}} \right), \ \ s+t+u=4M^2_\pi, \\ & i=j=k=l=3 :\  B_{\text{LO} }(s,t,u)=\frac{M^{2}_{\pi}}{F^{2}_{\pi}} \left( \frac{t}{M^{2}_{\pi}} -1\right)  ,\  \  s+t+u=4M^{2}_{\pi,3}, \\
    & i=j\neq3: B_{\text{LO} }(s,t,u)=\frac{M^{2}_{\pi,3}}{F^{2}_{\pi}} \left( 1-\frac{t}{2M^{2}_{\pi,3}} \right)  ,\  \  s+t+u=2M^{2}_{\pi}+2M^{2}_{\pi,3}\label{eq:ii33}.
\end{align}
\end{comment}
Other amplitudes involving a single pair of third-flavor pions can be derived from~\eqref{eq.LOii33} and~\eqref{eq.NLOii33} using crossing relations. All the formulas for scattering amplitudes are coded in the Mathematica files available in the repository~\cite{kolesova_2025_14864550}.

%\hk{Connection to the discussion about relation between $L_7$ contribution to NLO mass and amplitude? Claim in one the old G\&L papers that the NLO amplitude in two-flavor SM ChPT is unchanged for split masses.}
%\dk{I have not fully understood what they meant there so Im not sure what to write...}

\section{NLO results for \soth theories}\label{sec:SO}
In this section, we summarize the NLO results for the theories with quarks in a real representation of the gauge group. For the details on the general strategy for computing the different observables, see the previous \sec\ref{sec:Sp}.

\subsection{Masses}\label{sec:SOMass}
%When the dark quarks are in the real representation of the gauge group, e.g. fundamental representation of $SO(N_c)$, then the coset space of the flavor symmetry is $SU(2N_F)/SO(2N_F)$, where $N_F$ is the number of dark flavors. For $N_F=2$ there are $N_\pi=9$ dark pions. 
For the \soth theory with non-degenerate quark masses, the nine pions split into three groups with equal LO masses where again the pion numbering follows the numbering of the generators in \app\ref{appendixSU4gen}:
\begin{align}
M^2_1&\equiv M^{2}_{\text{LO},1}=M^{2}_{\text{LO},2}=M^{2}_{\text{LO},3}=M^{2}_{\text{LO},4}=M^{2}_{\text{LO},5}=B_{0}\left(m_{u}+m_{d}\right), \label{NLOso.1} \\
M^2_{6}&\equiv M^{2}_{\text{LO},6}=M^{2}_{\text{LO},7}=2B_{0}m_{u},\label{NLOso.2} \\ 
 M^2_{8}&\equiv M^{2}_{\text{LO},8}=M^{2}_{\text{LO},9}=2B_{0}m_{d}\label{NLOso.3}.  
\end{align}
This LO mass splitting affects the loop calculations, since different masses appear in the propagators unlike in the \spth case. Consequently, the NLO formulas become more complicated in the \soth case. For the NLO contribution to the meson masses, we obtain:
%Note that in this case as the LO masses are different this should be taken into account when calculating the loops because now in the propagators different masses should appear. 
%At the NLO:
\begin{align}
M_{\text{NLO},1}^2 &= M_{\text{NLO},2}^2=M_{\text{NLO},4}^2=M_{\text{NLO},5}^2\nonumber \\
&=    \frac{M_1^4}{F^2}  \left( -8 \left( 4 L_4^r + L_5^r - 8 L_6^r - 2 L_8^r \right) - \frac{1}{64 \pi^2} \ln \left(\frac{M_1^2}{\mu^2}\right) \right), \label{NLOso.4} \\
M_{\text{NLO},3}^2 &=  - \frac{M_1^4}{F^2} 
8 \left( 4 L_4^r + L_5^r - 8 L_6^r \right) + 
\frac{32M^{4}_{1}}{F^{2}} \left( 2L^{r}_{7}\frac{\left( r-1\right)^{2}  }{\left( r+1\right)^{2}  } +L^{r}_{8}\frac{r^{2}+1}{\left( r+1\right)^{2}  } \right)  \nonumber \\
& + \frac{1}{ 64 \pi^2 F^2 } \left( 3 M_1^4 \ln \left(\frac{M_1^2}{\mu^2}\right) - 2 M_6^4 \ln\left( \frac{M_6^2}{\mu^2} \right)- 2 M_8^4 \ln\left( \frac{M_8^2}{\mu^2}\right) \right), \label{NLOso.5} \\
M_{\text{NLO},6}^2 &=M_{\text{NLO},7}^2\nonumber\\
&=  - \frac{M_6^2}{F^2} \left( 32 \left( L_4^r - 2 L_6^r \right) M_1^2 + 8 \left( L_5^r - 2 L_8^r \right) M_6^2 + \frac{M_1^2}{64 \pi^2} \ln \left(\frac{M_1^2}{\mu^2}\right) \right), \label{NLOso.6} \\
M_{\text{NLO},8}^2 & =M_{\text{NLO},9}^2  \nonumber \\
&=  - \frac{M_8^2}{F^2} \left( 32 \left( L_4^r - 2 L_6^r \right) M_1^2 + 8 \left( L_5^r - 2 L_8^r \right) M_8^2 + \frac{M_1^2}{64 \pi^2} \ln \left( \frac{M_1^2}{\mu^2}\right) \right). \label{NLOso.7}
\end{align}
where again $r=m_d/m_u$. The multiplet structure here corresponds to the results presented in \cite{Pomper:2024otb}.

\subsection{Condensates}
At leading order the condensates are the same:
\begin{align}
\left< \bar{u} u\right>_\text{LO}  &=\left< \bar{d} d\right>_\text{LO} =-2B_{0}F^{2},\label{condSO1}
 \end{align}
but they get a splitting at the NLO:
\begin{align}
\langle \bar{u} u \rangle_\text{NLO}  &= \langle \bar{u} u \rangle_\text{LO} \bigg(  \frac{64 M_1^2 L_6^r}{F^2} + \frac{8 B_0 m_u}{F^2} (H_2^r + 2 L_8^r) \nonumber \\ 
& \quad - \frac{5}{64 \pi^2} \frac{M_1^2}{F^2} \log\left(\frac{M_1^2}{\mu^2}\right) - \frac{4 M_6^2}{64 \pi^2 F^2} \log\left(\frac{M_6^2}{\mu^2}\right) \bigg), \label{condSO2} \\ 
\langle \overline{d} d \rangle_\text{NLO} &= \langle \bar{d} d \rangle_\text{LO} \bigg(  \frac{64 M_1^2 L_6^r}{F^2} + \frac{8 B_0 m_d}{F^2} (H_2^r + 2 L_8^r) \nonumber \\ 
& \quad - \frac{5}{64 \pi^2} \frac{M_1^2}{F^2} \log\left(\frac{M_1^2}{\mu^2}\right) - \frac{4 M_8^2}{64 \pi^2 F^2} \log\left(\frac{M_8^2}{\mu^2}\right) \bigg). \label{condSO3}
\end{align}

\subsection{Decay constants}

At LO all 9 mesons have the same decay constant $F_{\text{LO}}=F$ as in the \spth case. However, an important difference with the \spth arises at NLO where the pion decay constants split, following the same pattern as the meson masses:
\begin{align}
F_{\text{NLO},1} &= F_{\text{NLO},2}=F_{\text{NLO},4}=F_{\text{NLO},5}= \frac{M^{2}_{1}}{F} \left( 16L^{r}_{4}+4L^{r}_{5}\right) \\
&-\frac{1}{F} \frac{1}{32\pi^{2} } \left( M^{2}_{1}\ln \frac{M^{2}_{1}}{\mu^{2} } +\frac{1}{2} M^{2}_{6}\ln \frac{M^{2}_{6}}{\mu^{2} } +\frac{1}{2} M^{2}_{8}\ln \frac{M^{2}_{8}}{\mu^{2} } \right)    , \label{NLOso.8} \\
F_{\text{NLO},3} &=  \frac{M^{2}_{1}}{F} \left( 4\left( 4L^{r}_{4}+L^{r}_{5}\right)  -\frac{1}{16\pi^{2} } \ln \frac{M^{2}_{1}}{\mu^{2} } \right), \label{NLOso.9} \\
F_{\text{NLO},6} &=F_{\text{NLO},7}\nonumber\\
&= \frac{1}{F} \left( 16M^{2}_{1}L^{r}_{4}+4M^{2}_{6}L^{r}_{5}-\frac{1}{32\pi^{2} } M^{2}_{1}\ln \frac{M^{2}_{1}}{\mu^{2} } -\frac{1}{32\pi^{2} } M^{2}_{6}\ln \frac{M^{2}_{6}}{\mu^{2} } \right)   , \label{NLOso.10} \\
F_{\text{NLO},8} & =F_{\text{NLO},9} \nonumber \\
&= \frac{1}{F} \left( 16M^{2}_{1}L^{r}_{4}+4M^{2}_{8}L^{r}_{5}-\frac{1}{32\pi^{2} } M^{2}_{1}\ln \frac{M^{2}_{1}}{\mu^{2} } -\frac{1}{32\pi^{2} } M^{2}_{8}\ln \frac{M^{2}_{8}}{\mu^{2} } \right)    . \label{NLOso.11}
\end{align}

\subsection{Meson scattering}\label{secScatSO}

The formulas for the scattering amplitudes for the \soth theories are even more lengthy than in the \spth case due to LO pion mass splitting. Consequently, we present the full formulas for $\mathcal{M}^{ij\to kl}$ with general indices in the supplementary Mathematica package ~\cite{kolesova_2025_14864550} and we restrict ourselves to a subset of these expressions in this text. If we choose $m_d>m_u$, the lightest pions at LO are pions 6 and 7 according to formulas \eqref{NLOso.1}-\eqref{NLOso.3} (and this hierarchy is likely kept also at NLO, see \sec\ref{sec:apSO}). Within the pion dark matter setup, the relic abundances of the heavier species would be Boltzmann-suppressed in today's Universe, consequently, the self-interactions of the lightest species 6 and 7 would be most important for the dark matter phenomenology and we give the corresponding scattering amplitudes below.
When we consider only a subset of particles $\{6,7\}$, the amplitude can be expressed in terms of a single independent function $A(s,t,u)$ as 
\begin{align}\label{eqSO4AmplitudeNew}
\mathcal{M}^{ab\rightarrow cd} \left( s,t,u\right)  =\delta^{ab} \delta^{cd} A\left( s,t,u\right)  +\delta^{ac} \delta^{bd} A\left( t,u,s\right)  +\delta^{ad} \delta^{bc} A\left( u,s,t\right)
\end{align}
with $A(s,t,u)=A(s,u,t)$ and $a,b,c,d=6,7$.
\begin{comment}
In this case instead of 9 invariants (3.8) we will only have $3$ of them: $\delta^{ab} \delta^{cd} $, $\delta^{ac} \delta^{bd} $, $\delta^{ad} \delta^{bc}$, because the traces of products of 4 generators $\text{Tr}\left( X^{a}X^{b}X^{c}X^{d}\right)$ with $a,b,c,d=6,7$ can be rewritten in terms of Kronecker deltas. Crossing symmetries further simplify the problem and we are left with just one function: 

\end{comment}
The LO part of the function $A(s,t,u)$ reads
\begin{align}\label{eqSO4AmplitudeNew1}
A_{\text{LO}}\left( s,t,u\right)  =\frac{M^{2}_{\pi,6}+2s-t-u}{3F^{2}_{\pi,6}}
\end{align}
while the NLO correction is given by
\begin{align}\label{eqSO4AmplitudeNew2}
&F^{4}_{\pi,6}\times A_{\text{NLO}}\left( s,t,u\right)= \nonumber\\
&\frac{1}{1152} \bigg( 2M^{2}_{\pi,1}\left( 6M^{2}_{\pi,6}+2M^{2}_{\pi,8}+5s-t-u\right)  +M^{2}_{\pi,6}\left( -6M^{2}_{\pi,8}-101s+37(t+u)\right)  \nonumber\\
&-5sM^{2}_{\pi,8}+tM^{2}_{\pi,8}+uM^{2}_{\pi,8}-4M^{4}_{\pi,1}  \nonumber\\
& -38M^{4}_{\pi,6}-M^{4}_{\pi,8}-6s^{2}+6st+6su-12t^{2}+18tu-12u^{2}\bigg)\nonumber \\ &+4\left( 4M^{2}_{\pi,6}(s-t-u)+4M^{4}_{\pi,6}-s^{2}+t^{2}+u^{2}\right)  L^{r}_{0}+8\left( s-2M^{2}_{\pi,6}\right)^{2}  L^{r}_{1}\nonumber\\
&+4\left( -4M^{2}_{\pi,6}(t+u)+8M^{4}_{\pi,6}+t^{2}+u^{2}\right)  L^{r}_{2}+4\left( s-2M^{2}_{\pi,6}\right)^{2}  L^{r}_{3}\nonumber\\
&+\frac{16}{3} \bigg( M^{2}_{\pi,6}\left( -2M^{2}_{\pi,8}+s+t+u\right)  +M^{2}_{\pi,8}(-2s+t+u) \nonumber \\ &+M^{2}_{\pi,1}\left( 4M^{2}_{\pi,6}+4s-2(t+u)\right)  -8M^{4}_{\pi,6}\bigg)  L^{r}_{4}\nonumber\\
&+8M^{2}_{\pi,6}\left( s-2M^{2}_{4,6}\right)  L^{r}_{5}+\frac{32}{3} M^{2}_{\pi,6}\left( -2M^{2}_{\pi,1}+4M^{2}_{\pi,6}+M^{2}_{\pi,8}\right)  L^{r}_{6}+16M^{4}_{\pi,6}L^{r}_{8}\nonumber\\
&+\frac{1}{72} \bigg( -2M^{2}_{\pi,1}\left( 6M^{2}_{\pi,6}+2M^{2}_{\pi,8}+5s-t-u\right)  +\left( M^{2}_{\pi,8}+3s\right)  \left( M^{2}_{\pi,8}+2s-t-u\right) \nonumber\\
& +M^{2}_{\pi,6}\left( 6M^{2}_{\pi,8}+17s-t-u\right)  +4M^{4}_{\pi,1}+14M^{4}_{\pi,6}\bigg)  \bar{J} \left( M^{2}_{\pi,1},M^{2}_{\pi,1},s\right) \nonumber \\
& +\frac{1}{24} (s-u)\left( t-4M^{2}_{\pi,1}\right)  \bar{J} \left( M^{2}_{\pi,1},M^{2}_{\pi,1},t\right) \nonumber\\
& +\frac{1}{24} (s-t)\left( u-4M^{2}_{\pi,1}\right)  \bar{J} \left( M^{2}_{\pi,1},M^{2}_{\pi,1},u\right) \nonumber\\
& -\frac{1}{6} M^{2}_{\pi,6}\left( 2M^{2}_{4,6}-5s+t+u\right)  \bar{J} \left( M^{2}_{\pi,6},M^{2}_{\pi,6},s\right)\nonumber\\
&  +\frac{1}{6} \left( -M^{2}_{\pi,6}(s+3t-3u)+2M^{4}_{\pi,6}+t(t-u)\right)  \bar{J} \left( M^{2}_{\pi,6},M^{2}_{\pi,6},t\right)  \nonumber\\
&+\frac{1}{6} \left( -M^{2}_{\pi,6}(s-3t+3u)+2M^{4}_{\pi,6}+u(u-t)\right)  \bar{J} \left( M^{2}_{\pi,6},M^{2}_{\pi,6},u\right) \nonumber \\
&+\frac{1}{5760\pi^{2} } \bigg( -5\bigg( M^{2}_{\pi,6}\left( 6M^{2}_{\pi,8}+17s-t-u\right)  +M^{2}_{\pi,8}(5s-t-u)+14M^{4}_{\pi,6} +M^{4}_{\pi,8} \nonumber \\
& +6\left( s^{2}-tu\right)  \bigg)  +M^{2}_{\pi,1}\left( 56M^{2}_{\pi,6}+32M^{2}_{\pi,8}+54s-6t-6u\right)  -44M^{4}_{4,1}\bigg)  \log \left( \frac{M^{2}_{\pi,1}}{\mu^{2} } \right)\nonumber \\
& +\frac{1}{1440\pi^{2} } \left( M^{2}_{\pi,6}(49(t+u)-101s)-46M^{4}_{\pi,6}-15(t-u)^{2}\right)  \log \left( \frac{M^{2}_{\pi,6}}{\mu^{2} } \right). 
\end{align}
%The full amplitudes for both \spth and \soth theories for all species are provided in the supplementary Mathematica package \cite{kolesova_2025_14864550}.

\section{NLO LECs for the \spth theory}\label{sec:fit}

In this section, we determine the NLO LECs for the case of $Sp(N_c=4)$ gauge theory with $N_F=2$ quarks in the fundamental representation, described by the \spth EFT at low energies. To this end, we use the lattice data for non-degenerate pion masses and decay constants presented in \cite{Kulkarni:2022bvh}, and the data on pion scattering lengths in the mass-degenerate case from \refe\cite{Dengler:2024maq}, see \sec\ref{sec:FitData} for details. In \sec\ref{sec:FitFit}, we fit the LEC entering the NLO formulas presented in \sec\ref{sec:Sp} to the output of these lattice calculations. Let us note that a similar fit for the $SU(N_c=2)\cong Sp(N_c=2)$ gauge theory was performed in~\cite{Arthur:2016dir} while for the real-world QCD based on the $SU(N_c=3)$ gauge group, the fits of LEC are based both on experimental and lattice data, see~\cite{Bijnens:2014lea} for a review.

%\subsection{Fitting of low-energy constants in other works}
%\hk{This definitely won't be a separate subsection, let's see if we'll put it somewhere.}
%The low-energy constants for Standard Model 2-flavour ChPT were fitted from experimental data and lattice calculations. In particular, the values of $\bar l_{1,2,4}$ were determined by matching of $\chi$PT to experimental input at intermediate energies via Roy equations (see refs.~\cite{Ananthanarayan:2000ht,Colangelo:2001df} for details). On the other hand, the error of $\bar{l}_3$ can be substantially reduced by taking into account lattice simulations, see~\cite{Baron:2011sf,Aoki:2019cca}.

%Another relevant reference is~\cite{Amoros:2001cp} where LEC in Standard Model 3-flavour ChPT were fitted to give correct predictions for meson masses and decay properties, taking into account also the isospin breaking (i.e., more LEC appear, as in our mass-split case). When stating the results for LEC, they say: ``Errors are fitting errors as quoted by MINUIT'', so perhaps they are using this software?

%Finally, in~\cite{Bennett:2019jzz}, fit of LEC in hidden-local-symmetry Lagrangian to lattice data was performed. The seem to employ ``constrained bootstrapped $\chi^2$ minimisation''.

%\dk{Maybe one paper which we need to read and cite here is this review by Bijnens \cite{Bijnens:2014lea} . I have not read it but it seems to be an important one (even by number of citations)}

\subsection{Lattice data}\label{sec:FitData}

%In order to get an estimate of the values of the $Sp(4)$ LECs we use the data presented in \cite{Kulkarni:2022bvh} for non-degenerate pion masses   and decay constants and \cite{Dengler:2024maq} for the scattering length of the pions with degenerate masses
%\hk{The content of this footnote and the next sentence should be moved somewhere below.}\footnote{
%The connection between the dimensionful scattering length $a_0$ calculated in  \cite{Dengler:2024maq} and  the  dimensionless scattering length $a_0^{MS}$ from \cite{Bijnens:2011fm} is $a_0m_\pi=-a_0^{MS}$ }. 
\subsubsection{Pion masses and decay constants}
In \cite{Kulkarni:2022bvh}, lattice data are presented for three different values of the ratio of the vector meson mass to the pion mass in the case of degenerate quark masses,  $M_{\rho} / M_{\pi} \approx 1.14, 1.24, 1.46$. We use the last dataset, corresponding to  $M_{\rho} / M_{\pi} \approx 1.46$, as it is closest to the chiral limit. Each dataset provides values for the pion and vector meson masses, as well as the corresponding decay constants. In our notation, the mass of the singlet pion is denoted by $M_{\pi,3}$, while the pions in the four-plet have mass $M_{\pi,1}$, these two masses correspond to $m_C$ and $m_A$ from~\cite{Kulkarni:2022bvh}, respectively.

The results are given for three different values of the inverse gauge coupling: $\beta = 6.9, 7.05, 7.2$, with larger $\beta$  corresponding to smaller lattice spacing. %closer proximity to the continuum limit. 
The continuum limit is not performed; however, the lattice systematics are discussed, concluding that the results obtained at finite lattice spacing give a good approximation to the continuum values.\footnote{The continuum extrapolation for the meson masses in the mass-degenerate case was performed in~\cite{Bennett:2019jzz}. However, the PCAC quark mass was not measured in this study, hence, these results do not provide additional information for our fits. On the other hand, parameters entering other type of low-energy EFT including also vector mesons were fitted in~\cite{Bennett:2019jzz}.} We work with the data for each value of $\beta$ separately and comparison of the results later allows us to asses the discretization errors. Further, Ref.~\cite{Kulkarni:2022bvh} presents data for different lattice volumes which allowed us to perform the infinite volume limit for most of the data points and we use these extrapolated data in our fits, see \app\ref{AppendixLatticeData} for details.

The lattice data of \cite{Kulkarni:2022bvh} show that quark mass non-degeneracy leads not only to a splitting in the pion masses, but also to a splitting in the decay constants. The authors attempt to reproduce this behavior in their effective theory by allowing for different chiral condensates for the two quark flavors already at leading order. While it may capture the behaviour at LO, this approach makes it more difficult to account for the effects of non-zero quark masses within a consistent power counting scheme. Within the standard chiral effective theory used in this work, the splitting of decay constants is expected to appear only at NNLO. %Since we employ NLO expressions that do not feature the splitting of the decay constant, we use an average value of $F_\pi$ from~\cite{Kulkarni:2022bvh} in our fit (see \fig\ref{figFitDec} for the example of the $\beta = 7.2$ data for $F_\pi$ that we employ).
Since we employ NLO expressions that do not feature the splitting of the decay constant, we construct a single effective dataset by averaging the two values of $F_\pi$ provided in \cite{Kulkarni:2022bvh} and assign it an uncertainty that combines the quoted statistical error with a systematic contribution given by half the observed channel difference (see \fig\ref{figFitDec} for an example of the $\beta = 7.2$ data). %This procedure allows us to consistently use the NLO expression while accounting for the missing higher-order splitting.

To characterize the quark mass splitting, the so-called PCAC quark masses are used. These are unrenormalized quark masses extracted from lattice correlation functions using the partially conserved axial current relation. They are commonly used in lattice QCD, as they can be computed without knowledge of the renormalization constants. Importantly, the renormalization factors cancel in the ratio of PCAC masses $m_u^{\text{PCAC}}/m_d^{\text{PCAC}}$, and we identify this ratio with the quantity $r$ entering the NLO formula for the pion mass as defined in equation ~\eqref{eq.Sp4.42.1}.  %refer to the PCAC masses of the up- and down-type quarks, respectively. 

For all values of $\beta$, when $ r \gtrsim 5.5 $, the mass of the lightest vector meson becomes very close to the mass of the heavier pion. We expect that in this regime our EFT is no longer applicable. Therefore, we impose a cutoff at  $r = 5.5$, retaining only data points with  $r \leq 5.5$. This leaves us with five usable values of $ r $ for  $\beta = 6.9$  and $ \beta = 7.2$, and four values for $\beta = 7.05$ (see \fig\ref{figFitMass} for the $ \beta = 7.2$ data).

Note that the results of~\cite{Kulkarni:2022bvh} for dimensionful quantities are given in terms of the inverse lattice spacing that varies for different values of $\beta$. Consequently, the values of pion masses and decay constants cannot be directly compared across datasets with different $\beta$; on the other hand, any dimensionless ratio of these observables should be independent of $\beta$ if the discretization artifacts are small enough. Similarly, the NLO LECs are dimensionless and should, in principle, be independent of $\beta$ in absence of discretization errors. %In practice, however, discretization artifacts can introduce a mild dependence. %\dk{Is this phrase correct? Probably not}

%\dk{Add here about that fact that the continuum limit was not performed. Also comment on this for the Maas paper}. \hk{Also speak about the general issue of scale setting.}

\subsubsection{Scattering lengths}
In \cite{Dengler:2024maq}, the authors present results for the scattering length of two mass-degenerate pions (i.e. $M_{\pi ,1}=M_{\pi ,3}\equiv M_{\pi }$) obtained using the L\"{u}scher’s method \cite{Luscher:1986pf}. The two-pion operators are constructed from interpolating fields of the form $\bar{u}(x)\gamma_5d(x)$, and the scattering is analyzed in the 14-dimensional irreducible representation of the flavor $Sp(4)$ symmetry group, sometimes referred to as the ``isospin-2'' channel in analogy to QCD. This particular channel does not contain single-vector-meson states that live in the 10-dimensional representation of $Sp(4)$.
The values of the scattering length $a_0M_\pi$ are given as a function of $M_\pi/F_\pi$ and this scattering length can be related to our results on scattering amplitudes as shown in \app\ref{AppendixChannels}. 

In \cite{Dengler:2024maq}  there are several datasets presented (see \cite{dengler_2024_12920978} for the data release). In our analysis, we use the scattering length data extracted from the bottom panel of figure~3 in \cite{Dengler:2024maq}. This dataset corresponds to a subset of lattice ensembles selected by the authors to reduce systematic uncertainties due to discretization and finite-volume effects. In particular, they exclude: (i) ensembles with small spatial volume ($N_L = 8$), (ii) ensembles where the finite-volume pion mass deviates significantly from its infinite-volume value, (iii) ensembles with too small two-pion energy, %smaller then $0.95$ in lattice units 
which may be more affected by discretizations artifacts. 
%\begin{itemize}
    %\item  ensembles with small spatial volume ($N_L = 8$),
    %\item ensembles where the finite-volume pion mass deviates significantly from its infinite-volume value,
    %\item ensembles with two-pion energy smaller then $0.95$, which may be more affected by discretizations artifacts.
%\end{itemize}
These restrictions, described in their appendix~A, define a cleaner subset of data from which more reliable values for the scattering length can be extracted although the continuum limit is again not performed.
In this dataset, there is only one data point for $\beta=6.9$  while for both values $7.05$ and $7.2$ there are two points available (see \fig\ref{figFitScat} for the $\beta=7.2$ data).

Let us note that the infinite volume limit is performed as an integral part of the L\"{u}scher’s method, hence, the infinite-volume values of pion masses and scattering lengths can be directly read off from~\cite{Dengler:2024maq}. Although the values of $F_\pi$ are given at finite volume, the lattices are large enough, hence, the corresponding systematic errors are expected to be smaller than the statistical ones, see \app\ref{AppendixLatticeData} for details.

\subsection{Fitting the LEC}\label{sec:FitFit}
We perform a global fit using the NLO formulas~\eqref{eq.Sp4.42}, \eqref{eq.Sp4.42.1} for the pion masses, formula~\eqref{eq.Sp4.48} for the pion decay constant, and finally, formula~\eqref{eq.Sp4.Fit.5} for the scattering length that can be obtained from the scattering amplitudes presented in \sec\ref{secScat} as explained in \app\ref{AppendixChannels}.

For the fit, it is useful to replace the $\mu$-dependent LECs $L_i^r(\mu)$ with the $\mu$-independent LECs $\bar{L}_{i}$~\cite{Gasser:1983kx,Colangelo:2001df}:
%NB: scale-independent LEC were fitted also in~\cite{Necco:2008ish}, on the other hand, Sannino is for some reason referring to scale-dependent LEC, both in~\cite{Arthur:2016dir} and~\cite{Hansen:2015yaa}.
\begin{align}
L^{r}_{i}(\mu )=\frac{\Gamma_{i} }{32\pi^{2} } \left( \bar{L}_{i} +\log \frac{M^{2}_{\pi,1}}{\mu^{2} } \right)\label{eq.Sp4.Fit.1}\,,\quad i\neq7,8
\end{align}
where the coefficients $\Gamma_i$ are given in \tableTag\ref{tab:LECs_table}.\footnote{The $\mu$-independent LECs can be also viewed as LECs at the scale $\mu=M_\pi$. Let us note that in~\cite{Colangelo:2001df}, where NNLO calculations were performed, LECs at scales $\mu = M$ and $\mu = M_\pi$ are distinguished, however, when working at NLO order, these two definitions are equivalent.} %Notice also that \refe\cite{Hansen:2015yaa} refers to scale-dependent LECs and chooses the scale $\mu = 2M_\pi$ when describing the pion self-interactions. 
A key advantage of this definition is that it eliminates chiral logarithms from the NLO expressions for observables.
\begin{comment}
    Notice also that the LECs $\bar{L}_{i}$ assume $\mathcal{O}(1)$ values while, e.g., \refes\cite{Amoros:2001cp,Hansen:2015yaa} work with the scale-dependent LECs $L^{r}_{i}(\mu)$ with $\mathcal{O}(10^{-4})$ values.
\end{comment}
For $L_7^r$ and $L_8^r$, $\Gamma_7=\Gamma_8=0$,\footnote{For $\Gamma_7$ this holds for any $N_F$, while for $\Gamma_8$ this is specific to $N_F=2$.} but we use following rescaling so that all LECs are of the same order of magnitude:
\begin{equation}\label{eq:rescaledLECs}
    L^{r}_{i}=\frac{1}{128\pi^{2} } \bar{L}_{i} ,\  i=7,8.
\end{equation}

The available lattice data do not allow to fit all the LECs individually, we are limited to estimating following linear combinations:
\begin{align}
\tilde{L}_{1} &\equiv\bar{L}_{4} +\bar{L}_{5},\label{eqLtilde1}\\
\tilde{L}_{2} &\equiv\frac{1}{4}\bar{L}_{0} -\frac{3}{4}\bar{L}_{1}-\frac{3}{2}\bar{L}_{2} -\bar{L}_{3},\label{eqLtilde2}\\
\tilde{L}_{3} &\equiv\bar{L}_{6} +\frac{8}{5}\bar{L}_{8},\label{eqLtilde3}\\
\tilde{L}_{4} &\equiv\bar{L}_{8} +4\bar{L}_{7}.\label{eqLtilde4}
\end{align}
Using these definitions, the formulas~\eqref{eq.Sp4.42}, \eqref{eq.Sp4.42.1}, \eqref{eq.Sp4.48} and~\eqref{eq.Sp4.Fit.5} (summed with the corresponding LO contributions) can be rewritten as follows:
\begin{align}
M^2_{\pi,1}&=M^{2}\left( 1+\frac{1}{64\pi^{2} } \frac{M^{2}}{F^{2}} \left[ -4 \tilde{L}_{1} +5\tilde{L}_3 \right]  \right)  ,\label{eq.Sp4.Fit.mA.new.new} \\
   M^2_{\pi,3}&=M^{2}\left( 1+\frac{1}{64\pi^{2} } \frac{M^{2}}{F^{2}}\left[ -4\tilde{L}_{1} +5\tilde{L}_{3} +8\frac{\left( 1-r\right)^{2}  }{\left( 1+r\right)^{2}  } \tilde{L}_{4} \right]    \right)  ,\label{eq.Sp4.Fit.mC.new}\\
  F_{\pi}&=F\left( 1+\frac{1}{32\pi^{2} } \frac{M^{2}}{F^{2}} \tilde{L}_{1} \right) , \label{eq.Sp4.Fit.fpi}\\
a^{MS}_{0}&=-\frac{1}{32\pi } \frac{M^{2}_{\pi }}{F^{2}_{\pi }} -\frac{1}{6144\pi^{3} } \left( \frac{M^{2}_{\pi }}{F^{2}_{\pi }} \right)^{2}  \left( 3+24\tilde{L}_{1} +16\tilde{L}_{2} -15\tilde{L}_{3}  \right) \,. \label{eq.Sp4.Fit.a0}
\end{align}
Recall that $r \equiv m_d/m_u$, hence, the sign of $\tilde{L}_4$ determines the pion mass hierarchy in the case of split quark masses $r\neq 1$: $\tilde{L}_4 < 0$ implies $M^2_{\pi,3} < M^2_{\pi,1}$, while $\tilde{L}_4 > 0$ implies $M^2_{\pi,3} > M^2_{\pi,1}$. Let us note that our strategy is to compare the lattice data with the observables~\eqref{eq.Sp4.Fit.mA.new.new}-\eqref{eq.Sp4.Fit.a0} calculated at NLO precision. This corresponds to first fits of LECs within SM QCD~\cite{Gasser:1983kx,Gasser:1983yg} while with more data available, fits using NNLO formulas were performed later, see, e.g.,~\cite{Bijnens:2014lea}. Given the limited amount of data available in our case, we restrict ourselves to fitting the NLO formulas stated above.

Let us note that in eq. \eqref{eq.Sp4.Fit.1}, the LECs $\bar{L}_i$ would in principle acquire the dependence on
quark masses through $M_{\pi,1}$. To keep $\bar{L}_i$ independent of quark masses, one may instead fix $M_{\pi,1}$ 
 to a reference value $M_{\pi,0}$. In that case, the chiral logarithms in eqs. \eqref{eq.Sp4.Fit.mA.new.new}–\eqref{eq.Sp4.Fit.a0} do not cancel, but instead become
 $\log \left( M^{2}_{\pi ,1}/M^{2}_{\pi ,0}\right)$ . We have verified that, 
 for  $M^2_{\pi,0}\approx0.18a^{-2}$, this
logarithm is smaller or comparable than the statistical errors on $\bar{L}_i$ obtained in the fits below
and was therefore neglected. This effect will be included in future work.

In total, there are 6 parameters to fit. They are the two parameters of the LO Lagrangian: $B_{0} m_u$ (from $M^2=B_{0}m_{u}\left( 1+r\right)$) and $F$, as well as the 4 linear combinations of the LECs: $\tilde{L}_{1}$, $\tilde{L}_{2}$, $\tilde{L}_{3}$, $\tilde{L}_{4}$.
Let us anticipate that the fully non-relativistic $2\to2$ cross section for the mass-degenerate case averaged over all different scattering channels 
%($s\to4M_{\pi}^2$, $t,u\to0$) 
only depends on the same linear combination of LECs as $a_0^{MS}$, which will allow us to use the estimated values of LECs for the dark matter phenomenology in \sec\ref{sec:Sp4.Applications}.

\begin{figure}[htb]
\centering
\includegraphics[width=0.6\textwidth]{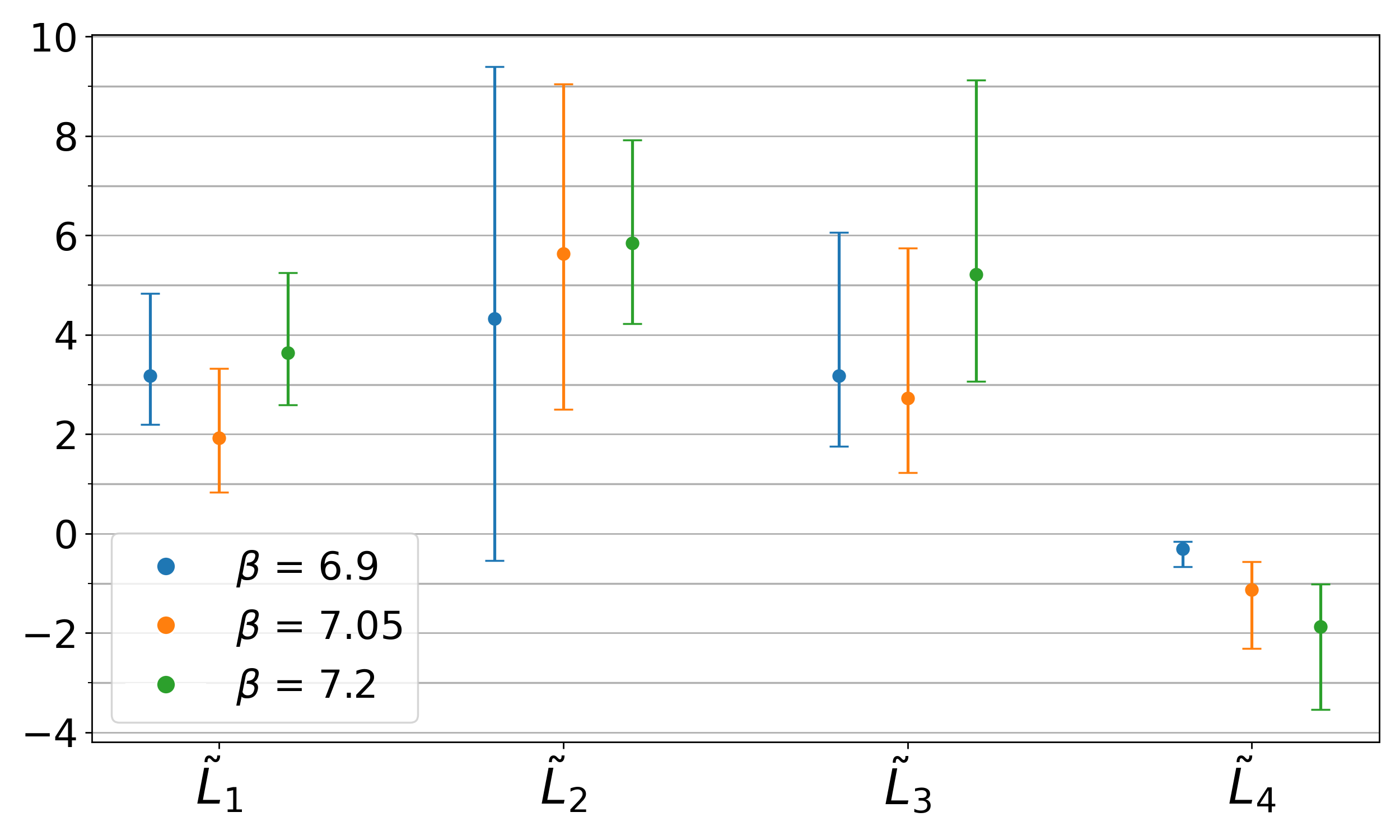}
\caption{Linear combinations of LECs from \eqs\eqref{eqLtilde1}-\eqref{eqLtilde4} for different values of $\beta$. Best-fit values together with 1$\sigma$-uncertainties are displayed.}
\label{fig:LECs_vs_beta}
\end{figure}

We performed a Bayesian fit using Markov Chain Monte Carlo (MCMC) sampling to explore the posterior distributions of the model parameters. The sampling was carried out with the \texttt{emcee}  package \cite{Foreman-Mackey:2012any}. Both datasets from~\cite{Kulkarni:2022bvh} and \cite{Dengler:2024maq} include significant uncertainties in the horizontal axis variables, specifically, the PCAC mass ratio $r$
and the dimensionless ratio $M_\pi/F_\pi$. Therefore, in our Bayesian analysis, the true values of $r$ and $M_\pi/F_\pi$  corresponding to each individual data point were treated as latent variables and marginalized over. This approach ensures that uncertainties in both the dependent and independent variables are properly propagated.
The likelihood function for each dataset was assumed to be Gaussian. In the case of the scattering length data from Ref.~\cite{Dengler:2024maq}, where asymmetric errors are reported, we conservatively adopted the larger of the two uncertainties as an effective symmetric error. This allowed us to model the likelihood using a standard Gaussian form.
For all the linear combinations of LECs, we assumed Gaussian priors centered at zero with a standard deviation of $10$. This choice is motivated by the observed size of the QCD NLO LECs, which are of order
$ L_i^r \sim \mathcal{O}(1)/(16\pi^2)$ \cite{Scherer:2012xha}. We have verified that the fit results are robust under moderate variations of this prior width.
For the parameters  $F$ and $B_0 m_u$ we adopted uniform priors over the interval $\left[ 0,0.1\right]$ in the units of inverse lattice spacing.

\begin{figure}[htb]
\centering
\includegraphics[width=0.6\textwidth]{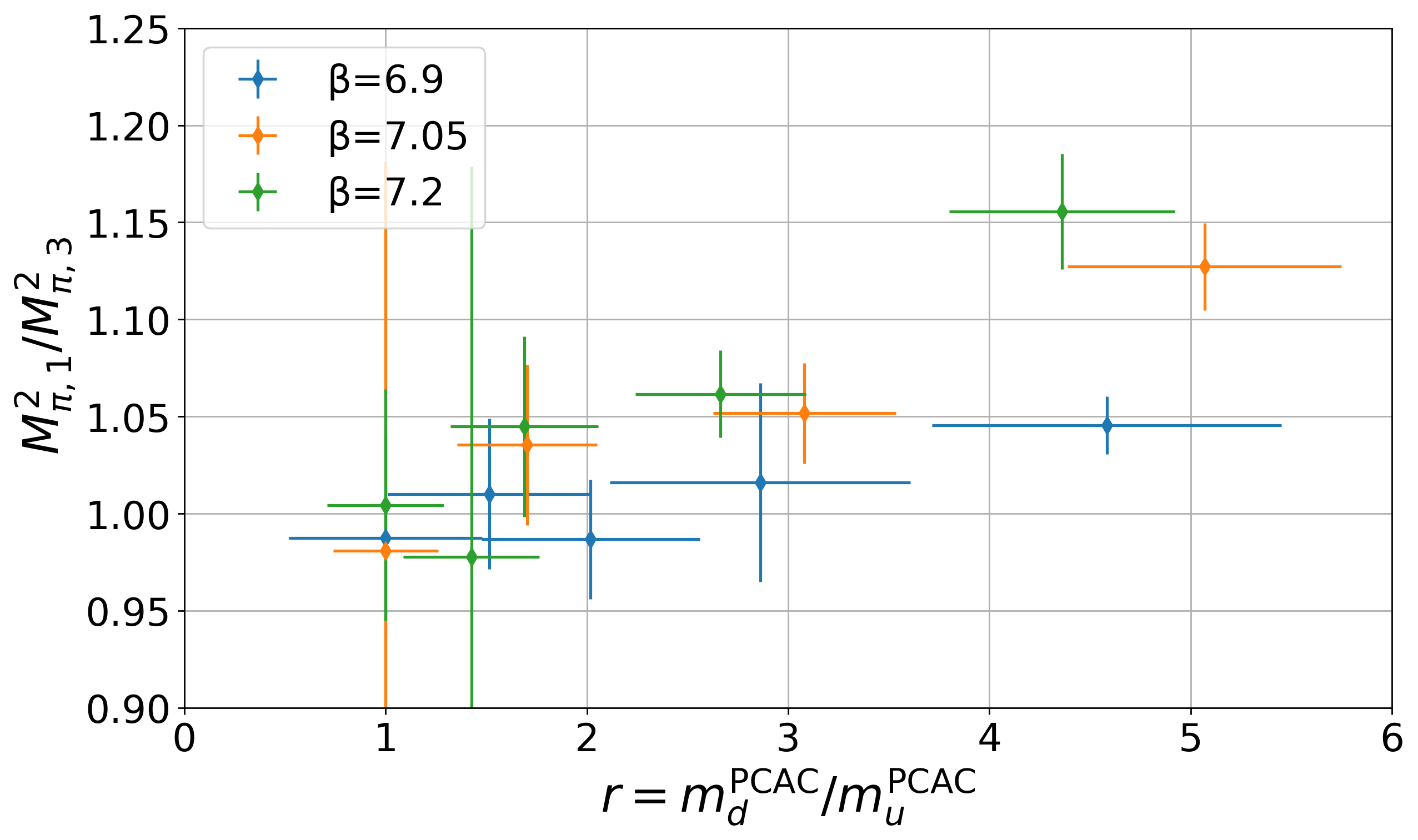}
\caption{The ratio of pion masses obtained by lattice calculations of \refe\cite{Kulkarni:2022bvh}. This dimensionless ratio should be in principle independent of the lattice spacing and, hence, of the value of the inverse gauge coupling $\beta$. In contrast, variation of $M_{\pi,1}^2/M_{\pi,3}^2$ with $\beta$ is observed that may point to discretization artifacts. This behavior then also leads to $\beta$-dependence of the fitted value of the $\tilde{L}_4$ LEC that controls the pion mass hierarchy.}
\label{fig:ratioOfMasses}
\end{figure}

In order to estimate the uncertainties related to lattice discretization, we performed the fits to the data for different values of $\beta$. The results for the fitted LECs are presented in figure \ref{fig:LECs_vs_beta}, including $1\sigma$ errors.\footnote{Here and throughout the paper, $n\sigma$ intervals refer to central credible regions of the posterior distribution containing the same probability mass as $n\sigma$ intervals of a one-dimensional Gaussian. In practice, these are computed as the central quantile ranges: $1\sigma$ corresponds to the 16th–84th percentiles (68\% credible interval), $2\sigma$ to the 2.3rd–97.7th percentiles (95\% credible interval), and $3\sigma$ to the 0.15th–99.85th percentiles, without assuming Gaussianity of the underlying distributions.} We observe that the values of $\tilde{L}_1$, $\tilde{L}_2$ and $\tilde{L}_3$ are consistent among the different $\beta$ within errors, however, variation of $\tilde{L}_4$ is more pronounced. Recall that $\tilde{L}_4$ controls the amount of splitting between the dark pion masses and the difference in its values for different $\beta$ can be traced back to the fact that the lattice results for the ratio $M^2_{\pi,1}/M^2_{\pi,3}$ vary for different $\beta$, see figure \ref{fig:ratioOfMasses}. This points towards possible discretization artifacts; on the other hand, the prediction of $\tilde{L}_4$ being negative, i.e., the singlet pion being lighter than the four-plet ones, seems to be robust, independent of $\beta$.

\begin{figure}[htbp]
\centering
\includegraphics[width=0.6\textwidth]{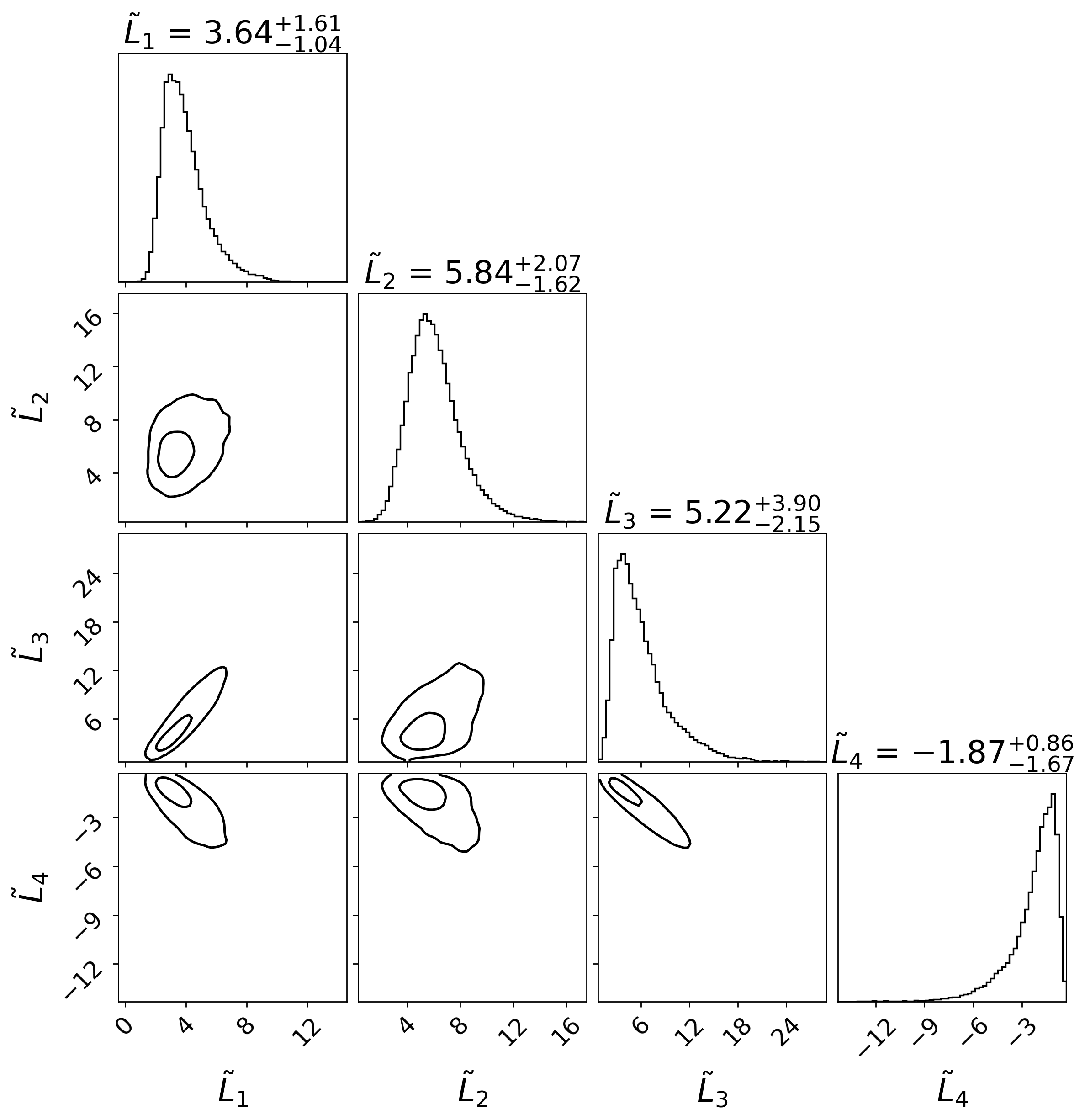}
\caption{
Triangle plot showing the posterior inference for the linear combinations of scale-independent LECs $\tilde{L}_1$, $\tilde{L}_2$, $\tilde{L}_3$, and $\tilde{L}_4$ for $\beta=7.2$. The diagonal panels display the marginalized one-dimensional posteriors, with the median and 68\% credible intervals (16th–84th percentiles) indicated above each histogram. Off-diagonal panels show the marginalized two-dimensional posteriors, with contours enclosing 39.3\% and 86.5\% of the probability mass. The sharp behavior observed for $\tilde{L}_4$ on the right-hand side is due to the fact that the data strongly suggests the singlet $\pi_3$ to be the lightest state.
}
\label{fig:posteriors}
\end{figure}

For our phenomenological analysis presented in \sec\ref{sec:Sp4.Applications}, we use the values of the dimensionless LECs for $\beta=7.2$ since larger values of $\beta$ imply being closer to the continuum limit. The fitted values of these LECs are summarized in Fig.~\ref{fig:posteriors}, which shows the marginalized posterior distributions for each parameter, together with their median values and associated $1\sigma$ uncertainties. Let us note that $\tilde{L}_4$ that might be affected by discretization errors does not enter our results on the dark matter self-interactions in the mass-degenerate case. %For the other dimensionless parameters, $\tilde{L}_1$, $\tilde{L}_2$ and $\tilde{L}_3$ we use the values obtained for $\beta=7.2$ since larger values of $\beta$ imply being closer to the continuum limit. \TableTag\ref{tableLECs} summarizes the fitted values of all dimensionless LECs for $\beta=7.2$,
Let us add for completeness that the  values of the dimensionful parameters obtained for $\beta=7.2$ read $B_0m_u=0.026(7)$ and $F=0.046(5)$ in units of inverse lattice spacing, however, these do not enter the phenomenology analysis in \sec\ref{sec:Sp4.Applications} either.

\begin{figure}[htbp]
    \centering
\includegraphics[width=0.6\textwidth]{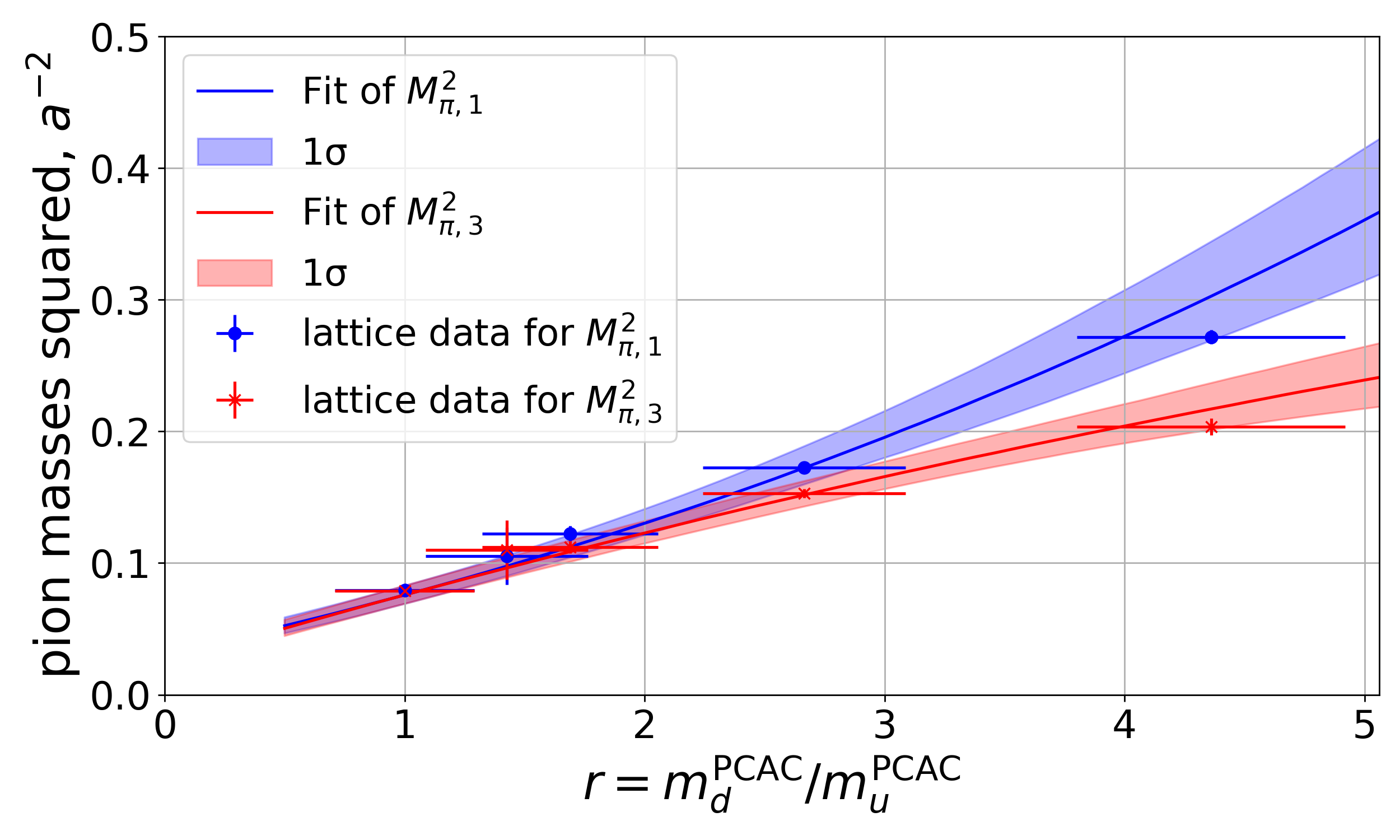}
    \caption{Lattice data \cite{Kulkarni:2022bvh} and fits for masses $M^2_{\pi,1}$ and $M^2_{\pi,3}$ as a function of $r$. The colored bands correspond to regions obtained when the LECs are varied within the 1$\sigma$ errors.}\label{figFitMass}
\end{figure}

\begin{figure}[htbp]
    \centering
    \includegraphics[width=0.6\textwidth]{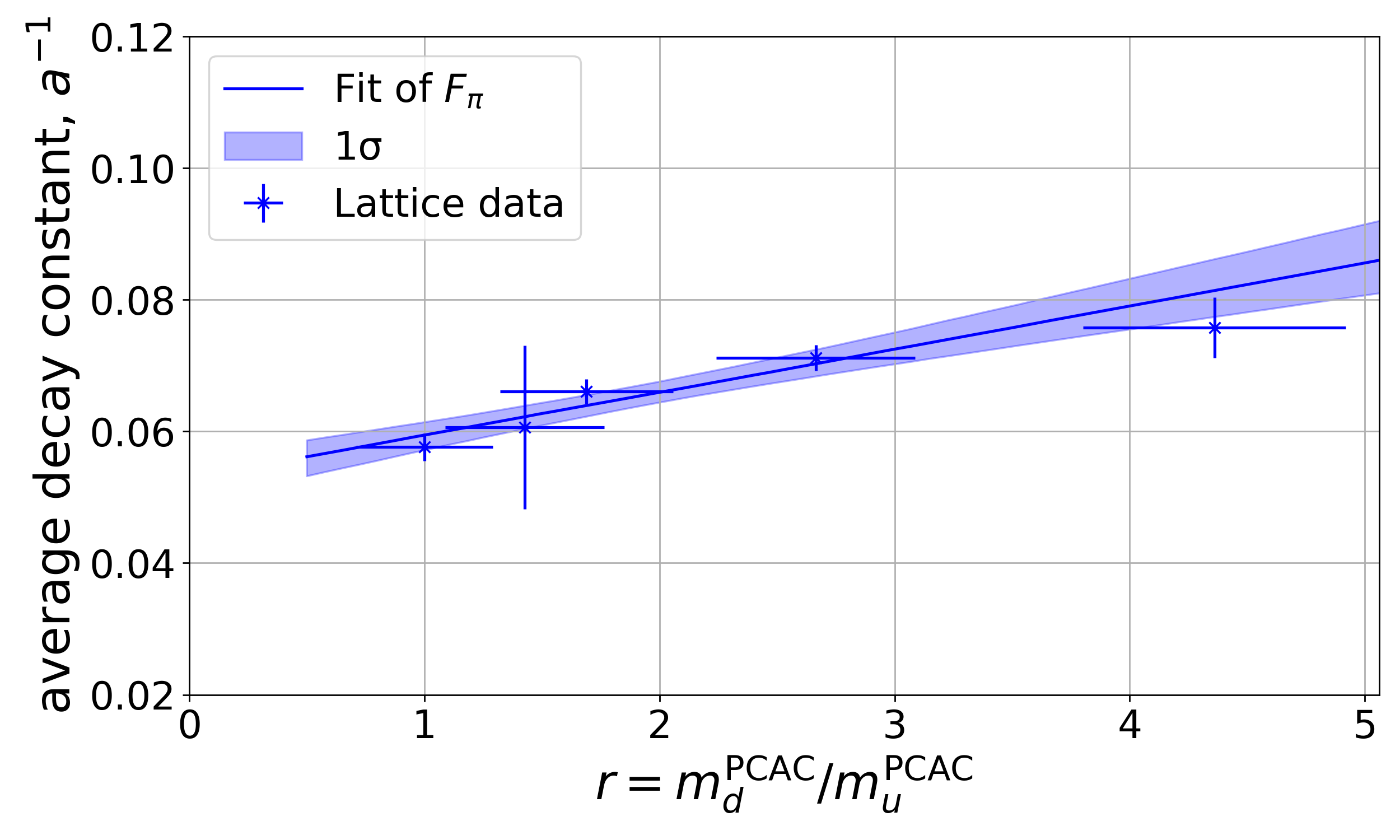}
    \caption{Lattice data \cite{Kulkarni:2022bvh} and fit for  a decay constant $F_\pi$ as a function of $r$. Since NLO calculations do not account for the splitting observed in the lattice results, we use the average of the two decay constants $F_{\pi,1}$ and $F_{\pi,3}$. The error bars shown combine the statistical uncertainties of the two channels with an additional systematic contribution equal to half of their difference, thereby reflecting the uncertainty due to neglecting the splitting.
}\label{figFitDec}
\end{figure}

\begin{figure}[htbp]
    \centering
    \includegraphics[width=0.6\textwidth]{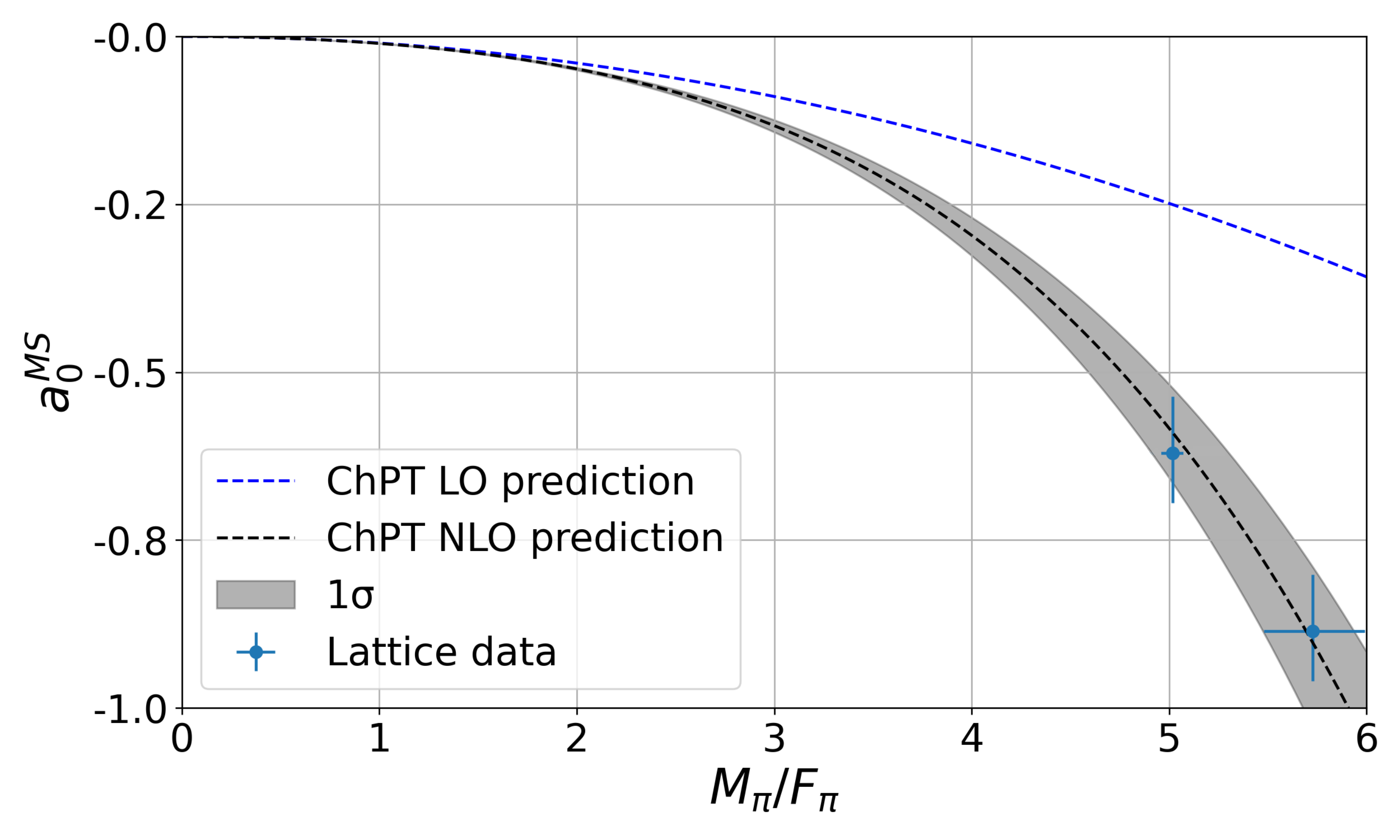}
    \caption{Lattice data \cite{dengler_2024_12920978} and a fit for the scattering length $a_0^\text{MS}$ as a function of $M_\pi/F_\pi$. The LO result as shown as a reference.}

    \label{figFitScat}
\end{figure}

Finally, in \figs\ref{figFitMass},~\ref{figFitDec} and~\ref{figFitScat}, we display the results of our fits for the pion masses, decay constant, and scattering length, respectively, together with the corresponding lattice data. The colored bands depict the regions obtained when the LECs are varied within the 1$\sigma$ uncertainties.

\begin{comment}
    \renewcommand{\arraystretch}{1.5}
\begin{table}[ht]
\centering
\begin{tabular}{ c c c c c c}
\hline
 $B_0 m_u$ & $F$ & $\tilde{L}_1$ & $\tilde{L}_2$ & $\tilde{L}_3$ & $\tilde{L}_4$ \\
\hline
$0.0310^{+0.0073}_{-0.0072}$ & 
$0.0445^{+0.0039}_{-0.0043}$ & 
$2.90^{+1.16}_{-0.76}$ & 
$5.57^{+1.60}_{-1.40}$ & 
$3.64^{+2.56}_{-1.38}$ & 
$-1.25^{+0.51}_{-1.04}$ \\
\hline
Goodness of fit & $\chi^2$/ndof \approx 1.3 \\
\hline
\end{tabular}
\caption{Parameter estimates with asymmetric 1$\sigma$ uncertainties, $\beta=7.2$.}
\label{tableLECs}
\end{table}
\renewcommand{\arraystretch}{1.0}
\end{comment}
\begin{comment}
\renewcommand{\arraystretch}{1.5}
\begin{table}[ht]
\centering
\begin{tabular}{ c c c c}
\hline
  $\tilde{L}_1$ & $\tilde{L}_2$ & $\tilde{L}_3$ & $\tilde{L}_4$ \\
\hline
$2.89^{+1.37}_{-0.81}$ & 
$5.54^{+1.62}_{-1.43}$ & 
$3.61^{+3.07}_{-1.44}$ & 
$-1.25^{+0.53}_{-1.24}$ \\
\hline
\end{tabular}
\caption{Parameter estimates with asymmetric 1$\sigma$ uncertainties for the $\beta=7.2$ dataset. These uncertainties are related to the statistical errors of the lattice calculations.}
\label{tableLECs}
\end{table}
\renewcommand{\arraystretch}{1.0}
\end{comment}

\section{Application: pion dark matter self-interactions at NLO}\label{sec:Sp4.Applications}

As mentioned in \sec\ref{sec:intro}, dark pions serve as dark matter candidates in different compelling scenarios (see, e.g., \refes~\cite{Ryttov:2008xe,Essig:2009nc, Bai:2010qg, Buckley:2012ky, Frigerio:2012uc, Bhattacharya:2013kma, Cline:2013zca, Hochberg:2014kqa, Carmona:2015haa, Kopp:2016yji, Beauchesne:2018myj, Beauchesne:2019ato,Bernreuther:2019pfb,Contino:2020god,Davighi:2024zip,Alfano:2025non,Davighi:2025awm%,Khlopov:2007ic}
}) 
and low-energy EFT can help study their phenomenology. In particular, the freeze out of pion dark matter typically happens when the dark-matter particles become non-relativistic and, hence, their typical momenta are much smaller than their mass. Similarly, dark-matter particles in today's universe are typically non-relativistic. If, moreover, the pion mass is smaller compared to the cut-off scale $4\pi F_\pi$, the chiral EFT is applicable to the description of both the dark matter freeze-out in the early Universe and also the self-interactions in the late times. 

We choose to demonstrate the effect of the NLO corrections on the example of the dark matter self-scattering in the late Universe; however, let us note that these corrections may play a role also in the early Universe. For example, in works like~\cite{Buckley:2012ky, Kopp:2016yji, Beauchesne:2018myj, Beauchesne:2019ato}, dark matter is formed by one particular pion species, while other pion states are unstable. The relic abundance is then set by the $2\to 2$ interactions interchanging the pion type described by the results of sections~\ref{secScat} and~\ref{secScatSO}. We believe that NLO corrections may alter the phenomenology results in some of these cases; however, a quantitative study is beyond the scope of this work. 

Sizable dark-matter self-interactions are an interesting feature of pion dark matter models since they can potentially contribute to explanation of the small-scale structure puzzles~\cite{Tulin:2017ara}. On the other hand, they may constrain the viable parameter space since too strong self-interactions are excluded by measurements of galaxy cluster collisions (e.g. the Bullet Cluster~\cite{Randall:2008ppe,Robertson:2016xjh,Wittman:2017gxn}). This constraint is crucial, e.g., for determining the viable parameter space of SIMP dark matter \cite{Hochberg:2014kqa,Hansen:2015yaa,Lee:2015gsa,Hochberg:2015vrg,Kamada:2017tsq,Hochberg:2018rjs,Hochberg:2018vdo,Berlin:2018tvf,Choi:2018iit,Katz:2020ywn,Kulkarni:2022bvh,Zierler:2022uez,Bernreuther:2023kcg,Braat:2023fhn,Pomper:2024otb} and the necessity of including the higher-order corrections in this context was pointed out in~\cite{Hansen:2015yaa}. This reference demonstrated important effect of the NLO corrections on the viability of different SIMP scenarios, however, their results were presented with a large uncertainty due to unknown NLO LECs.\footnote{More precisely, \refe\cite{Hansen:2015yaa} presented their results using random values of the scale-dependent LECs $L_i^r$ at $\mu = 2M_\pi$, sampled according to Gaussian distribution with zero mean and standard deviation of $5\times 10^{-4}$.} Below, we will update these NLO results using the fitted LECs from \sec\ref{sec:fit}. 

We choose to postpone the detailed discussion of the particular example of SIMP dark matter to \app\ref{AppendixSIMP} and present here only the general results on dark matter self-interactions that can be applied to any pion dark matter model. In subsection~\ref{sec:apSp}, we concentrate on the scenario based on the $Sp(4)$ gauge group with $N_F=2$ fermions in fundamental representation for which the lattice data are available and present the $2\to 2$ scattering cross section in the non-relativistic limit for the mass-degenerate case including the knowledge of NLO LECs. For completeness also the formula for the non-degenerate case is derived. In subsection~\ref{sec:apSO}, we present analogous formulas for the mass-degenerate and non-degenerate case for the \soth theory. We also examine the NLO mass hierarchy in the \soth case, so that we identify the lightest pion multiplet that would form most of dark matter in today's Universe in mass non-degenerate case.

\subsection{\spth theory}\label{sec:apSp}
To obtain the self-interaction cross section in the \spth theory we use the expression for the $2\to2$ amplitude for the degenerate case:\footnote{Let us note that this simplified shape of the scattering amplitude is not applicable in the non-degenerate case due to broken $Sp(4)$ symmetry.} %Writing the formula for the general scattering matrix element in terms of only two functions $B$ and $C$ is possible due to different crossing symmetries that are absent if the pion masses are non-degenerate.} 
\begin{equation}\label{eq:pheno1}
\begin{split}
    \mathcal{M}^{ab\rightarrow cd} \left( s,t,u\right)  &=\xi^{abcd} B\left( s,t,u\right)  +\xi^{acdb} B\left( t,u,s\right)  +\xi^{adbc} B\left( u,s,t\right)  \\ &+\delta^{ab} \delta^{cd} C\left( s,t,u\right)  +\delta^{ac} \delta^{bd} C\left( t,u,s\right)  +\delta^{ad} \delta^{bc} C\left( u,s,t\right) 
\end{split}
\end{equation}
with the group-theoretical factor
\begin{equation}\label{eq:pheno2}
\xi^{abcd} =\frac{1}{2} \left( \delta^{ab} \delta^{cd} -\delta^{ac} \delta^{bd} +\delta^{ad} \delta^{bc} \right).
\end{equation}
The functions $B(u,s,t)$ and $C(u,s,t)$ in the non-relativistic limit are given in \cite{Hansen:2015yaa}.\footnote{We agree with~\cite{Hansen:2015yaa} on the shape of functions $B$ and $C$, but we disagree on the coefficient of $B(t,u,s)$ which is $\xi^{acbd}$ there.} The self-scattering cross section in the center-of-mass frame evaluated at the threshold 
($s\to4M_\pi$, $t\to0$, $u\to0$) then
reads \cite{Hansen:2015yaa}:
\begin{align}\label{eq:pheno11}
\sigma_{2\rightarrow 2} &=\frac{1}{128\pi N^{2}_{\pi }M^{2}_{\pi }} \sum^{N_{\pi }=5}_{a,b,c,d=1} \left| \mathcal{M}^{ab\rightarrow cd} \right|^{2} \\&= \frac{9}{512\pi } \frac{M^{2}_{\pi }}{F^{4}_{\pi }} +\frac{M^{4}_{\pi }}{81920\pi^{3} F^{6}_{\pi }} \Bigg( 383+19456\pi^{2} \left( 4L^{r}_{4}+L^{r}_{5}\right)  +28672\pi^{2} \left( L^{r}_{1}+L^{r}_{2}+L^{r}_{6}\right)  \nonumber \\&+7168\pi^{2} \left( L^{r}_{0}+L^{r}_{3}+L^{r}_{8}\right)  -451\log \left( \frac{M^{2}_{\pi }}{\mu^{2} } \right)  \Bigg)  +\mathcal{O}\left( \frac{M^{6}_{\pi }}{F^{8}_{\pi }} \right) \label{Self_Sp_all}
\\ &=\frac{9}{512\pi } \frac{M^{2}_{\pi }}{F^{4}_{\pi }} +\frac{1}{245760\pi^{3} } \frac{M^{4}_{\pi }}{F^{6}_{\pi }} \left( 1149+456\tilde{L}_{1} -112\tilde{L}_{2} +105\tilde{L}_{3} \right) \nonumber
\\&+\mathcal{O}\left( \frac{M^{6}_{\pi }}{F^{8}_{\pi }} \right),
\end{align} 
where it is taken into account that in the non-relativistic limit the amplitude is isotropic.\footnote{In studies of self-interacting dark matter it is common to quote the momentum-transfer cross section, $\sigma_T \equiv \int d\Omega\,(1-\cos\theta)\,\frac{d\sigma}{d\Omega}$, which suppresses the contribution from small-angle scattering and highlights large-angle deflections, see e.g. \cite{Tulin:2017ara,Girmohanta:2022dog}. For the isotropic scattering considered here, $\sigma_T$ is identical to the total cross section, so the distinction is irrelevant in this case.}

\begin{comment}
  where $E_{\mathrm{CM}}$ is the energy of the scattering particles in the center-of-mass frame, hence, $E_{\mathrm{CM}} \to 2M_\pi$ in the highly non-relativistic limit.   
\end{comment}
The LO part here reproduces the self-scattering expression in~\cite{Hochberg:2014kqa} if one accounts for the fact that the decay constant $f_\pi$ used there is twice as large as the one adopted in this work, $F_\pi = f_\pi/2$.\footnote{Our convention for $F_\pi$ is followed in \cite{Scherer:2012xha, Hansen:2015yaa, Bijnens:2009qm, Bijnens:2011fm, Kulkarni:2022bvh, Dengler:2024maq, Pomper:2024otb, Katz:2020ywn, Gasser:1983yg, Kamada:2022zwb, Bernreuther:2023kcg}. The convention of~\cite{Hochberg:2014kqa} is used also in~\cite{Heikinheimo:2018esa}. Other definitions appear as well: for example, $F_\pi/\sqrt{2}$ in \cite{Kogut:2000ek} and $\sqrt{2}F_\pi$ in \cite{Ioffe:2001bn}.}

\begin{figure}[t]
    \centering
    \includegraphics[width=0.6\textwidth]{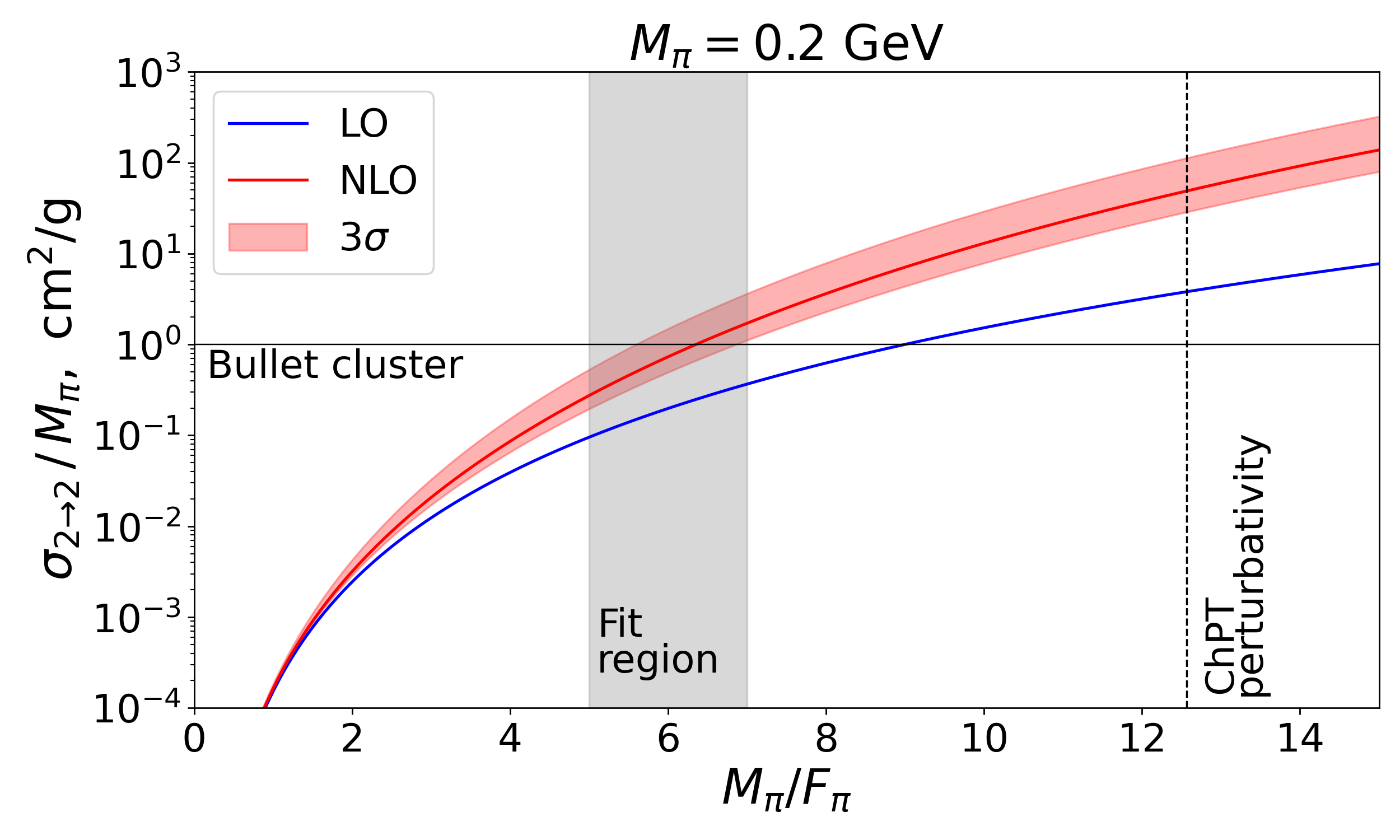} 
    \caption{Dependence of the non-relativistic $2\to 2$ pion cross section \eqref{eq:pheno11} on $M_\pi/F_\pi$ for a fixed (typical SIMP) mass $M_\pi = 0.2\,$GeV. To obtain a plot for different pion mass, $M'_\pi$, the curves should be multiplied by a factor $(M_\pi/M'_\pi)^3$. We also depict the constraint on dark matter self-interaction by observation of the Bullet cluster and the bound on ChPT validity, $M_\pi/F_\pi = 4\pi$. Finally, we indicate the region of $M_\pi/F_\pi$ for which the lattice data are available, i.e., where our NLO treatment should be most accurate. Note that the NLO curve is plotted with $3\sigma$ error coming from the fits of LECs, if $1\sigma$ error was depicted, the width of the band would be comparable with the line width.
    }
    \label{fig:Self-scattering}
\end{figure}

We plot the resulting non-relativistic LO and NLO $2\to 2$ pion cross sections in \fig\ref{fig:Self-scattering} for the typical SIMP mass $M_\pi=0.2$ GeV. Let us note that the matrix element entering the cross section~\eqref{eq:pheno11} is a function of $M_\pi/F_\pi$ only, while the pion mass itself enters $\sigma_{2\to 2}$ solely through $1/s \to 1/(4 M_\pi^2)$. Consequently, $\sigma_{2\to 2}/M_\pi \propto 1/M_\pi^3$, and the plot for a different pion mass $M'_\pi$ is obtained by a simple rescaling of the curves in \fig\ref{fig:Self-scattering} by a factor $(\text{200\,MeV}/M'_\pi)^3$. The NLO curve depends on the values of the LECs fitted in \sec\ref{sec:fit}, and we depict the $3\sigma$ range of the NLO cross section given the errors in the LECs stated in \fig\ref{fig:posteriors}. We observe that the $2\to 2$ cross section quickly decreases as $M_\pi/F_\pi \to 0$. This is a manifestation of the Adler zero principle~\cite{Brauner:2024juy}, namely, that the scattering amplitudes of Goldstone bosons vanish in the limit $p \to 0$. In the present case, the cross-section is fully non-relativistic, and it is only non-zero due to the mass of the pseudo-Goldstone pions. As expected, the LO and NLO results coincide for small $M_\pi/F_\pi$. For large values of $M_\pi/F_\pi$, in turn, the NLO result differs considerably from the LO one. We observe that the LO cross-section  is modified by 20\% for $M_\pi/F_\pi\approx 1.6$, by 50\% for $M_\pi/F_\pi\approx 2.5$ and by 100\% for $M_\pi/F_\pi\approx 3.6$.

We compare our results with the constraint on dark matter self-interactions coming from observation of the Bullet Cluster $\sigma_{2\to 2}/M_\pi \lesssim 1\,$cm$^2$/g~\cite{Randall:2008ppe}, although the exact value of this bound is a subject of ongoing discussion \cite{Robertson:2016xjh,Wittman:2017gxn, cha2025highcaliberviewbulletcluster}. As another illustration of the role of the NLO corrections, let us add that for a typical SIMP pion mass of $M_\pi = 0.2\,$GeV, the LO self-scattering cross-section per mass, $\sigma^{\text{LO}}_{2\rightarrow 2}/M_\pi$, exceeds the Bullet Cluster constraint at $M_\pi/F_\pi \approx 9$, while the NLO cross section, $\sigma^{\text{NLO}}_{2\rightarrow 2}/M_\pi$, exceeds it already at $M_\pi/F_\pi \approx 6.5$. Let us note that due to the scaling of $\sigma_{2\rightarrow 2}/M_\pi$ with $M_\pi^3$, we observe that for pion masses larger than $\sim 0.75\,$GeV, the central value of the NLO self-interaction lies below the Bullet cluster bound up to $M_\pi/F_\pi = 4\pi$. Hence, this bound naively seems to be evaded for all relevant parameter points although it might be argued that the NLO formula is not applicable for the values of $M_\pi/F_\pi$ near $4\pi$.

In \fig\ref{fig:Self-scattering}, we also highlight the region of $M_\pi/F_\pi$ where the lattice data is available. In this region, our NLO result is expected to reproduce the self-interaction cross section most accurately since the LECs were fitted here.\footnote{Let us emphasize that this is because we used the NLO formulas for fitting the LECs. If we, instead, fitted the NNLO formulas (which, as discussed previously, would be difficult with the amount of lattice data available since further unknown NNLO LECs enter the calculation), the NLO LECs may acquire slightly different values, and this time the most accurate value of the cross section would be reproduced with NNLO expressions in the region where lattice data are available.}

%\hk{Shall we show for which $M_\pi/F_\pi$ the data were fitted?} Comment on NNLO: at least in the region where we have the data, NLO should correspond to reality!

%it was later shown that vector mesons play a substantial role for the dark matter freeze out if $M_\pi/F_\pi$ is large and this may re-open part of the SIMP parameter space~\cite{Berlin:2018tvf,Arthur:2016dir,Bernreuther:2023kcg}

%Although later works showed that the vector mesons start playing an important role for the dark pion freeze-out if $M_\pi/F_\pi$ is close to $4\pi$~\cite{Berlin:2018tvf,Arthur:2016dir,Bernreuther:2023kcg}, this story shows that NLO corrections may play an important role for dark matter phenomenology. Let us note that there are pion dark matter scenarios where smaller $M_\pi/F_\pi$ are achieved like, hence, in the resonant and secluded case, plus other dark pion scenarios.
Further, let us comment on the role of the heavier states present in the QCD-like theories, in particular the vector mesons similar to the $\rho$ meson in SM QCD. Lattice simulations show that the mass of such dark vector mesons can become close to the mass of dark pions when $M_\pi/F_\pi$ is large~\cite{Hietanen:2014xca,Bennett:2019jzz,Kulkarni:2022bvh}. Consequently, these states introduce new processes that may, e.g., alter the freeze-out dynamics in the context the SIMP scenarios~\cite{Berlin:2018tvf,Arthur:2016dir,Bernreuther:2023kcg}. In this sense, the phenomenology analysis of the early SIMP works~\cite{Hochberg:2014kqa,Hansen:2015yaa} is actually somewhat incomplete. On the other hand, the non-relativistic self-interaction cross section presented in this section is \emph{not} affected even if the dark vector mesons become lighter, with the only exception of the resonant regime $M_\rho \sim 2 M_\pi$ that corresponds roughly to $M_\pi/F_\pi \sim 4$ according to the lattice results of~\cite{Bennett:2019jzz}. In fact, the effect of the vector mesons on pion scattering is captured in the lattice calculations, hence, also in our fitted values of LECs.\footnote{Moreover, even if the vector mesons are included in the effective description, their contribution to the pion scattering cross section vanishes in the $v\to 0$ limit, see~\cite{Choi:2018iit}. Let us also note that the presence of resonances and the consequent velocity-dependent dark matter self-interactions might be interesting for explaining the small-scale structure puzzles~\cite{Chu:2018fzy,Tsai:2020vpi}, however, this is beyond the scope of our work.} In conclusion, at least in the ``fit region'' depicted in \fig\ref{fig:Self-scattering}, $5\lesssim M_\pi/F_\pi \lesssim 7$, our prediction of the pion self-interaction should be fully trustable.

%TODO: more comments on the fact that the singlet meson is probably lightest (as the lattice data suggest), but in presence of a portal to SM, it might become unstable!
For completeness, we also present the formula for the mass non-degenerate case. The lattice calculations suggest that, at least for the case of $Sp(N_c=4)$ gauge theory with $N_F=2$ flavors of fundamental fermions, the singlet pion is lighter than the four-plet ones (see the discussion in \sec\ref{sec:FitFit}). Consequently, the singlet pion would form dark matter if it is stable enough, such a scenario was considered, e.g., in~\cite{Frigerio:2012uc}.
The self-scattering cross section for this singlet pion reads:
\begin{align}
\sigma^{\prime}_{2\rightarrow 2} &=\frac{1}{128\pi M^{2}_{\pi,3 }}  \left| \mathcal{M}^{33\rightarrow 33} \right|^{2} \nonumber\\
&=\frac{M^{2}_{\pi ,1}}{512\pi F^{4}_{\pi }} +\frac{7\left( M^{2}_{\pi ,3}-M^{2}_{\pi ,1}\right)  }{512\pi F^{4}_{\pi }}  \nonumber\\
&+\frac{M^{4}_{\pi ,1}}{16384\pi^{3} F^{6}_{\pi }} \Bigg( 27+3072\pi^{2} \left( L^{r}_{0}+L^{r}_{3}+L^{r}_{8}\right) +12288\pi^{2} \left( L^{r}_{1}+L^{r}_{2}+L^{r}_{6}\right)  \nonumber\\&-1024\pi^{2} \left( 4L^{r}_{4}+L^{r}_{5}\right)  -47\log \left( \frac{M^{2}_{\pi ,1}}{\mu^{2} } \right)  \Bigg)  + \mathcal{O}\left(\frac{M^{6}_{\pi ,1}}{F_\pi^8}\right).
\end{align} 
The LO part here agrees with the result from \refe\cite{Hochberg:2014kqa}. As discussed in this reference, the self-scattering cross section in the case with only one pion species forming dark matter is significantly smaller than in the degenerate one, since the kinetic term then does not contribute to $2\to 2$ interactions. The viable parameter space for SIMPs is correspondingly broadened. On the other hand, in the presence of a portal to SM (like, e.g., a dark photon), the singlet pion may become unstable since it is not protected by the flavor symmetry in the presence of additional symmetry breaking by quark mass splitting.

\subsection{\soth theory}\label{sec:apSO}
%In this appendix, we provide the formulas for the self-scattering cross section of pions analogous to the \spth formula \eqref{eq:pheno11}.
Let us now turn to theories with quarks in a real representation. In the mass-degenerate case, the NLO formula for the $2\to 2$ cross section in the non-relativistic limit can be obtained from the results of~\cite{Bijnens:2011fm}, but we present it here explicitly for convenience: 
%In the mass-degenerate case, the result for the self-scattering cross section of pions analogous to the \spth formula \eqref{eq:pheno11} reads
\begin{align}\label{eq:ScSO1}
\sigma_{2\rightarrow 2}& =\frac{145}{4608\pi } \frac{M^{2}_{\pi }}{F^{4}_{\pi }}\nonumber\\ & +\frac{M^{4}_{\pi }}{147456\pi^{3} F^{6}_{\pi }} \Bigg( 551+1024\pi^{2} ( 51L^{r}_{0}+76L^{r}_{1}+76L^{r}_{2}+51L^{r}_{3}\nonumber\\& +60L^{r}_{4}+47L^{r}_{5}+76L^{r}_{6}+51L^{r}_{8})  -1227\log \left( \frac{M^{2}_{\pi }}{\mu^{2} } \right) \Bigg)+\mathcal{O}\left( \frac{M^{6}_{\pi }}{F^{8}_{\pi }} \right)  .  
\end{align}
The LO part here corresponds to the result in \cite{Hochberg:2014kqa}.

%%%
For the case of split quark masses, a pair of pions is expected to form the lightest states in the spectrum according to the LO formulas~\eqref{NLOso.1}–\eqref{NLOso.3}. In particular, in our convention $m_u < m_d$, the 6th and 7th pions would be the lightest. However, for sufficiently large values of the dimensionless ratio $M_{\mathrm{LO,deg}}/(4\pi F)$, where $M_{\mathrm{LO,deg}}$ is the LO pion mass in the degenerate limit $r \equiv m_d/m_u = 1$, NLO corrections can potentially alter the mass hierarchy. Since no lattice data are currently available to our knowledge for the \soth theory with $m_u\neq m_d$, we explore this possibility by modeling the effect of NLO terms. Specifically, we sample the relevant $\mu$-dependent LECs $L_i^r$ from a normal distribution with zero mean and standard deviation $5\times 10^{-4}$ that is motivated by the order of magnitude of the LECs we obtained in the \spth case (see \eqs\eqref{eq.Sp4.Fit.1}-\eqref{eqLtilde4} for the relation between $L_i^r$ and the $\mu$-independent LECs $\tilde{L}_i$ that we fitted and note the difference in normalization). Any samples that produced negative squared pion masses were rejected. This approach allows us to estimate the statistical spread of pion mass ratios induced by NLO corrections.

\begin{figure}[t]
    \centering
    \includegraphics[width=0.6\textwidth]{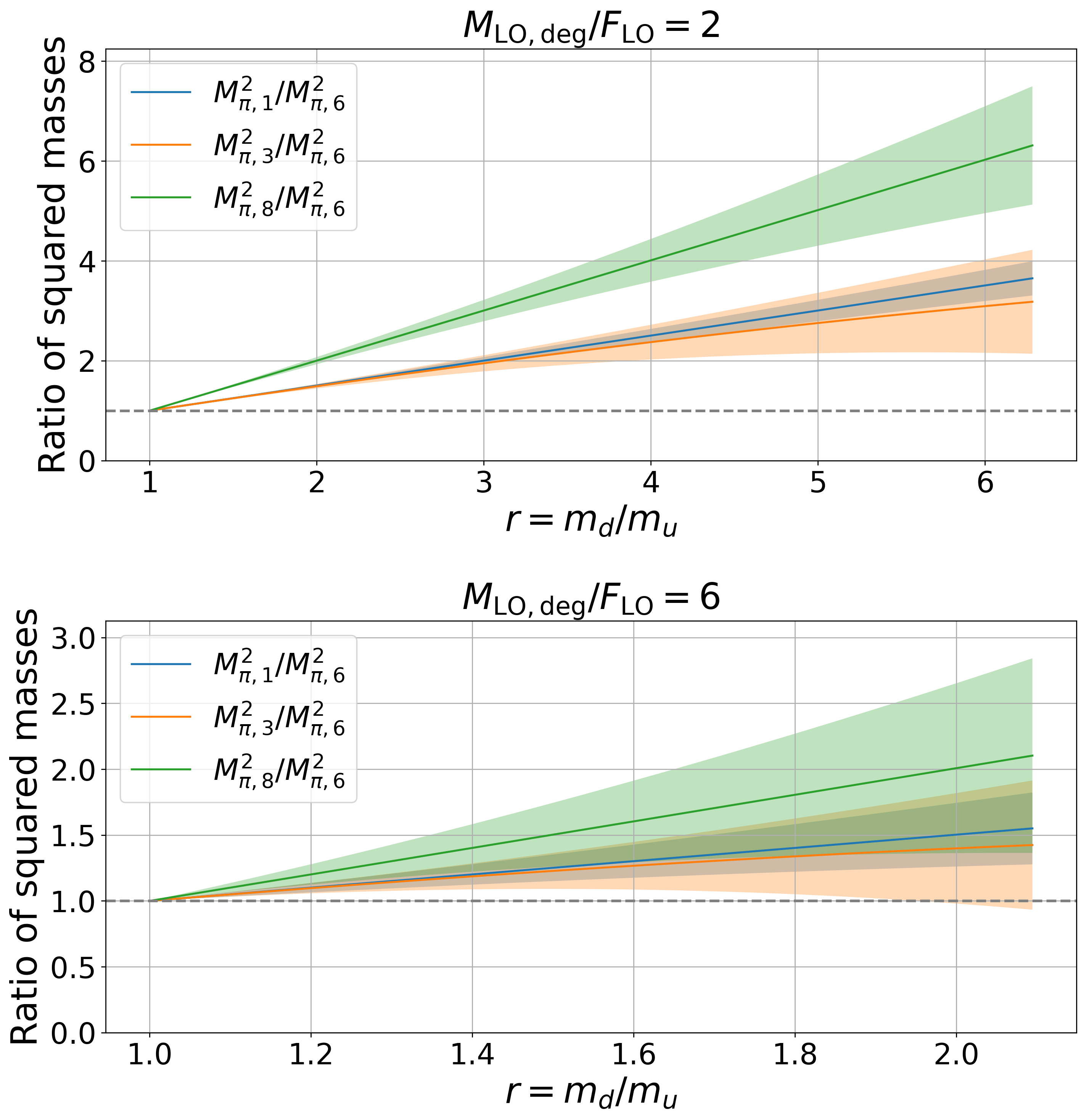} 
    \caption{
Ratios of different NLO pion masses to $M_{\pi,6}$, where $\pi_6$ is the lightest species at LO, as functions of the quark mass ratio $r$. The LECs entering the NLO masses are sampled from a Gaussian distribution with zero mean and standard deviation $5\times 10^{-4}$, and we plot the resulting $1\sigma$ spread. The upper and lower plots correspond to leading order ratio $M_\text{LO,deg}/(4\pi F_\text{LO})$ for degenerate quark masses being approximately equal to $0.16$ and $0.48$ respectively. Notice that while in the upper case, practically all LEC configurations imply $\pi_6$ being the lightest, in the lower case, $\approx8.9\%$ of configurations lead to lightest $\pi_3$ and $\approx 1.8\%$ of configurations lead to lightest $\pi_8$ for $r=1.7$. }
    \label{massHierarchySO}
\end{figure}

The results of this sampling are shown in figure \ref{massHierarchySO}, where we plot the ratios of the squared pion masses to the squared mass of the 6th pion (see formulas \eqref{NLOso.4}–\eqref{NLOso.7}) as a function of $r = m_d/m_u$. The renormalization scale was chosen to be $\mu=M_\text{LO,deg}$.
The range of $r$ is restricted to values for which the expansion parameter $M_{\pi,8}/(4\pi F)$ remains below unity, ensuring that perturbation theory is applicable. Our modeling indicates that the LO prediction for the mass hierarchy remains valid for the majority of sampled LEC combinations, meaning that $M_{\pi,6}$ is typically the smallest mass and thus the $(\pi_6, \pi_7)$ doublet remains the lightest, and, hence, dominate the dark matter abundance in the Universe today.

Consequently, we average only over these pions in formula~\eqref{eq:pheno11} and obtain the following expression for the dark matter self-interactions:
\begin{align}
\sigma^{\prime}_{2\rightarrow 2}& =\frac{3M^{2}_{\pi,6}}{64\pi F^{4}_{\pi ,6}}  +\frac{1}{36864\pi^{3} F^{6}_{\pi ,6}} \Bigg( 4M^{4}_{\pi ,1}-52M^{2}_{\pi ,1}M^{2}_{\pi ,6}+250M^{4}_{\pi ,6}-4M^{2}_{\pi ,1}M^{2}_{\pi ,8}\nonumber \\ 
&+26M^{2}_{\pi ,6}M^{2}_{\pi ,8} +M^{4}_{\pi ,8} +6144M^{2}_{\pi ,6}\Big( 3\left( L^{r}_{0}+2L^{r}_{1}+2L^{r}_{2}+L^{r}_{3}+L^{r}_{5}+L^{r}_{8}\right)  M^{2}_{\pi ,6} \nonumber\\
&+2L^{r}_{4}\left( 10M^{2}_{\pi ,1}-2M^{2}_{\pi ,6}-5M^{2}_{\pi ,8}\right)   +2L^{r}_{6}\left( -2M^{2}_{\pi ,1}+4M^{2}_{\pi ,6}+M^{2}_{\pi ,8}\right)  \Big)  \pi^{2} \Bigg)\nonumber\\
&-\frac{\delta }{147456\pi^{3} F^{6}_{\pi ,6}M^{2}_{\pi ,6}} \Bigg( 4M^{4}_{\pi ,1}+178M^{4}_{\pi ,6}+26M^{2}_{\pi ,6}M^{2}_{\pi ,8}+M^{4}_{\pi ,8} \nonumber\\
&-4M^{2}_{\pi ,1}\left( 13M^{2}_{\pi ,6}+M^{2}_{\pi ,8}\right)  \Bigg)  \log \left( \frac{M^{2}_{\pi ,1}-2M^{2}_{\pi ,6}-\delta }{M^{2}_{\pi ,1}-2M^{2}_{\pi ,6}+\delta } \right)  \nonumber \\
&-\frac{1}{184320\pi^{3} F^{6}_{\pi ,6}} \Bigg( \Big( 44M^{4}_{\pi ,1}-16M^{2}_{\pi ,1}\left( 17M^{2}_{\pi ,6}+2M^{2}_{\pi ,8}\right)   \nonumber\\
&+5\left( 178M^{4}_{\pi ,6}+26M^{2}_{\pi ,6}M^{2}_{\pi ,8}+M^{4}_{\pi ,8}\right)  \Big)  \log \left( \frac{M^{2}_{\pi ,1}}{\mu^{2} } \right)  +1800M^{4}_{\pi ,6}\log \left( \frac{M^{2}_{\pi ,6}}{\mu^{2} } \right)  \Bigg)\nonumber \\
&+\mathcal{O} \left( \frac{M^{6}_{\pi ,6}}{F^{8}_{\pi ,6}} \right)\,,\label{SO_self_67}
\end{align}
where $\delta =2\sqrt{M^{2}_{\pi ,6}\left( M^{2}_{\pi ,6}-M^{2}_{\pi ,1}\right)  } $. 

For comparison, the mass non-degenerate scenario studied in \cite{Hochberg:2014kqa} assumes that a single pion species is the lightest and therefore alone controls dark-matter self-interactions. Our sampling analysis, however, indicates that this pattern is unlikely. In the doublet case, the LO contribution to Eq.\eqref{SO_self_67} is six times larger than the self-scattering cross section obtained in the single-pion scenario of \cite{Hochberg:2014kqa}.  Moreover, if dark matter were composed of the singlet pion, it may become unstable in the presence of a portal to the SM, as discussed in the \spth case.

On the other hand, radiative corrections from couplings to additional fields (e.g., a dark photon) could in principle reshuffle the mass hierarchy in certain models, consequently, the NLO self-interaction cross section for the case when $\pi_3$ is the lightest is for completeness given in the repository~\cite{kolesova_2025_14864550} (where also the formulas~\eqref{Self_Sp_all}-\eqref{SO_self_67} are typed for convenience).
%%%%

\section{Conclusions}\label{sec:conclusions}

%Strongly-interacting massive particle (SIMP) paradigm marks 
Dark pions represent an important class of dark matter models, which deserve a careful analysis. These scenarios are challenging due to their inherent non-perturbative nature and the corresponding low-energy description written in terms of leading-order chiral perturbation theory needs to be carefully examined. In this context, there are two important aspects, first, the validity of leading-order chiral theory expansion, which may not be adequate depending on the application at hand, and second, its description for theories containing mass non-degenerate quarks brings additional complexity. 

Keeping both these aspects in mind, in this work, we employed the chiral theory formalism at next-to-leading order for 2-flavor theory with quarks in (pseudo)real representation. We used recent lattice computations for mass spectrum as well as pion scattering to fit the resulting LECs for the case of $Sp(N_c=4)$ theory with 2 flavors of quarks in fundamental representation. For this theory, the mass spectrum results were available for mass non-degenerate theory~\cite{Kulkarni:2022bvh}, while the scattering calculations were performed only in the mass degenerate limit~\cite{Dengler:2024maq}. Never-the-less, it was possible to use the scattering calculations since the LECs entering the scattering length do not depend on the quark mass splitting. Let us note that, to the best of our knowledge, such detailed lattice data do not exist for the case of quarks in real representations,\footnote{\Refe\cite{DeGrand:2015lna} provides the spectroscopy data for the case of $SU(N_c=4)$ gauge theory with two flavors of quarks in real sextet representation. However, since only the mass-degenerate case is considered, we expect that very limited subset of LECs could be obtained from these data.} consequently, we do not have the corresponding LECs at hand. 

While our procedure is rooted in theoretical analysis, limited lattice data forced us to perform certain approximations. First, the limited data means we can not fit all the LECs individually but only certain linear combinations.  Second, the lattice data used are not extrapolated to continuum %or chiral 
limit. This means that our LECs are susceptible to discretizations artifacts. We address possible effects by fitting the LECs for the three lattice spacings and find consistent values for most of the linear combinations that we consider. However, one of our LECs seems to be prone to lattice artifacts, and we traced this back to the $\beta$-dependence of the amount of pion mass splitting. %To tackle the absence of chiral extrapolations, 
Third, the lattice data on pion masses and decay constants~\cite{Kulkarni:2022bvh} included only values at finite lattices, however, for most of the parameter points, we performed the infinite-volume extrapolation as a part of this work to avoid the related systematic errors. Fourth, the lattice simulations can typically be performed for relatively large quark masses only, where our effective theory may no longer be valid. Consequently, we restricted ourselves to a subset of data with smallest $M_\pi/F_\pi$, thus being as close to chirality as possible. Finally, the lattice data for pion - pion scattering amplitudes include only ``isospin - 2'' channel, however, it was still possible to extract sufficient number of LECs in order to determine the non-relativistic cross section averaged over all channels relevant for dark matter self-interactions in today's Universe. %The ``isospin - 2'' leads to the largest contribution to the self-scattering cross-section, we therefore believe that lack of lattice results for additional channels should not introduce a large error in our LEC estimates. In addition, using our LECs, it should be possible to compute the self-scattering cross-section by summing over all channels. This being the most dominant channel, the derive LECs with this data should be as close to reality as possible. 
Let us also comment on the NLO versus NNLO calculations. While in principle it is theoretically possible to perform NNLO calculations, we do not believe that we can extract the relevant LECs with the available lattice data. 

Finally, we apply our results to dark matter scenarios where pion - pion scattering provides dark matter self-interactions. We demonstrate that NLO corrections provide important contributions to the relevant cross section and may influence the dark matter viable parameter space. Our calculations are therefore important when establishing the validity of dark matter parameter space of strongly-interacting dark matter theories.

%TODO: perhaps we could also mention here that NLO corrections might be important for dark matter freeze out in certain scenarios, but it's maybe not so important.

%Finally, we apply our results to SIMP DM scenario where pion - pion scattering provide DM self-interactions and the number changing process which drive relic density are governed by the WZW term. Our analysis shows that NLO corrections significantly suppresses the viable region of SIMP parameter space, in line with other studies. We also stress here that this analysis assumes spin-1 rho mesons do not play any significant role in DM relic density generation, while contrary examples exist in the literature. 

%Suchita's comment above, I think it suits nicely here to conclusions.
We point here in passing that the effective theories we discuss here are also widely used and established in the context of composite Higgs models. In particular, the \soth and \spth breaking patterns, which we discuss here provide relevant models for composite Higgs~\cite{Ferretti:2013kya, Witzel:2019jbe, Bellazzini:2014yua, Cacciapaglia:2020kgq}. The NLO corrections we consider in our work might be relevant for some of these theories, especially if the scale at which the theory is probed is comparable with the scale of the composite dynamics.

In conclusion, this analysis marks a step towards a more accurate description of the \soth and \spth theories, and may correspondingly improve previous phenomenology studies. %of pion-only dark matter processes. 
We provide not only the analytical results at NLO but also use the available lattice data to fit the resulting LECs. Due to limited lattice data our analysis is performed under controlled approximations, however, we hope that the importance of our calculations motivates further lattice calculations in similar directions, which will enable the phenomenology community to perform NNLO calculations or obtain more precise results also for the theories with real quarks.

%\hk{In the end, I think that a separate discussion section is not needed. Perhaps we can repeat in the conclusions following points:}
%\begin{itemize}
    %\item Our analysis restricted to NLO since NNLO would bring in further unknown LECs etc. and there is not enough lattice data to constrain these. However, in the region of $M_\pi/F_\pi$ where the lattice data are available, we believe our NLO predictions are close to truth. This region is actually interesting for dark matter phenomenology (see \fig\ref{fig:Self-scattering}).
    %\item Outlook: if more lattice data is available, we may look at NNLO fit. It would be good since the pion decay constants are non-denegerate in the lattice results, but our EFT does not capture this fact.
%\end{itemize}

\section{Acknowledgments}

We thank Tomas Brauner, Joachim Pomper, Francesco Sannino, Torsten Bringmann, Halvor Melkild, Tore Selland Kleppe, Josef Pradler,  Oleg Komoltsev, and Jimmy Huy Tran for useful discussions. We are especially grateful to Yannick Dengler, Axel Maas, and Fabian Zierler for providing additional lattice data and helpful explanations related to ref.~\cite{Dengler:2024maq}.

The work of H.K. and D.K. is supported by the Research Council of Norway under the FRIPRO Young Research Talent grant no.~335388. S.K. is supported by the FWF research group funding FG1 and FWF
project number P 36947-N.

\appendix

%\section{SO(4)-theory at NLO}\label{sec:so4-nlo}

\section{Generators of $SU(4)$}\label{appendixSU4gen}
We use the following set of Hermitian generators for $SU(4)$:
\begin{align}
&T^1 = \begin{pmatrix}
\frac{1}{2} & 0 & 0 & 0 \\
0 & \frac{1}{2} & 0 & 0 \\
0 & 0 & -\frac{1}{2} & 0 \\
0 & 0 & 0 & -\frac{1}{2}
\end{pmatrix},\quad
T^2 = \begin{pmatrix}
0 & \frac{1}{2} & 0 & 0 \\
\frac{1}{2} & 0 & 0 & 0 \\
0 & 0 & 0 & \frac{1}{2} \\
0 & 0 & \frac{1}{2} & 0
\end{pmatrix},\quad
T^3 = \begin{pmatrix}
0 & \frac{1}{2} & 0 & 0 \\
\frac{1}{2} & 0 & 0 & 0 \\
0 & 0 & 0 & -\frac{1}{2} \\
0 & 0 & -\frac{1}{2} & 0
\end{pmatrix},
\nonumber\\[10pt]
&T^4 = \begin{pmatrix}
0 & -\frac{i}{2} & 0 & 0 \\
\frac{i}{2} & 0 & 0 & 0 \\
0 & 0 & 0 & \frac{i}{2} \\
0 & 0 & -\frac{i}{2} & 0
\end{pmatrix},\quad
T^5 = \begin{pmatrix}
0 & -\frac{i}{2} & 0 & 0 \\
\frac{i}{2} & 0 & 0 & 0 \\
0 & 0 & 0 & -\frac{i}{2} \\
0 & 0 & \frac{i}{2} & 0
\end{pmatrix},\quad
T^6 = \begin{pmatrix}
\frac{1}{2} & 0 & 0 & 0 \\
0 & -\frac{1}{2} & 0 & 0 \\
0 & 0 & \frac{1}{2}  & 0 \\
0 & 0 & 0 & -\frac{1}{2}
\end{pmatrix},
\nonumber\\[10pt]
&T^7 = \begin{pmatrix}
\frac{1}{2} & 0 & 0 & 0 \\
0 & -\frac{1}{2} & 0 & 0 \\
0 & 0 & -\frac{1}{2} & 0 \\
0 & 0 & 0 & \frac{1}{2}
\end{pmatrix},\quad
T^8 = \begin{pmatrix}
0 & 0 & 0 & \frac{1}{2} \\
0 & 0 & \frac{1}{2} & 0 \\
0 & \frac{1}{2} & 0 & 0 \\
\frac{1}{2} & 0 & 0 & 0
\end{pmatrix},\quad
T^9 = \begin{pmatrix}
0 & 0 & 0 & \frac{i}{2} \\
0 & 0 & \frac{i}{2} & 0 \\
0 & -\frac{i}{2} & 0 & 0 \\
-\frac{i}{2} & 0 & 0 & 0
\end{pmatrix}, \nonumber\\[10pt]
&T^{10} = \begin{pmatrix}
0 & 0 & \frac{1}{\sqrt{2}} & 0 \\
0 & 0 & 0 & 0 \\
\frac{1}{\sqrt{2}} & 0 & 0 & 0 \\
0 & 0 & 0 & 0
\end{pmatrix},\quad
T^{11} = \begin{pmatrix}
0 & 0 & \frac{i}{\sqrt{2}} & 0 \\
0 & 0 & 0 & 0\\
-\frac{i}{\sqrt{2}} & 0 & 0 & 0 \\
0 & 0& 0 & 0
\end{pmatrix},\quad
T^{12} = \begin{pmatrix}
0 & 0 & 0 & 0 \\
0 & 0 & 0 & \frac{1}{\sqrt{2}} \\
0 & 0 & 0 & 0 \\
0 & \frac{1}{\sqrt{2}} & 0 & 0
\end{pmatrix},\nonumber\\[10pt]
&T^{13} = \begin{pmatrix}
0 & 0 & 0 & 0 \\
0 & 0 & 0 & \frac{i}{\sqrt{2}} \\
0 & 0 & 0 & 0 \\
0 & -\frac{i}{\sqrt{2}} & 0 & 0
\end{pmatrix},\quad
T^{14} = \begin{pmatrix}
0 & 0& 0 & \frac{1}{2}  \\
0 & 0 & -\frac{1}{2}  & 0 \\
0& -\frac{1}{2}  & 0 & 0 \\
\frac{1}{2} & 0 & 0 & 0
\end{pmatrix},\quad
T^{15} = \begin{pmatrix}
0 & 0& 0 & \frac{i}{2}  \\
0 & 0 & -\frac{i}{2}  & 0 \\ 0& \frac{i}{2}  & 0 & 0 \\ -\frac{i}{2} & 0 & 0 & 0 
\end{pmatrix}.
\end{align}
The generators are normalized as $\left< T^{a}T^{b}\right>  =\delta^{ab}$. For the \spth theory the broken generators corresponding to the pion fields are 
\begin{equation}
 X^{1}\equiv T^{2},\quad X^{2}\equiv T^{4},\quad X^{3}\equiv T^{6},\quad X^{4}\equiv T^{14},\quad X^{5}\equiv T^{15},
\end{equation}
while for the \soth theory they read
\begin{align}
X^{1}&\equiv T^{2},\quad X^{2}\equiv T^{4},\quad X^{3}\equiv T^{6},\quad X^{4}\equiv T^{8},\quad X^{5}\equiv T^{9},\quad \nonumber \\ X^{6}&\equiv T^{10},\quad X^{7}\equiv T^{11},\quad X^{8}\equiv T^{12},\quad X^{9}\equiv T^{13}. 
\end{align}

\section{Quark condensate at leading order}\label{AppCondensate}

In this appendix, we outline the derivation of the quark condensate at LO. At LO, $V_{\text{eff} }=-\mathcal{L}_{\text{LO} } \left( \pi =0\right)$, i.e. we consider $u_\mu=0$ and, consequently, 
\begin{align}
    V_{\text{eff} }=-\frac{F^{2}}{4} \left< \chi J^{T}+J\chi^{\dagger } \right>.  
\end{align}
Using $\chi=2B_0\hat{\mathcal{M}}$ and~\eqref{eq.EFT17b}, we get
\begin{align}
  V_{\text{eff} }&=-\frac{F^{2}}{4} 2B_{0}\left< \begin{pmatrix}0&\eta \mathcal{M} \\ \mathcal{M} &0\end{pmatrix} \begin{pmatrix}0&\mathbf{1}_{2} \\ \eta \mathbf{1}_{2} &0\end{pmatrix} +\begin{pmatrix}0&\eta\mathbf{1}_{2} \\ \mathbf{1}_{2} &0\end{pmatrix} \begin{pmatrix}0&\mathcal{M} \\ \eta \mathcal{M} &0\end{pmatrix} \right> \nonumber \\& =-\frac{F^{2}}{2} B_{0}\left< 2\begin{pmatrix}\mathcal{M} &0\\ 0&\mathcal{M} \end{pmatrix} \right>  =-2F^{2}B_{0}\left< \mathcal{M} \right>  =-2F^{2}B_{0}\left( m_{u}+m_{d}\right).      
\end{align}
The quark condensates are the same at LO and read
\begin{align}
\left< \bar{u} u\right>=\frac{\partial V_{\text{eff} }  }{\partial m_{u}}\left( \pi =0\right)=-2 B_0F^2,\quad \left< \bar{d} d\right>  =\frac{\partial V_{\text{eff} }  }{\partial m_{d}}\left( \pi =0\right)=-2 B_0F^2.  
\end{align}
The total condensate is then 
\begin{align}
    \left< \bar{q} q\right>  =\left< \bar{u} u\right>  +\left< \bar{d} d\right> = -4B_0F^2,
\end{align}
which differs by a factor of 2 from the result in~\cite{Bijnens:2009qm}.

\section{Channels and partial waves}\label{AppendixChannels}

In this appendix, we briefly introduce the notion of the scattering length that was determined by lattice methods in \refe\cite{Dengler:2024maq} and explain its connection to the scattering amplitude calculated in our \sec\ref{secScat}. The description here assumes $m_u=m_d$. The detailed discussion of the group theory of $Sp(2N_F)$ and $SO(2N_F)$ can be found in \cite{Kamada:2022zwb}.

The pseudoreal quarks transform in a fundamental 4-dimensional representation of the unbroken $Sp(4)$-flavor symmetry. The mesons composed of two fundamental fermions are either in the 5-dimensional,  10-dimensional, or singlet representation:
\begin{equation}\label{eq.Sp4.Scat.10}
    4 \otimes 4 = 1 \oplus 5 \oplus 10.
\end{equation}
The pions belong to the 5-dimensional irreducible representation of $Sp(4)$.
They can scatter in three different channels 
\begin{equation}
  5 \otimes 5 = 1 \oplus 10 \oplus 14.
\end{equation}
In \cite{Dengler:2024maq}, the lattice calculation is performed for the 14-dimensional channel where no single-vector-meson states can exist.

The amplitude in a given channel $I$, denoted $T^I$, can be obtained by projecting the full amplitude:
\begin{equation}\label{eq.Sp4.Scat.11.a} 
T^{I}P^{ijkl}_{I}=\sum^{{}N_{\pi }}_{m,n=1} P^{ijmn}_{I}\mathcal{M}^{mn\rightarrow kl} 
\end{equation}
where $P_I$ is the projection operator onto channel $I$ (no summation over $I$ implied). The full amplitude is a sum of all channels:
\begin{equation}\label{eq.Sp4.Scat.12} 
 \mathcal{M}^{ij\rightarrow kl} =\sum_{I} T^{I}P^{ijkl}_{I}.
\end{equation}
Expressions for the projectors are given in \cite{Bijnens:2009qm} and knowing the general amplitude allows us to calculate the necessary channel.
The dimension of representation $d_I$ can be calculated as \cite{Kamada:2022zwb}
\begin{equation}\label{eq.Sp4.Scat.12.1} 
    \sum^{N_{\pi }}_{i,j=1} P^{ijij}_{I}=d_{I}.
\end{equation}

Each channel amplitude $T^I$ can be decomposed via partial wave expansion:
\begin{equation}\label{eq.Sp4.Scat.13} 
T_{\ell}^{I}(s) = \frac{1}{64\pi} \int_{-1}^{1} d(\cos\theta) P_{\ell}(\cos\theta) T_I(s, t).
\end{equation}
Near threshold ($s\rightarrow 4M^{2}_{\pi },\  t\rightarrow 0$), $T_{\ell}^{I}(s)$ can be expanded in terms of the small three-momentum $q$ in the center-of-mass frame:
\begin{equation}\label{eq.Sp4.Scat.14} 
\Re T_{\ell}^{I}(s) = q^{2\ell} \left[ a_{\ell}^{I} + q^2 b_{\ell}^{I} + O(q^4) \right],
\end{equation}
where $a_{\ell}^{I}$ is the scattering length, and $b_{\ell}^{I}$ is the slope. For $\ell = 0$ (s-wave), the scattering length is given by:
\begin{equation}\label{eq.Sp4.Scat.15} 
a^{I}_{0}=T^{I}_{0}\left( s\rightarrow 4M^{2}_{\pi },\  t\rightarrow 0\right).  
\end{equation}
The formula for the 14-dimensional channel reads \cite{Bijnens:2011fm}
\begin{equation}\label{eq.Sp4.Fit.5}
\begin{split}
a^{MS}_{0}&=-\frac{1}{32\pi } \frac{M^{2}_{\pi}}{F^{2}_{\pi }} +\left( \frac{M^{2}_{\pi }}{F^{2}_{\pi }} \right)^{2}  \frac{1}{\pi } \Bigg( -\frac{1}{2048\pi^{2} } +\frac{1}{2} L^{r}_{0}+2L^{r}_{1}+2L^{r}_{2}\\&+\frac{1}{2} L^{r}_{3}-2L^{r}_{4}-\frac{1}{2} L^{r}_{5}+2L^{r}_{6}+\frac{1}{2} L^{r}_{8}-\frac{5}{2048\pi^{2} } \log \frac{M^{2}_{\pi }}{\mu^{2} } \Bigg).
\end{split}
\end{equation}
We note that the relation between the dimensionful scattering length $a_0$ determined by lattice calculations in~\cite{Dengler:2024maq} and the dimensionless scattering length $a_0^{MS}$ above is 
\begin{equation}\label{eq:aTranslate}
    a_0 M_\pi=-a_0^{MS}.
\end{equation}

\section{Lattice data}\label{AppendixLatticeData}

In this appendix, we present the lattice data based on refs.~\cite{Kulkarni:2022bvh} and~\cite{Dengler:2024maq} that were used to determine the NLO low-energy constants of the \spth theory (see Sec.~\ref{sec:fit}). Since we discuss the finite-volume effects, we denote the infinite-volume values by the superscript $\infty$ in this appendix, while this superscript is omitted in the main text.

\begin{figure}[t]
    \centering
    \includegraphics[width=0.7\textwidth]{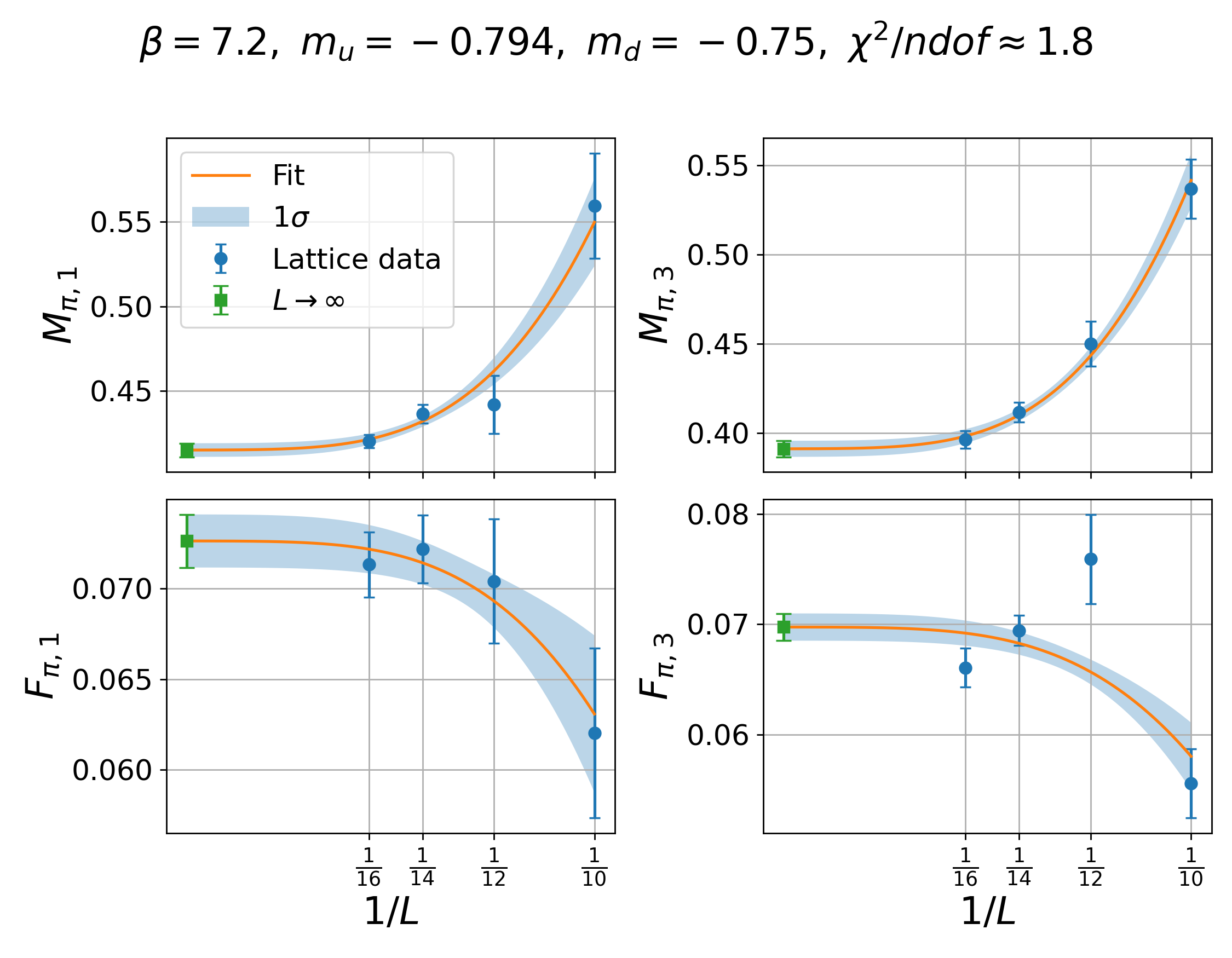} 
    \caption{An example of an infinite volume fit for a single ensemble. %\dk{What to add here?}
    }
    \label{fig:InfVol}
\end{figure}

Ref.~\cite{Kulkarni:2022bvh} provides data at multiple lattice volumes for most ensembles defined by a given inverse bare gauge coupling $\beta$ and bare quark masses $a m_u^0$, $a m_d^0$, supplemented by additional data from the dataset~\cite{zierler_2026_18716220}.   Using the available multi-volume data, we perform explicit infinite-volume extrapolations where feasible. In practice, we follow the approach of \cite{Bennett:2019jzz} and perform fits to the following expressions:
\begin{align}
    M_{\pi ,1}(L)&=M_{\pi ,1}^\infty\left( 1+A_{M_{1}}\frac{e^{-M_{\pi ,3}^\infty L}}{\left( M_{\pi ,3}^\infty L\right)^{3/2}  } \right) \nonumber ,\\
    M_{\pi ,3}(L)&=M_{\pi ,3}^\infty\left( 1+A_{M_{3}}\frac{e^{-M_{\pi ,3}^\infty L}}{\left( M_{\pi ,3}^\infty L\right)^{3/2}  } \right) \nonumber ,\\
    F_{\pi ,1}(L)&=F_{\pi ,1}^\infty\left( 1+A_{F_{1}}\frac{e^{-M_{\pi ,3}^\infty L}}{\left( M_{\pi ,3}^\infty L\right)^{3/2}  } \right) \nonumber ,\\
    F_{\pi ,3}(L)&=F_{\pi ,3}^\infty\left( 1+A_{F_{3}}\frac{e^{-M_{\pi ,3}^\infty L}}{\left( M_{\pi ,3}^\infty L\right)^{3/2}  } \right).
\end{align}
Here, $A_{M_{1}}$, $A_{M_{2}}$, $A_{F_{1}}$, and $A_{F_{3}}$ are treated as free parameters, while $M_{\pi,1}^\infty$, $M_{\pi,3}^\infty$, $F_{\pi,1}^\infty$, and $F_{\pi,3}^\infty$ denote the infinite-volume values. The functional dependence on the product $M_{\pi,3} L$, with $M_{\pi,3}$ corresponding to the lightest state according to the lattice data, is motivated by ChPT calculations \cite{Sharpe:2006pu}. The results are summarized in \tableTag\ref{tab:LECs_table_infinite_volume_limit}, while an illustrative example of the fit is displayed in \fig\ref{fig:InfVol}. % \dk{Do we comment on the quality of the fits}?

For several parameter points the data at smaller lattices were missing or the value of $\chi^2/$ndof for the fit of the volume dependence was too large (see the data with no value for $\chi^2/$ndof in \tableTag\ref{tab:LECs_table_infinite_volume_limit}). In such cases, we adopt the result from the largest available volume and inflate its uncertainty based on the consideration of the finite-volume effects in~\cite{Bennett:2019jzz}. This reference considered the $Sp(N_c=4)$ theory with two degenerate quark flavours and observed that for the lightest state and largest lattice extent satisfying $M_\pi^\infty L \sim 5$, the infinite volume masses deviated from those at the largest volume by 4-5\%, while for $M_\pi^\infty L \sim 6.3$ this deviation decreased to the level of 1-2\% and for $M_\pi^\infty L \gtrsim 7.5$ the largest observed deviation was $\sim 0.3$\%, i.e., smaller than the statistical errors. Further, \Refe\cite{Kulkarni:2022bvh} considered also parameter points with $M_{\pi,3}^\infty L \lesssim 5$ at the largest lattice and reported the shift at infinite volume by up to 10\%. Based on these results, we assign an additional 10\% systematic error to the values of the pion masses at the largest lattice for the data points $(\beta, am_u^0, am_d^0) = (7.2,-0.794,-0.78)$ and $(7.05, -0.85, -0.85)$ and an extra 2\% error in case of the point $(6.9, -0.92, -0.88)$. For the decay constants, we double this error since within the ChPT calculation, the leading volume correction is predicted to be twice as big compared to the contribution to masses~\cite{Sharpe:2006pu}.

\begin{table}[h]
\centering
\begin{tabular}{c c c c c c c c c c c}
\hline
$\beta$ & $am^0_u$ & $am^0_d$ & $r$  & $a^2(M_{\pi,1}^\infty)^2$   & $a^2(M_{\pi,3}^\infty)^2$ & $aF_\pi^\infty$ & $\chi^2/\text{ndof}$& \\
\hline
6.9  & -0.92 & -0.85  & 4.6(9) &  0.433(6)  &  0.396(6)      & 0.117(5) & 1.6 & \\
6.9  & -0.92 & - 0.88 & 2.9(7) &   0.301(16) & 0.291(14)     & 0.100(6) & - &\\ % 2\% 
6.9  &-0.92 & -0.9 & 2.0(5)&   0.229(9)   & 0.235(6)    & 0.093(3) & 0.6 &\\
6.9  & -0.92 & -0.91 & 1.5(5)&  0.194(9)  & 0.190(6)     & 0.091(4) & 2.0 &\\
6.9  & -0.92 & -0.92 & 1.0(5)&  0.135(8)  & 0.138(7)   & 0.083(4) & 0.4 &\\
7.05 & -0.85 & -0.75  & 5.1(7)&  0.387(6)  & 0.305(7)     & 0.095(8) & 1.0 &\\
7.05 & -0.85  &  -0.8 & 3.1(5)&  0.248(6)  & 0.224(6)      & 0.085(4) & 1.2 &\\
7.05 & -0.85 & -0.83 & 1.7(3)&   0.181(8) & 0.169(6)     & 0.080(3) & 0.6 &\\
7.05 & -0.85 & -0.85  & 1.0(3)& 0.129(27)   & 0.135(27)      & 0.069(14) & - & \\ %10\%
7.2  & -0.794 &  -0.7 & 4.4(6)&  0.272(5)  & 0.204(6)     & 0.076(5) & 1.4 &\\
7.2  & -0.794 & -0.75 & 2.7(4)&  0.172(3)   & 0.153(4)    & 0.071(2) & 1.8 &\\
7.2  & -0.794 & -0.77 & 1.7(4)&  0.123(6)   & 0.112(5)   & 0.066(2) & 3.4 &\\
7.2  & -0.794 & -0.78 & 1.4(3)&  0.105(22)   & 0.110(22)    &   0.061(12) & - &\\ %10\%
7.2  & -0.794 & -0.794 & 1.0(3)&  0.080(5)    & 0.079(5)   &   0.058(2) & 4.5 & \\
\hline
\end{tabular}
\caption{Infinite-volume extrapolation of the lattice data from ref.~\cite{Kulkarni:2022bvh}, in the region $M_{\rho}/M_{\pi} \approx 1.46$, corresponding to the ensembles listed in their table~7 (which reports the results for the largest available volumes for each $(\beta, a m_u^0, a m_d^0)$).
For ensembles where the infinite-volume fit is performed, we report the corresponding $\chi^2/\mathrm{dof}$. In the few cases where such a fit is not feasible, we assign conservative systematic uncertainties to the largest-volume value, see the text. We calculated the infinite-volume limits of the two species-dependent decay constants, $F_{\pi,1}$ and $F_{\pi,2}$ that are provided by \refe\cite{Kulkarni:2022bvh}; however, we present here their average, $F_\pi = (F_{\pi,1}+F_{\pi,2})/2$, that is used for the fits of LECs as discussed in the main text.}
\label{tab:LECs_table_infinite_volume_limit}
\end{table}

\begin{comment}
\begin{table}[h]
\centering
\begin{tabular}{c c c c c c c c }
\hline
$\beta$ &  $T$ &  $L$ &  $a_0M_{\pi}^\infty$ & $\Delta a_0M_{\pi}^\infty$ & $M_\pi^\infty/F_\pi$ & $\Delta M_\pi^\infty/F_\pi$\\
\hline
 6.9 & 32 & 24 &   0.523 &        0.228 &   4.658 &  0.050 \\
7.05 & 36 & 20 &   0.698 &        0.252 &   5.344 &         0.211 \\
7.05 & 36 & 24 &   0.553 &        0.240 &   4.814 &         0.088 \\
 7.2 & 36 & 24 &   0.886 &        0.085 &   5.728 &         0.261 \\
 7.2 & 36 & 28 &   0.620 &        0.110 &   5.018 &         0.055 \\
\hline
\end{tabular}
\caption{\dk{add caption. comment on what $M^\infty $ is. delete the column with the $am_0$ (done) }}
\label{tab:LECs_table2}
\end{table}
\end{comment}

Lattice data from \refe\cite{Dengler:2024maq} are in turn summarized in \tableTag\ref{tab:LECs_table4}. As mentioned in the main text, the values of the pion masses and scattering lengths already correspond to the infinite-volume limit, while for the pion decay constant we consider the value at the largest lattice. As can be seen from \tableTag\ref{tab:LECs_table4}, $M_\pi^\infty L > 7.5$ is always satisfied at the largest available lattice; therefore, we expect that the finite-volume corrections to the values of $F_\pi$ are smaller than the statistical errors.

\begin{table}[h]
\centering
\begin{tabular}{c c c c c c}
\hline
$\beta$ & $T$ & $L$ & $a M_{\pi}^\infty \times 10^4 $ &
$a_0M_{\pi}^\infty$ &
$M_\pi^\infty/F_\pi$ \\
\hline
6.9  & 32 & 24 & $3845^{+20}_{-29}$ & 0.52(23)  & 4.66(5) \\
7.05 & 36 & 20 & $4361^{+11}_{-11}$ & 0.70(25)  & 5.34(21)  \\
7.05 & 36 & 24 & $3298^{+12}_{-14}$ & 0.55(24)  & 4.81(9) \\
7.2  & 36 & 24 & $3675^{+8}_{-8}$ & 0.89(9) &  5.73(26)  \\
7.2  & 36 & 28 & $2852^{+4}_{-4}$ & 0.62(11) &  5.02(6) \\
\hline
\end{tabular}
\caption{
Lattice data determining the scattering length $a_0$ from \refe\cite{Dengler:2024maq}, corresponding to the bottom panel of their figure~3. $M_{\pi}^\infty$ is the pion mass in the infinite volume limit, $a$ is the lattice spacing an $T$ and $L$ correspond to temporal and spacial extents of the largest lattice considered.}
\label{tab:LECs_table4}
\end{table}

\section{Application: parameter space of SIMP dark matter}\label{AppendixSIMP}

In this appendix, we briefly introduce the model of SIMP dark matter~\cite{Hochberg:2014kqa,Hansen:2015yaa,Lee:2015gsa,Hochberg:2015vrg,Kamada:2017tsq,Hochberg:2018rjs,Hochberg:2018vdo,Berlin:2018tvf,Choi:2018iit,Katz:2020ywn,Kulkarni:2022bvh,Zierler:2022uez,Bernreuther:2023kcg,Braat:2023fhn,Pomper:2024otb} and then apply our results on dark matter self-interactions (see \sec\ref{sec:Sp4.Applications}) to this example scenario. The necessity and effect of NLO corrections for SIMP dark matter have been discussed already in~\cite{Hansen:2015yaa}, however, we refine these results due to our knowledge of the NLO LECs.

In the SIMP models, the relic abundance of dark matter is determined by the freeze-out mechanism, i.e., the dark matter particles are initially assumed to be in thermal equilibrium; however, as the Universe expands, the interaction rate becomes insufficient to further deplete the dark matter abundance and its comoving density is fixed. Within the traditional Weakly Interacting Massive Particles (WIMP) paradigm, the chemical equilibrium is kept via $2 \rightarrow 2$ annihilations into Standard Model particles, while for SIMPs this happens primarily through number-changing $3 \rightarrow 2$ self-interactions within the dark sector~\cite{Hochberg:2014dra}. In case of the SIMP scenario where dark pions play the role of dark matter~\cite{Hochberg:2014kqa}, the $3\to2$ interactions are naturally provided by the Wess-Zumino-Witten (WZW) term, a topological term in the effective action that encodes anomaly-induced interactions. The original work~\cite{Hochberg:2014kqa} showed that the observed relic abundance is reproduced and the constraints on dark-matter self-interactions are simultaneously satisfied only for $M_\pi/F_\pi$ close to the perturbative limit $4\pi$, unless $N_c$ is large ($\gtrsim 10$). However, the dark matter self-interactions were calculated within LO ChPT in~\cite{Hochberg:2014kqa} while the follow-up \refe\cite{Hansen:2015yaa} argued that since the WZW term is NLO in the chiral power-counting scheme and since the values of $M_\pi/F_\pi$ are large, the $2\to2$ self-scattering should also be considered at NLO in a consistent treatment. The higher-order corrections were shown to introduce significant deviations from the LO results that increased tension with phenomenological constraints. 

Let us note that it was later shown that vector mesons may play an important role for the dark pion freeze-out if $M_\pi/F_\pi$ is large~\cite{Berlin:2018tvf,Arthur:2016dir,Bernreuther:2023kcg}, in this sense, the original works~\cite{Hochberg:2014kqa,Hansen:2015yaa} are somewhat incomplete. However, we would still like to demonstrate the impact of the NLO corrections on dark matter phenomenology by redoing the analysis of~\cite{Hansen:2015yaa} with the values of NLO LECs from our fits. To this end, we briefly review the calculation of the SIMP relic abundance and show the constraints on the parameter space of the SIMP scenario based on $Sp(N_c=4)$ gauge group with $N_F=2$ degenerate flavors of dark quarks in fundamental representation.

%and also the values of $M_\pi/F_\pi$ are large, requiring the $2\to 2$ scattering to be treated at NLO precision. The resulting increase of the self-interaction cross section closed the SIMP parameter space for moderate $N_c$~\cite{Hansen:2015yaa}. 

%Although we explained in \sec\ref{sec:Sp4.Applications} that the original works~\cite{Hochberg:2014kqa,Hansen:2015yaa} were somewhat incomplete due to missing effect of the dark vector mesons, 

As mentioned above, the $3\to2$ pion interactions are provided by the WZW term~\cite{Wess:1971yu,Witten:1983tw,Witten:1983tx} that appears when the fifth homotopy group of the coset space is non-trivial ($\pi_5(G/H) = \mathbb{Z}$). The WZW action cannot be written as a four-dimensional integral but rather as a five-dimensional one~\cite{Hansen:2015yaa}:\footnote{We give here the WZW term with no external gauge fields. For a discussion of the gauged WZW term see \cite{Duan:2000dy} or \cite{Brauner:2018zwr}.}
\begin{equation}\label{eq:SIMP1}
S_{\text{WZW}} = \frac{\frac{1}{2}N_c}{240\pi^2} \int_0^1 d\alpha \int d^4x\, \epsilon^{abcde} \langle u_a u_b u_c u_d u_e \rangle,
\end{equation}
were $\alpha$ is the fifth space-time coordinate and $u_a$ is a five-dimensional analogue of \eqref{eq.EFT14}:
\begin{equation}\label{eq:SIMP1.a}
 \quad u_a = i(u^\dagger \partial_a u - u \partial_a u^\dagger), \quad u = \exp\left( \frac{i \alpha}{\sqrt{2} F_\pi} X^a \phi^a \right).
\end{equation}
\Eq\eqref{eq:SIMP1} differs from the corresponding formula in~\cite{Hansen:2015yaa}  by a factor of $1/2$ which should appear for the case of $Sp(N_c)$ (and $SO(N_c)$) gauge groups as shown in \cite{Kamada:2017tsq}.
Let us stress that the strength of the $3\to2$ interaction is controlled not only by the number of fermion flavors $N_F$, pion mass and decay constant, but also by the number of colors describing the underlying gauge symmetry $N_c$.

%Let us consider particle dark matter which consists of dark pions of the same mass $m_\pi$.  The dominant process which defines the relic abundance of dark matter is a $3\to2$ process due to WZW term \eqref{eq:SIMP1} and the evolution of the total dark pions density $n_\pi(t)$ is given by the Bolzmann equation \cite{Hochberg:2022jfs}:

As in~\cite{Hochberg:2022jfs}, we assume that the dark and visible sectors are thermalized, however, the coupling between these sectors is weak enough, so that the $3\to2$ process can dominate the number-changing interactions. The evolution of the total dark pion density $n_\pi(t)$ is then given by the following Boltzmann equation \cite{Hochberg:2022jfs}:
\begin{equation}\label{eqAppB1}
\dot{n}_\pi + 3H(t)n_\pi = -\left(n_\pi^3 - n_\pi^2 n_{\pi,\text{eq}}\right) \langle \sigma v^2 \rangle_{3 \to 2}\,.
\end{equation}
Here $H(t)$ is the Hubble parameter, $n_{\text{eq}}$ is the equilibrium particle density defined by the temperature of the thermal bath $T$ and the thermally averaged cross section reads
\begin{equation}\label{eqAppB3}
\langle \sigma v^2 \rangle_{3 \to 2} =\frac{1}{12}\,\frac{75 \sqrt{5}}{512 \pi^5 x^2} \frac{M_\pi^5 N_c^2}{F_\pi^{10} N_\pi^3},
\end{equation}
where $x=M_\pi/T$ and this formula differs by a factor of $1/12$ from~\cite{Hochberg:2014kqa,Hansen:2015yaa}. An additional factor of $1/3$ is due to improved kinematics considerations~\cite{Kamada:2022zwb} while the factor of $1/4$  is related to the correction in the WZW action~\eqref{eq:SIMP1}.

We assume that the freeze-out happens during the radiation domination, i.e.
\begin{equation}\label{eqAppB6} 
H(T)^{2}=\sqrt[]{\frac{8\pi }{3} } \frac{1}{M^{2}_{\text{Pl} }} \rho \left( T\right)  
\end{equation}
where $ M_{\text{Pl}}\approx 1.22\times 10^{19}\text{ GeV} $ is the Planck mass and the energy and entropy densities of the heat bath read
\begin{equation}\label{eqAppB7} 
\rho = \frac{\pi^2}{30} \, g_e(T) \, T^4, \quad s = \frac{2\pi^2}{45} \, h_e(T) \, T^3\,.
\end{equation}
Here $g_e(T)$ and $h_e(T)$ are the effective numbers of the relativistic degrees of
freedom contributing to the energy and entropy densities, respectively~\cite{Husdal:2016haj,Laine:2015kra}. Then the equation \eqref{eqAppB1} can be rewritten using the yield $Y=n_\pi/s$:
\begin{equation}\label{eqAppB8} 
\frac{dY}{dx} = -\sqrt{ \frac{4\pi^5  }{91125}   }M_{\text{Pl}}\sqrt{g_\ast}\frac{M_\pi^4 }{x^5} \left( Y^3 - Y^2 Y_{\text{eq}} \right) \langle \sigma v^2 \rangle_{3 \to 2},
\end{equation}
where  $g_\ast$ reads
\begin{equation}\label{eqAppB9} 
\sqrt{g_\ast}\equiv\frac{h^{2}_{e}(x)}{\sqrt[]{g_{e}(x)} } \left( 1-\frac{x}{3} \frac{d}{dx} \log h_{e}(x)\right) 
\end{equation}
and the non-relativistic equilibrium yield is given by \cite{Pereira:2024vdk}
\begin{equation}\label{eqAppB10} 
Y_{\text{eq}} = \frac{45 N_\pi x^2}{4\pi^4 h_e(M_\pi / x)}  K_2(x) ,
\end{equation}
where where $K_2(x)$ is the modified Bessel function of the second kind of order 2.  At early times, the yield  $Y$  closely follows its equilibrium value $Y_{\text{eq}}$, since interactions are frequent enough to maintain chemical equilibrium. As the universe expands and cools, the interaction rates drop, and the annihilation process becomes inefficient compared to the Hubble rate. Around  $x \sim 10-20$ , the system undergoes freeze-out, where  $Y$  departs from  $Y_{\text{eq}}$ and effectively becomes constant. This asymptotic value is required to agree with the relic abundance observed today $Y_\infty = 4.1\times10^{-10} \left( \text{GeV}/M_{\pi } \right)$. This fixes the value of $F_\pi$ in the numerical simulations for a given $M_\pi$.

\begin{figure}[t]
    \centering
    \includegraphics[width=0.6\textwidth]{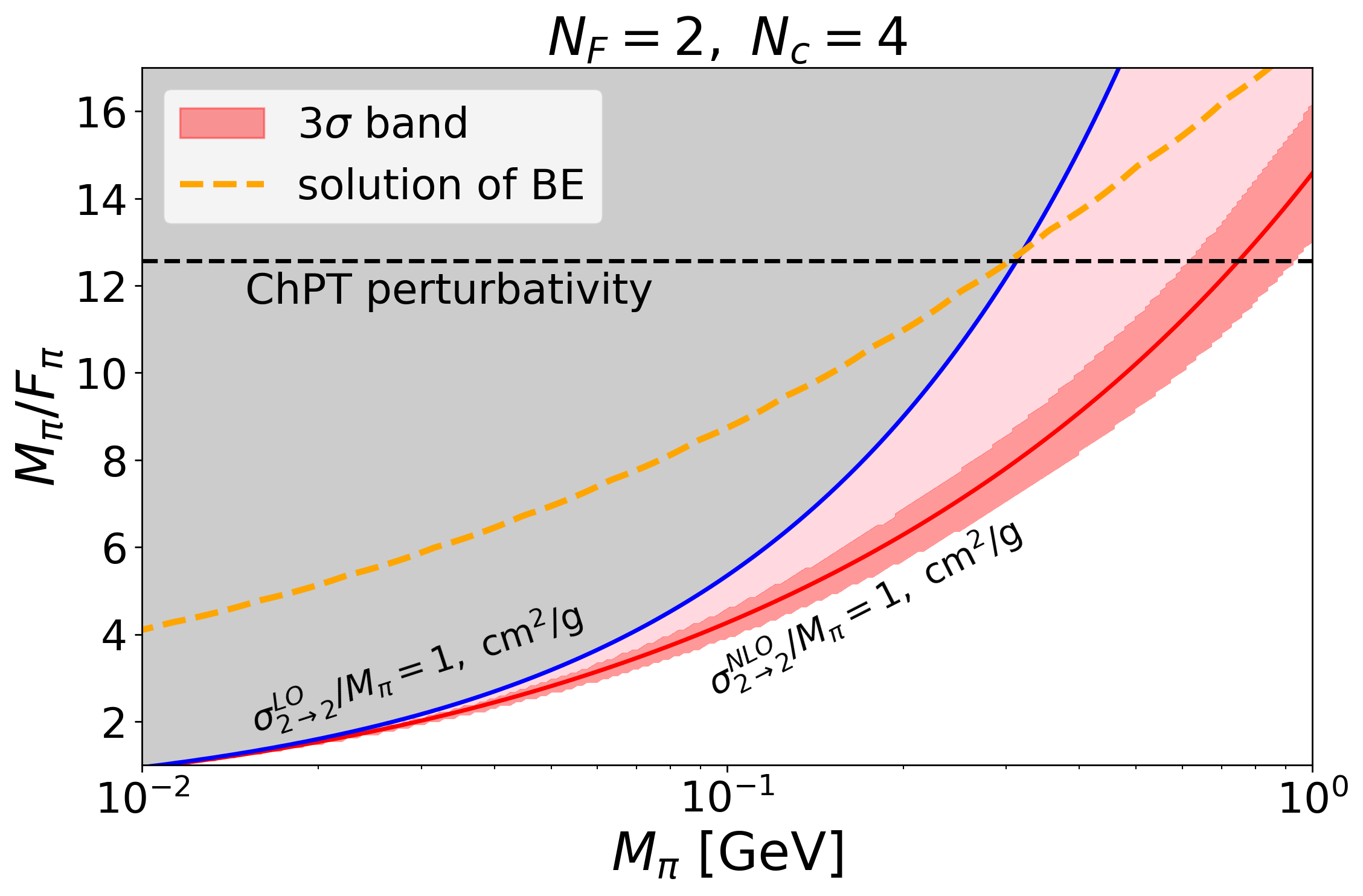} 
    \caption{Parameter space for SIMPs with LO and NLO self-scattering cross-sections. The orange line shows the required $M_\pi / F_\pi$ for a given $M_\pi$, as determined from the solution of the Boltzmann equation (BE). The value of $M_\pi / F_\pi$ must remain below $4\pi$; otherwise, the perturbativity of the EFT breaks down. In the grey (pink) regions, the LO (NLO) self-scattering cross-section exceeds the limits set by the Bullet Cluster constraints.}
    \label{fig:SIMPregions}
\end{figure}

Knowing the value of $F_\pi$ for a given $M_\pi$ one can calculate the self-scattering $2\to2$ cross section $\sigma_{2\to2}$ which is bounded by the Bullet cluster observations~\cite{Randall:2008ppe}. \Fig\ref{fig:SIMPregions} shows the resulting SIMP parameter space. The orange curve represents the value of $M_\pi/F_\pi$ for a given $M_\pi$ determined from solving the Boltzmann equation. The blue and red curves correspond to the Bullet Cluster constraints for the LO and NLO self-scattering cross-sections, respectively. Finally, for $M_\pi/F_\pi \geq 4\pi$, the use of ChPT is no longer reliable, hence, our calculation of the dark matter relic abundance is not valid. Due to the changes with respect to \refes\cite{Hochberg:2014kqa,Hansen:2015yaa} mentioned below \eqref{eqAppB3}, we observe that there is no open parameter space left already with LO formulas. On the other hand, if the bound on self-interaction cross section is slightly loosened, the LO formula could still allow for a small open parameter space whereas at NLO, no viable parameter space would remain even if only $\sigma_{2\to2}/M_\pi \lesssim 15 \ \text{cm}^2$/g is required.  We confirm the conclusions of \cite{Hansen:2015yaa} that NLO corrections have a significant impact on SIMP phenomenology and substantially reduce the viable parameter space.

Let us remark that at NNLO, further corrections to both $2\to2$ and $3\to 2$ scatterings appear which leads to small quantitative shifts but no qualitative changes. Ref.~\cite{Hansen:2015yaa} reports that for $N_c=2$, no parameter space remains open for both NLO and NNLO treatment, while for $N_c=16$ the open parameter space gets slightly smaller if NNLO corrections are taken into account.

As anticipated above, these results should be understood as an illustration of the impact of the NLO correction and this example was chosen to enable comparison with the previous work~\cite{Hansen:2015yaa}. In reality, vector mesons in the $Sp(N_c=4)$ theory with two degenerate quark flavors in fundamental representation are lighter than twice the pion mass for $M_\pi/F_\pi \gtrsim 4$ according to later lattice results of~\cite{Bennett:2019jzz}. This mass hierarchy makes further pion-number-changing processes efficient in the early Universe and the freeze-out curve in \fig\ref{fig:SIMPregions} will likely be correspondingly altered~\cite{SIMP_VM} similarly to the case of quarks in complex representations discussed in \refes\cite{Berlin:2018tvf,Arthur:2016dir,Bernreuther:2023kcg}. On the other hand, for example, in the secluded SIMP scenario of~\cite{Heikinheimo:2018esa}, correct relic abundance might also be obtained for $M_\pi/F_\pi \lesssim 4$ where the vector mesons do not bring any significant change.

\addcontentsline{toc}{section}{References}
\bibliographystyle{JHEP}
\bibliography{biblioNew}
\end{document}